\documentclass[3p,times]{elsarticle}

\usepackage{latexsym}

\usepackage{indentfirst}
\usepackage[english]{babel}
\usepackage{cancel}

\usepackage{lipsum}
\usepackage{array}
\usepackage{color} 
\usepackage{tabularx}
\usepackage{graphicx} 
\usepackage{amsmath}
\usepackage{amssymb}
\usepackage{amsfonts}	
\usepackage{moreverb}
\usepackage{dsfont}
\usepackage{tipa}
\usepackage{upgreek}
\usepackage{bm}
\usepackage{multirow}
\usepackage{soul}
\usepackage{ textcomp }
\usepackage[dvipsnames]{xcolor}
\usepackage{mathtools}
\usepackage{setspace}
\usepackage{algorithm}
\usepackage{algorithmic}
\usepackage{caption}
\usepackage{subcaption}
\allowdisplaybreaks 
\DeclareMathOperator\erf{erf}

\newcommand\BibTeX{{\rmfamily B\kern-.05em \textsc{i\kern-.025em b}\kern-.08em
T\kern-.1667em\lower.7ex\hbox{E}\kern-.125emX}}

\usepackage{hyperref} 
\hypersetup{
    colorlinks=true,                          
    linkcolor=blue, 
    citecolor=red, 
    urlcolor=blue  } 

\newcommand{\x}{\mbf{x}}

\newcommand{\csi}{\mbf{\xi}}

\newcommand{\mbf}[1]{\mathbf{#1}}			%

\newcommand{\Q}{\mathbf{Q}}
\renewcommand{\S}{\mathbf{S}}
\renewcommand{\u}{\mathbf{u}}

\newcommand{\q}{\mathbf{q}}
\newcommand{\F}{\mathbf{F}}
\newcommand{\f}{\mathbf{f}}

\newcommand{\B}{\mathbf{B}}

\newcommand{\A}{\AAA}

\newcommand{\halb}{\frac{1}{2}}

\newcommand{\bdm}{\begin{displaymath}}
\newcommand{\edm}{\end{displaymath}}

\newcommand{\bea}{\begin{eqnarray} }
\newcommand{\eea}{\end{eqnarray} }

\newcommand{\AAA}{{\boldsymbol{A}}}

\newcommand{\tu}{{^{^{(4)}}}\hspace{-0.5mm}}

\newfont{\numerikEleven}{ecrm1000}
\newfont{\numerikTen}{cmss10}
\newfont{\numerikNine}{cmss9}
\newfont{\numerikEight}{cmss8}

\journal{Journal of Computational Physics}

\begin{document} 
\begin{frontmatter}
\title{A well-balanced discontinuous Galerkin method for the first--order Z4 formulation of the Einstein--Euler system} 
\author[UniTN]{Michael Dumbser}
\ead{michael.dumbser@unitn.it}

\author[UniTN]{Olindo Zanotti \corref{cor1}}
\ead{olindo.zanotti@unitn.it}
\cortext[cor1]{Corresponding author} 

\author[INRIA]{Elena Gaburro}
\ead{elena.gaburro@inria.fr}

\author[UniTN]{Ilya Peshkov}
\ead{ilya.peshkov@unitn.it}

\address[UniTN]{Laboratory of Applied Mathematics, DICAM, University of Trento, via Mesiano 77, 38123 Trento, Italy} 
\address[INRIA]{INRIA, Univ. Bordeaux, CNRS, Bordeaux INP, IMB, UMR 5251, 200 Avenue de la Vieille Tour, 33405 Talence cedex, France}


\begin{abstract} \color[rgb]{0,0,0}
In this paper we develop a new well-balanced discontinuous Galerkin (DG) finite element scheme with subcell finite volume (FV) limiter for the 	numerical solution of the  Einstein--Euler equations of general relativity based on a first order hyperbolic reformulation of the Z4 formalism. The first order Z4 system, which is composed of 59 equations, is analysed and proven to be strongly hyperbolic for a general metric.  
The well-balancing is achieved for arbitrary but \textit{a priori} known equilibria by subtracting a discrete version of the equilibrium solution from the discretized time-dependent PDE system. Special care has also been taken in the design of the numerical viscosity so that the well-balancing property is achieved.    
As for the treatment of low density matter, e.g. when simulating massive compact objects like neutron stars surrounded by vacuum,  we have introduced a new filter in the conversion from the conserved to the primitive variables, preventing superluminal velocities when the density drops below a certain threshold, and being potentially also very useful for the numerical investigation of highly rarefied relativistic astrophysical flows.

Thanks to these improvements, all standard tests of numerical relativity are successfully reproduced, reaching three achievements: 
(i) we are able to obtain stable long term simulations
of stationary black holes, including Kerr black holes with extreme spin, which after an initial perturbation return perfectly back to the equilibrium solution up to machine precision;
(ii) a (standard) TOV star under perturbation is evolved in pure vacuum ($\rho=p=0$) up to $t=1000$ with no need to introduce any artificial atmosphere around the star; and, 
(iii) 
we solve the head on collision of two punctures black holes, that was previously considered un--tractable within the Z4 formalism.  

\textcolor{black}{Due to the above features, we consider that our new algorithm can be particularly beneficial for the numerical study of
quasi normal modes of oscillations, both of black holes and of neutron stars.}
\end{abstract}

\end{frontmatter}


%
\section{Introduction} 
\label{sec.introduction}
In spite of considerable progress made in the last two decades, the stable and accurate numerical solution of the Einstein field equations
still remains an extremely challenging task to be tackled. Among recent achievements, we highlight the results obtained in~\cite{Meringolo2021, Palenzuela2018, Mewes2020,ExaHype2020,Olivares2022,Tichy2023}. 
One of the primary obstacles for numerical discretization of the Einstein equations is the fact that these equations are not immediately well-posed in their original four-dimensional form, and a well-posed 3+1 formulation is required. On a mathematical ground, the well-posedness of a 3+1 formulation of the Einstein equations would be guaranteed if one could prove that such a system of time-dependent partial differential (PDE) equations is unconditionally symmetric hyperbolic~\cite{Choquet83,Reula98a,Choquet2009,Ripley2021}. However, there are a number of reasons which prevent from reaching a simple conclusion in this respect. 
First, the Einstein equations arise as nonlinear second order PDEs in the metric coefficients, and reducing them to a first--order system
from which the required mathematical properties can emerge more clearly is far from trivial~\cite{Gundlach:2005ta}. 
Second, the gauge freedom which is inherent to the Einstein equations is quite often 
a rather delicate issue, as it can substantially affect hyperbolicity~\cite{Friedrich85}. Finally, as the Einstein equations include a set of stationary nonlinear second order differential  
constraints which must be satisfied during the evolution, their proper treatment can also have important implications on the mathematical nature of the overall PDE system. 

Despite the symmetric hyperbolicity being necessary for strictly proving well-posedness of a given first--order PDE system, from 
the computational view point this condition might be slightly relaxed as it is well known that for stable numerical computations, 
in fact, a strongly hyperbolic formulation is usually enough. In particular, the first-order strongly hyperbolic 3+1 formulation 
used in this paper does not have an obvious symmetric hyperbolic reformulation, at least to the best of our knowledge. Yet, it provides the 
possibility to perform stable computations of the Einstein field equations. We note that several symmetric hyperbolic formulations of the Einstein's equations in 3+1 split are known~\cite{Friedrich85,Abrahams1995,Estabrook1997,Anderson1999,Buchman2003,Hilditch2012}, but  the applicability of most of these formulations in numerical general relativity (GR) has yet to be tested.

One can notice that, after the first detection of gravitational waves recorded in 2015~\cite{Abbott2016}, the vast majority of research groups 
performing numerical simulations of the gravitational signal from astrophysical sources have been adopting the so called 3+1 formalism~\cite{Alcubierre:2008} in its various formulations. 
Some representative examples include~\cite{Baiotti2017,Hanauske2017,Jimenez2017,Mewes2020,Nedora2021,Carmiletti2022,Reetika2022}. 
In many of these codes the amount of physical effects that are currently taken into account is really impressive (see~\cite{Baiotti2019} for a review).
The most popular and successful implementations using  the 3+1 foliation of spacetime include the 
BSSNOK (Baumgarte-Shapiro-Shibata-Nakamura-Oohara-Kojima) formulation
\cite{Shibata95,Baumgarte99,Nakamura87,Brown09}; the Z4 formulation of~\cite{Bona:2003fj,Bona:2003qn,Alic:2009}, which has the advantage of
incorporating the treatment of the Einstein constraints through the addition of a four vector $z^\mu$;
the Z4c formulation~\cite{Bernuzzi:2009ex}, which adds a conformal transformation to the metric; the CCZ4 formulation of~\cite{Alic:2011a,Alic2013}, where
suitable coefficients are added to  damp the violation of the Einstein constraints and it is particularly suitable for treating binary systems.
Finally, in recent work~\cite{Dumbser2017strongly,Dumbser2020GLM} a first--order version of CCZ4 was proposed, namely FO-CCZ4, which consists of a system of 59 equations,
it is strongly hyperbolic for a particular choice of gauges and it incorporates a curl-cleaning technique for the treatment of internal curl-free conditions.
As a proper mathematical formulation of the Einstein equations must
be accompanied by a good numerical scheme in order to obtain stable and accurate numerical simulations, in~\cite{Dumbser2017strongly,Dumbser2020GLM} 
a numerical scheme based on discontinuous Galerkin methods 
combined with finite volume subcell limiter~\cite{DGLimiter1} 
was used. 

In spite of their attractive features in terms of accuracy and scalability on parallel computers,
DG methods are far from  common in the relativistic framework. After the pioneering investigations 
of
\cite{DumbserZanotti,Radice2011}, and apart from a slightly better popularity
for treating relativistic flows in stationary spacetimes, with or without magnetic fields~\cite{Bugner2016,Kailiang2014,Duan2020,Deppe2022,Forrest2022}, their
usage in full numerical relativity remains rather limited, with only a few groups 
investing on them around the world~\cite{Teukolsky2015,Miller2017,Kidder2017,Hebert2018,Tichy2023}. While in the just mentioned works the time evolution
is performed via Runge--Kutta schemes at various orders, the approach followed by our group over the years has been to resort to ADER (arbitrary high order derivatives) 
schemes~\cite{Titarev2002,Titarev2005}, which incorporate the solution of a Generalized Riemann Problem (GRP) at the cell boundaries.
After the modern reformulation of ADER provided by~\cite{DumbserEnauxToro,dumbser2008unified}, where the 
approximate solution of the GRP is obtained by evolving the
data inside each cell through a local space-time discontinuous Galerkin predictor, ADER schemes have been successfully implemented to solve 
the relativistic hydrodynamics and magnetohydrodynamics equations in stationary spacetimes~\cite{Zanotti2015,Zanotti2015d,Zanotti2016,ADERGRMHD,Gaburro2021PNPMLimiter}.
With the present work, we resume our investigations in full numerical relativity with DG methods, by revisiting the original Z4 formulation of the Einstein equations,
which, as we clarify below, does not show any inconvenience with respect to the CCZ4 formulation and is significantly simpler.

In addition, when one performs numerical simulations of (nearly) stationary configurations, 
a crucial property that ought to be achieved is the ability 
to preserve equilibria exactly at the discrete level over long time scales. 
Indeed, this capability, besides guaranteeing long-time stable simulations of the equilibrium profiles themselves, 
allows to capture with increased accuracy small physical perturbations around them 
that otherwise would be hidden by spurious numerical oscillations. 
For instance, this is particularly relevant when studying normal modes of oscillations in relativistic astrophysical sources~\cite{Lockitch2003,friedman_stergioulas_2013}.
Thus, in this work we endow our high order finite volume and discontinuous Galerkin schemes with so-called {\emph{well-balanced}} (WB) techniques.
Such techniques were originally introduced in computational fluid dynamics for the shallow water equations, see e.g.~\cite{Bermudez1994,leveque1998balancing,gosse2001well,bouchut2004nonlinear,audusse2004fast,Castro2006,Noelle1,Noelle2}, 
and then
successfully employed for many different applications with a number of relevant
results over the last two decades~\cite{Castro2008,VMDansac,Gaburro2017CAFNonConf,Gaburro2018DiffuseInterface,arpaia2020well,castro2020well,Pimentel}.
In particular, there has been a major interest for well-balancing in astrophysical applications, 
starting from their use joint to the classical Newtonian Euler equations with gravity and more recently even for the MHD system, see for example~\cite{BottaKlein,Kapelli2014,KM15_630,Klingenberg2015,bermudez2016numerical, Gaburro2018MNRAS,Klingenberg2015,desveaux2016well,klingenberg2019arbitrary,thomann2019second,Thomann2020,Thomann2020b,grosheintz2019high,berberich2021high} and \cite{KlingenbergWBMHD1,BirkeBoscheri,HybridHexa1},  
to the more recent work of~\cite{Gaburro2021WBGR1D}, where WB has been applied for the first time to the general relativistic framework
allowing the (1D) numerical simulations of the coupled evolution of matter and spacetime for small perturbations of neutron star equilibrium configurations.  
In this work we propose a new, simple but rather efficient approach to obtain the well-balanced 
property inside an existing three-dimensional general purpose code for numerical general relativity 
that is based on finite volume and discontinuous Galerkin finite element schemes and which includes 
also adaptive mesh refinement (AMR) with time-accurate local time stepping (LTS), 
see~\cite{AMR3DCL,Zanotti2015,Dumbser2017strongly}. Our new kind of well-balancing can be easily 
applied even to very complex hyperbolic PDE systems, such as the Einstein field equations, for 
which the original WB algorithm of~\cite{Castro2006,Gaburro2021WBGR1D} becomes more cumbersome, in 
particular when combining DG and FV schemes inside a 3D AMR framework with LTS.  
\textcolor{black}{Since our work develops along different directions joining together various aspects concerning the formulation of the equations, the numerical scheme and potential astrophysical applications, we list here the major achievements attained in this paper.
\begin{enumerate}
	\item We provide a \textit{novel} first--order reformulation of the Einstein equations in their Z4 version, showing the \textit{hyperbolicity} of the resulting PDE system by
	the explicit computation of all the eigenvalues and eigenvectors for a general metric.
	\item We solve the \textit{full} Einstein--Euler equations written as a \textit{single} monolithic first order hyperbolic system applying the \textit{same} numerical scheme to all equations; the method employed in this paper is a very high order accurate and robust Discontinuous Galerkin (DG) scheme with adaptive mesh refinement (AMR), time-accurate local time stepping (LTS) and \textit{a posteriori} sub-cell finite volume limiter.
	\item We present a simple but at the same time very general \textit{well-balanced} version of the overall algorithm, capable of preserving any general but \textit{a priori} known equilibrium solution on arbitrarily long timescales. This opens the door to a wide field of potential applications in the numerical study of quasi normal modes of oscillations, both of black holes and of neutron stars.
	\item We propose a major improvement in the conversion from the conserved to the primitive variables (of the matter part) in the presence of \textit{vacuum}, which, at least in the simple case of an ideal gas equation of state, allows to treat physical regimes with
	$p=\rho=0$, thus avoiding any use of artificial low density atmospheres outside high density objects.
	\item We show that even the Z4 formulation of the Einstein equations, which does not contain a conformal factor in the spatial metric, can successfully treat \textit{binary black holes}, provided a ``non--shifting--shift" version of the Gamma driver is adopted and
	 a special filtering is applied to the metric terms, to avoid the formation of spikes.
\end{enumerate}
 }

The structure of the paper is the following: in Sect.~\ref{sec.Z4eqs} we present the original Z4 
formulation provided by~\cite{Bona:2003fj,Bona:2003qn,Bona-and-Palenzuela-Luque-2005:numrel-book} 
with only minor modifications. Sect.~\ref{sec.numerical} is devoted to the description of the new 
well-balanced ADER-DG scheme with subcell finite volume limiter, while Sect.~\ref{sec.tests} 
contains the results of our investigations. Finally, we conclude
our analysis in Sect.~\ref{sec.conclusions} with a few indications for further progresses.

Throughout this paper we assume a signature $(-,+,+,+)$ for the
spacetime metric and we will use Greek letters (running from $0$ to
$3$) for four-dimensional spacetime tensor components, while Latin
letters (running from $1$ to $3$)  for
three-dimensional spatial tensor components. Moreover, we adopt a geometrized system of units by 
setting $c=G=1$,
in such a way that the most convenient unit of lengths is $r_g=G\,M/c^2=M$.
We just recall that for a one solar mass black hole, this choice corresponds to $r_g=1.476\times 10^3\,{\text{m}}$ as a  unit of length 
and to $r_g=4.925\times 10^{-6}\,{\text{s}}$ as a  unit of time.

\section{Damped Z4 formulation of the Einstein equations}
\label{sec.Z4eqs}
%
\subsection{The 3+1 splitting of spacetime}
\label{sec:3+1}
According to the 3+1 formalism, the  spacetime can be foliated through $\Sigma_t=const$ hypersurfaces as
\begin{equation}
\label{eq:ds2_3p1}
ds^{2} = -(\alpha^{2}-\beta_{i}\beta^{i}) dt^{2}+ 
2 \beta_{i} dx^{i} dt + \gamma_{ij} dx^{i}dx^{j} \,,
\end{equation}
where $\alpha$ is the lapse, $\beta^i$ is the shift and $\gamma_{ij}$ is the metric of the three dimensional space, see~\cite{Alcubierre:2008,Rezzolla_book:2013,Baumgarte2010,Gourgoulhon2012} for an extended discussion.
An Eulerian observer is then introduced, with four velocity defined by $n^\mu=\frac{1}{\alpha}(1,-\beta^i)$ everywhere orthogonal to the hypersurface $\Sigma_t$, and with respect to whom
all physical quantities are measured. 
%
%
We recall that the original Z4 formulation of the Einstein equations was not meant to be restricted to the 3+1 formalism.
In fact, it   was
specifically devised by~\cite{Bona:2003fj,Bona:2003qn}
to hyperbolize the elliptic Einstein constraints in a general covariant framework, after introducing an additional quantity $z^\mu$ whose role
is analogous to the scalar $\Psi$ in the divergence cleaning approach of~\cite{MunzCleaning,Dedner:2002} for the Maxwell and magnetohydrodynamics equations.
On the other hand, the damped version of the Z4 formulation, first proposed by~\cite{Gundlach2005:constraint-damping}, 
was intrinsically  linked to the 3+1 framework, since it dragged the four vector $n^\mu$ directly into the Einstein equations, in combination with
two additional constant coefficients $\kappa_1$ and $\kappa_2$, which were introduced to allow for 
the damping of the four vector $z^\mu$ as it propagates constraint violations away. An alternative rigorous treatment of the constraints is obtained via so-called fully-constrained formulations, see e.g.~\cite{cordero2008mathematical,cordero2009improved} and references therein. 

Here we introduce a slightly different version with respect to~\cite{Gundlach2005:constraint-damping},  
where the coefficients $\kappa_1$ and $\kappa_2$ are never multiplied among each other and thus produce effects that are clearly separated. Hence the augmented Einstein equations with damped Z4 cleaning 
read
\begin{equation}
\label{Trento1-Z4-1}
G_{\mu\nu} + \nabla_\mu z_\nu + \nabla_\nu z_\mu  -  \nabla_\pi z^\pi g_{\mu\nu} - \kappa_1(n_\mu z_\nu + n_\nu z_\mu) - \kappa_2  n_\pi z^\pi g_{\mu\nu}
=8\pi T_{\mu\nu}\,,
\end{equation}
or, equivalently,
\begin{equation}
\label{Trento1-Z4-2}
\tu R_{\mu\nu} + \nabla_\mu z_\nu + \nabla_\nu z_\mu  - \kappa_1(n_\mu z_\nu + n_\nu z_\mu -   n_\pi z^\pi g_{\mu\nu}) + \kappa_2 n_\pi z^\pi g_{\mu\nu}  
=8\pi \left(T_{\mu\nu}-\frac{1}{2}T g_{\mu\nu}\right)\,,
\end{equation}
where $G_{\mu\nu}$ and $\tu R_{\mu\nu}$ are the Einstein and  the Ricci tensors\footnote{\textcolor{black}{In what follows we use the left superscript $^{(4)}$ 
to distinguish  between four-dimensional tensors and three dimensional ones, in those cases when confusion may arise (the Ricci and the Riemann tensor). 
Moreover, $\nabla_\mu$ denotes the four dimensional covariant derivative, while $D_\mu := 
\gamma^\nu_{\mu} \nabla_\nu = (n^\nu n_\mu + \delta^\nu_\mu)\nabla_\nu$ is used for the spatial 
covariant derivative. This is the same  convention of \cite{Alcubierre:2008,Rezzolla_book:2013}.}}, 
while $T^{\mu\nu}$ is the energy--momentum tensor of matter. 
In this paper we limit our attention to a perfect fluid with no magnetic fields, such that
\begin{equation}
T^{\mu\nu}=(e+p)u^\mu u^\nu + p g^{\mu\nu}= \rho h u^\mu u^\nu + p g^{\mu\nu}\,,
\end{equation}
with $e$, $p$, $\rho$ and $h$ being the energy density, the pressure, the rest mass density and the specific enthalpy, respectively, 
each of them measured in the comoving frame of the fluid with four velocity $u^\mu$.
We notice that the  wave equation for the four vector $z^\mu$ corresponding to~\eqref{Trento1-Z4-1} is 
\begin{equation}
\label{Trento1-Z4-3}
\nabla^\mu \nabla_\mu z_\nu + \tu R_{\mu\nu}z^\mu=\kappa_1 \nabla^\mu (n_\mu z_\nu + n_\nu z_\mu) + \kappa_2\nabla_\nu(n_\rho z^\rho)\,,
\end{equation}
which is obtained after taking the four divergence of~\eqref{Trento1-Z4-1}. 
Within the 3+1 decomposition, 
all vectors and tensors are split in their components parallel and perpendicular (or mixed, depending on the rank) to $n^\mu$. So, for instance, we have
\begin{eqnarray}
 u^{\,\mu} & = & W\, n^{\,\mu} + W\, v^{\,\mu}, 
\label{eq:u} \\ 
T^{\mu\nu} & = & S^{\mu\nu} + S^{\mu}n^{\nu}+ n^{\,\mu}S^{\nu} + E n^{\,\mu}n^{\nu},  
\label{eq:T} \\
z^\mu&=&\Theta n^\mu + Z^\mu,
\label{eq:Z} 
\end{eqnarray}
where $W=-u^\mu n_\mu=1/\sqrt{1-v^2}$ is the Lorentz factor of the fluid, 
$S_{\mu\nu} = \gamma^{\alpha}_{\mu} \, \gamma^{\beta}_{\nu} T_{\alpha\beta}$ 
is the spatial part of the energy--momentum tensor,
$S_{\mu}=-\gamma^{\alpha}_{\mu}\, n^\beta T_{\alpha\beta}$ is the momentum density, 
$\gamma^\mu_{\nu} = n^\mu n_\nu + \delta^\mu_\nu$ is the spatial projector tensor, 
$\delta^\mu_\nu$ is the Kronecker delta,
$E=n^\alpha \, n^\beta T_{\alpha\beta}$ is the energy density, $Z^\mu=\gamma^\mu_{ \nu}z^\nu$ 
is the purely spatial part of the four vector $z^\mu$
and $\Theta=-z^\mu n_\mu=\alpha z^0$, each of which is measured in the Eulerian observer frame.
In terms of the primitive variables they read
\begin{align}
S^{\mu\nu} &= \rho h W^2 v^\mu v^\nu + p \gamma^{\mu\nu}\,, \\
S^\mu &= \rho h W^2 v^\mu\,, \\
E &= \rho h W^2 - p \,.
\end{align}
There are also vectors and tensors which are intrinsically spatial, namely without any component along $n^\mu$, such as the four
acceleration of the Eulerian observer
\begin{equation}
a_\mu = n^\nu \nabla_\nu n_\mu = \gamma^{\nu}_{\mu}\nabla_\nu\ln\alpha=D_\mu\ln\alpha\,,
\end{equation}
or the {\emph {extrinsic curvature}} of the hypersurface $\Sigma_t$, a
symmetric tensor defined as 
\begin{equation}
K_{\mu\nu}=-\gamma^\alpha_\mu  \nabla_\alpha n_\nu=-\nabla_\mu n_\nu - n_\mu a_\nu\,,
\end{equation}
which plays a fundamental role as a dynamical set of quantities, representing the opposite of the (non--trace-free) shear tensor of the Eulerian four velocity $n^\mu$.
We notice that the purely spatial part of the Ricci tensor $R_{\mu\nu}$ is
not simply given by the full spatial projection of the four dimensional Ricci tensor $\tu R^{\mu\nu}$, but rather is obtained from the 
so-called {\emph {contracted Gauss relations}}, i.e.
\begin{equation}
R_{\mu\nu}=\gamma^{\alpha}_{\mu}\gamma^{\beta}_{\nu}\,\, \tu R_{\alpha\beta} + \gamma^{\alpha}_{\mu}\gamma^{\beta}_{\nu}n^\sigma n^\pi \,\, \tu R_{\alpha\sigma\beta\pi} - K K_{\mu\nu}   +  K_{\mu\pi} K^\pi_{\,\,\nu}\,,
\end{equation}  
where  $K=\gamma^{ij}K_{ij}=-\nabla_\mu n^\mu$ is the trace of the extrinsic curvature, also equal to the opposite of the Eulerian observer expansion.
\subsection{The second order Z4 system}
\label{sec:second}

The second order PDE system that governs the evolution of the gravitational field in the presence of matter
is given by (see also ~\cite{Bona:2003fj} for a comparison)
\begin{eqnarray}
\label{dtgamma}
  (\partial_t -{\cal L}_{\beta})~ \gamma_{ij}
  &=& - {2\,\alpha}\,K_{ij}
\\
\nonumber
   (\partial_t - {\cal L}_{\beta})~K_{ij} &=& -D_iD_j\alpha
    + \alpha\,   \bigg[R_{ij}
    + D_i Z_j+D_j Z_i  -\kappa_1\Theta\gamma_{ij} - \kappa_2\Theta\gamma_{ij} - 2\Theta K_{ij}
\\
\label{dtK}
    &-& 2K_{im}K^m_j + K\,K_{ij}
    - 8\pi \left(S_{ij}-\frac{1}{2}\,T\,\gamma_{ij}\right)\,\bigg]
\\
\label{dtTheta} (\partial_t -{\cal L}_{\beta})~\Theta &=&
\frac{\alpha}{2}\,
 e^2 \left[R + K^2 - K_{ij}\,K^{ij} -16\pi \,E\right] + \alpha
  \left[D_k Z^k 
  -  Z^k \frac{D_k \alpha}{\alpha} -\Theta( 2\kappa_1 + \kappa_2) - K\Theta  \right]
\\
\label{dtZ}
 (\partial_t -{\cal L}_{\beta})~Z_i &=& \alpha\, \left[D_j\,{K_i}^j
  -D_i \, K - 8\pi S_i\right] + \alpha[\partial_i \Theta
 - 2\, {K_i}^j\, Z_j  -  \Theta\, D_i\ln\alpha
-\kappa_1 Z_i]\,.
\end{eqnarray}
Furthermore, we stress the following facts about each of the above equations\footnote{While deriving the equations~\eqref{dtK}--\eqref{dtTheta} 
one uses the fact $\gamma^{\alpha}_{\mu}\gamma^{\beta}_{\nu}\nabla_\alpha z_\beta=-\Theta K_{\mu\nu}+D_\mu Z_\nu$\, 
and $\nabla_\mu z^\mu=-K\Theta - n^\mu n^\nu \nabla_\mu z_\nu + D_\mu Z^\mu=-K\Theta +n^\mu\partial_\mu\Theta + Z_\mu a^\mu + D_\mu Z^\mu$.}. 
Eq.~\eqref{dtgamma} is 
a pure relation coming from differential geometry and which can be derived without any reference to the Einstein field equations. It states that the dynamics of the spatial metric tensor
$\gamma_{ij}$ is determined by the extrinsic curvature.
 Eq.~\eqref{dtK} is obtained after inserting the four dimensional
Ricci tensor as given by the Einstein equation~\eqref{Trento1-Z4-2} into the so--called {\emph{ Ricci equation}} of differential geometry (see~\cite{Baumgarte2010,Rezzolla_book:2013} for an extended discussion). 
Finally, Eqs.~\eqref{dtTheta}--\eqref{dtZ} are the evolutionary version of the Einstein constraints within the Z4 formalism and
are obtained after contracting the Einstein equations~\eqref{Trento1-Z4-2} with $n^\mu n^\nu$ and 
$n^\mu\gamma^\nu_\pi$, respectively. 
In fact, the Hamiltonian constraint $H$ and the momentum constraints $M_i$, defined as
\begin{eqnarray}
\label{eqn.adm1}
H &=& R - K_{ij} K^{ij} + K^2 - 16\pi E\,, \\
\label{eqn.adm2}
M_i &=& \gamma^{jl} \left( \partial_l
K_{ij} - \partial_i K_{jl} - \Gamma^m_{jl} K_{mi} + \Gamma^m_{ji} K_{ml} 
\right)- 8 \pi S_i\,,
\end{eqnarray}
can be recognized on the right hand side of~\eqref{dtTheta} and~\eqref{dtZ}.
Assumed to be zero for proper initial data of the Einstein equations, on the discrete level such quantities can in fact increase, and the whole strategy of the Z4 approach 
is to keep their dynamics under control by transporting the numerical errors away from the 
computational domain at the velocity $e$, which is the so-called \textit{cleaning speed}.

\subsection{The first--order Z4 system with matter}
\label{sec:first}
Similarly to the standard approach of~\cite{Bona:2003fj,Bona:2003qn,Dumbser2017strongly}, we introduce 30 auxiliary variables involving first derivatives of the metric terms, namely
\begin{align}
\label{eq:Auxiliary}
A_i := \partial_i\ln\alpha = \frac{\partial_i \alpha }{\alpha}\,, \qquad
B_k^{\,\,i} := \partial_k\beta^i\,,
\qquad
D_{kij} := \frac{1}{2}\partial_k\gamma_{ij}\,. \qquad
\end{align}
In addition, we 
list the following expressions and identities, clarifying how second order spatial derivatives can be removed:
\begin{eqnarray}
{\gamma} &=& \textnormal{det}( {\gamma}_{ij} )\,, \\
\partial_k {\gamma}^{ij} & = &  - 2 {\gamma}^{in} {\gamma}^{mj} D_{knm}\,, \\
%
\Gamma_{ij}^k &=& {\gamma}^{kl} \left( D_{ijl} + D_{jil} - D_{lij} \right)\,, \\
\label{eqn.dchr}
\partial_k \Gamma_{ij}^m & = &   - 2 {\gamma}^{mn} {\gamma}^{pl} D_{knp}  \left( D_{ijl} + D_{jil} - D_{lij} \right)
												             + {\gamma}^{ml} \left( \partial_{(k} {D}_{i)jl} + \partial_{(k} {D}_{j)il} - \partial_{(k} {D}_{l)ij} \right) , \\ %
 R^m_{\,\,ikj} & = & \partial_k \Gamma^m_{ij} - \partial_j \Gamma^m_{ik} + \Gamma^m_{lk}\, \Gamma^l_{ij}  - \Gamma^m_{lj}\,\Gamma^l_{ik} , \\
 R_{ij} & = &   R^k_{\,\,ikj}=\partial_k \Gamma^k_{ij} - \partial_j \Gamma^k_{ik} + \Gamma^k_{lk} \,\Gamma^l_{ij}  - \Gamma^k_{lj}\,\Gamma^l_{ik} \,, \\
 R & = &  \gamma^{ij}\, R_{ij}\,, \\
D_i D_j \alpha &=& \alpha A_i A_j - \alpha\, \Gamma^k_{ij} A_k + \alpha \partial_{(i} {A}_{j)}\,, \\
{\Gamma}^i & = &  {\gamma}^{jk}\, {\Gamma}^i_{jk}\,,  \\
\partial_k {\Gamma}^i & = & -2 D_k^{jl} \, \Gamma^i_{jl} + {\gamma}^{jl} \, \partial_k \Gamma^i_{jl}\,. 
\end{eqnarray}
Having done that,  we can rephrase the system~\eqref{dtgamma}--\eqref{dtZ}  as a first--order system, augmented by the
matter part (see~\cite{DelZanna2007} for details). The full Z4 Einstein-Euler system is therefore given by
%
\begin{align}
\label{eqn.rho}
&\partial_t (\sqrt{\gamma}D)+\partial_i\left[\sqrt{\gamma}(\alpha v^i D - \beta^i D)\right]=0\,,\\ 
\label{eqn.S}
&\partial_t (\sqrt{\gamma}S_j)+\partial_i\left[\sqrt{\gamma}(\alpha S^i_{\,\,j} - \beta^i S_j)\right]=\sqrt{\gamma}\left[\alpha S^{ik}D_{jik}  +
S_i  B_j^{\,\,i} - \alpha E A_j   \right]\,,\\ 
\label{eqn.E}
&\partial_t (\sqrt{\gamma}E)+\partial_i\left[\sqrt{\gamma}(\alpha S^i - \beta^i E)\right]=\sqrt{\gamma}\left[\alpha S^{ij}K_{ij} - \alpha S^j A_j  \right]\,,\\ 
\label{eqn.gamma}
&\partial_t\gamma_{ij} - \beta^k\partial_k\gamma_{ij}=\gamma_{ik} B_{j}^{\,\,k}  + \gamma_{kj} B_{i}^{\,\,k} - 2\alpha K_{ij}\,,\\
\label{eqn.Kij}
&\partial_t K _{ij} - \beta^k \partial_k K_{ij}  + \alpha \partial_{(i} {A}_{j)}
- \alpha {\gamma}^{kl} \left( \partial_{(k} {D}_{i)jl}  - \partial_{(k} {D}_{l)ij} \right)
+ \alpha {\gamma}^{kl} \left( \partial_{(j} {D}_{i)kl}  - \partial_{(j} {D}_{l)ik} \right)
- 2\alpha  \partial_{(i} {Z}_{j)}
 =  K_{ki} B_j^{\,\,k} + K_{kj} B_i^{\,\,k}  \nonumber \\
&- \alpha A_i A_j + \alpha \Gamma^k_{ij} A_k +  
\alpha\bigg[  - 2 {\gamma}^{kn} {\gamma}^{pl} D_{knp}  \left( D_{ijl} + D_{jil} - D_{lij} \right)   
              + 2 {\gamma}^{kn} {\gamma}^{pl} D_{jnp}  \left( D_{ikl} + D_{kil} - D_{lik} \right) \nonumber \\
& + \Gamma^m_{lm} \Gamma^l_{ij} - \Gamma^m_{lj} \Gamma^l_{im}     \bigg] 
- 2\alpha\Gamma^k_{ij} Z_k     
-  \alpha\Theta\gamma_{ij}(\kappa_1 + \kappa_2)  
- 2 \alpha K_{il} \gamma^{lm} K_{mj} +  \alpha K_{ij}(K - 2 \Theta  ) \nonumber \\
& - 8\pi\alpha \left(S_{ij}-\frac{1}{2}\,T\,\gamma_{ij}\right)  \,, \\
\label{eqn.theta}
&\partial_t \Theta  - \beta^k\partial_k\Theta  -  \frac{1}{2}\alpha {e^2} \left[ \gamma^{ij} {\gamma}^{kl} \left( \partial_{(k} {D}_{i)jl}  - \partial_{(k} {D}_{l)ij} \right) - \gamma^{ij}{\gamma}^{kl} \left( \partial_{(j} {D}_{i)kl}  - \partial_{(j} {D}_{l)ik} \right) + 2 \gamma^{ij}\partial_{i} Z_{j}\right]
= \nonumber \\
& = \frac{\alpha}{2}\, e^2 \, 
 \bigg[- 2 \gamma^{ij}{\gamma}^{kn} {\gamma}^{pl} D_{knp}  \left( D_{ijl} + D_{jil} - D_{lij} \right)   
       + 2 \gamma^{ij}{\gamma}^{kn} {\gamma}^{pl} D_{jnp}  \left( D_{ikl} + D_{kil} - D_{lik} \right)\nonumber \\
&+\gamma^{ij}\left( \Gamma^m_{lm} \Gamma^l_{ij} - \Gamma^m_{lj} \Gamma^l_{im}   \right) + K^2 - K_{ij}\,K^{ij} -16\pi \,E\bigg] + 
\alpha\left[ -\gamma^{ij}\,\Gamma^k_{ij}Z_k 
  - Z^k A_k   \right] \nonumber  \\
&		- \alpha \Theta K  
				- \alpha\Theta(2\kappa_1+ \kappa_2)\,,  \\
\label{eqn.zeta}
&\partial_t Z_i  - \beta^k\partial_k Z_i - \alpha \partial_i\Theta -\alpha\left[ \gamma^{jm}\partial_jK_{mi} - \gamma^{mn}\partial_i K_{mn}  \right]
  = Z_k\,B_i^{\,\,k}  +  \alpha\, \bigg[-\gamma^{jm}(\Gamma^n_{jm} K_{ni}+\Gamma^n_{ji} K_{mn}) \nonumber \\
	&+\gamma^{mn}(\Gamma^l_{im} K_{ln}+\Gamma^l_{in} K_{ml}) - 8\pi S_i\bigg] + \alpha[ - 2\, {K_i}^j\, Z_j  -  \Theta\, A_i
-\kappa_1 Z_i]\,,
%
\end{align}
where we have written the principal part of the PDEs on the left hand side, while moving all algebraic source terms to the right.
In addition to the system~\eqref{eqn.rho}--\eqref{eqn.zeta}, we need to adopt specific gauge conditions, which we choose 
in the following way. For the lapse, we assume the standard form~\cite{Baumgarte2010}
\begin{equation}
\label{eqn.slicing}
\partial_t \ln\alpha - \beta^k{\partial_k\ln\alpha} = -g(\alpha)\alpha(K-K_0-2c\Theta)\,,
\end{equation}
which gives us the possibility to switch among 
the \emph{1+log gauge condition}, setting $g(\alpha)=2/\alpha$, and the
\emph{harmonic gauge condition}, setting $g(\alpha)=1$.
For the shift, on the other hand, we use
the \emph{gamma--driver} condition in those cases when the evolution of the shift is needed, and in particular
we adopt the so-called ``non--shifting--shift" version of~\cite{Faber2007}
\begin{eqnarray}
\label{g-driver1}
\partial_t \beta^i   &=& \frac{3}{4}b^i, \\
\label{g-driver2}
\partial_t b^i&=&\partial_t \hat\Gamma^i - \eta b^i\,,
\end{eqnarray}
where $\hat\Gamma^i=\Gamma^i + 2 \gamma^{ij}Z_j$ and $\Gamma^i=\gamma^{jk}\,\Gamma^i_{jk}$.
Note that the quantities $\hat\Gamma^i$ are not primary variables, and their time evolution can be deduced from the other
dynamical variables as specified below.
From the gauge conditions~\eqref{eqn.slicing}--\eqref{g-driver1} we can then obtain 
the PDEs for the auxiliary variables, namely
%
\begin{align}
\label{eqn.A}
& \partial_t A_{i} - {\beta^k \partial_k A_i} + \alpha g(\alpha) \left( \gamma^{mn}\partial_i K_{mn} - \partial_i K_0 - 2c \partial_i \Theta \right)
=-\alpha A_i \left( K - K_0 - 2 \Theta c \right) \left( g(\alpha) + \alpha g^\prime(\alpha)  \right) + \nonumber \\
& \textcolor{black}{+2\alpha g(\alpha) K^{jk}D_{ijk}}  +   B_i^{\,\,k} ~A_{k} \,,
\\
\label{eqn.B}
& \partial_t B_k^{\,\,i}  - s\left(  \frac{3}{4} \partial_k b^i  
- \alpha^2 \mu \, \gamma^{ij} \gamma^{nl} \left( \partial_k D_{ljn} - \partial_l D_{kjn} \right)  \right)
=		0 \,,
\\
\label{eqn.D}
&
\partial_t D_{kij} - {\beta^l \partial_l D_{kij}} 
         - \frac{1}{2} {\gamma}_{mi} \partial_{(k} {B}_{j)}^m
         - \frac{1}{2} {\gamma}_{mj} \partial_{(k} {B}_{i)}^m
				 +  \alpha \partial_k {K}_{ij} = B_k^{\,\,m} D_{mij} + B_j^{\,\,m} D_{kmi} + B_i^{\,\,m} D_{kmj}  \nonumber \\ 
&				- \alpha A_k  {K}_{ij}\,.
\end{align}
The following aspects ought to be emphasized about the whole system~\eqref{eqn.rho}--\eqref{eqn.D}
\begin{itemize}
\item The first five equations for the evolution of matter  are in conservative form, while the rest of the equations are in non conservative form.
\item The quantities $\hat\Gamma^i$ in Eq.~\eqref{g-driver2} are not primary variables. Their evolution in time is obtained from
\begin{eqnarray}
\partial_t \hat\Gamma^i=\Gamma^i_{jk}\,\,\partial_t \gamma^{jk} + \gamma^{jk}\,\,\partial_t \Gamma^i_{jk} + 2\left( Z_j\partial_t \gamma^{ij} + \gamma^{ij}\partial_t Z_j \right)\,,
\end{eqnarray}
which involve time derivatives of the already existing dynamical variables. In fact, we can write
\begin{eqnarray}
\partial_t\gamma^{ij}&=&-\gamma^{in}\gamma^{jm}\partial_t\gamma_{nm}\nonumber \\
&=&-2\gamma^{in}\gamma^{jm}\beta^k\,D_{knm}- \gamma^{jk}B_{k}^{\,\,i}- \gamma^{ik}B_{k}^{\,\,j}+2\alpha\gamma^{in}\gamma^{jm}K_{nm},\\
\partial_t \Gamma^i_{jk}&=&\partial_t \gamma^{im}\left(D_{jmk}+D_{kjm}-D_{mjk} \right)+\gamma^{im}\left(\partial_t D_{jmk}+\partial_t D_{kjm}-\partial_t D_{mjk}  \right)\nonumber \\
&=&\gamma^{im}\beta^r\left[\partial_r D_{jmk}+\partial_r D_{kjm}-\partial_r D_{mjk}\right] +\partial_{(j}B_{k)}^{\,\,i}-\alpha\gamma^{im}\left(\partial_j K_{mk}+\partial_k K_{jm}-\partial_mK_{jk}\right)+\nonumber \\
&&+\gamma^{im}\big[ D_{jmn}B_{k}^{\,\,n} + D_{nmk}B_{j}^{\,\,n} + D_{knm}B_{j}^{\,\,n} + D_{njm}B_{k}^{\,\,n} - D_{mjn}B_{k}^{\,\,n} - D_{mnk}B_{j}^{\,\,n} \big] \nonumber \\
&&-\alpha \gamma^{im}\left( A_j K_{mk} + A_k K_{jm} - A_m K_{jk} \right) + \nonumber \\
&&+\big[ -2\gamma^{ip}\gamma^{mq}\beta^r D_{rpq} - \gamma^{mr}B_r^{\,\,i} + 2\alpha \gamma^{ip}\gamma^{mq}K_{pq}  \big]\left( D_{jmk} + D_{kjm} - D_{mjk}  \right).
\end{eqnarray}
\item The binary parameter $s$ in Eq.~\eqref{eqn.B}, either 1 or 0, is introduced to switch the \emph{gamma--driver} on or off, depending on the test
being considered.
\end{itemize}
The equations~\eqref{eqn.rho}--\eqref{eqn.D} above
form a non-conservative first-order hyperbolic system, namely they can be written as
\begin{equation}
\label{eqn.pde.mat.preview}
\frac{\partial \u }{\partial t} + \frac{\partial \f_i(\u)}{\partial x_i} +
\B_i(\u) \frac{\partial \u}{\partial x_i} 
 = \S(\u),
 \qquad 
 \textnormal{or, equivalently,}
 \qquad 
\frac{\partial \u }{\partial t} + \nabla \cdot \F(\u) +
\B(\u) \cdot \nabla \u  
= \S(\u),
\end{equation}
where $\u$ is the state vector, composed of 59 dynamical variables\footnote{More specifically, 5 for the matter part, 10 for the lapse, the shift vector and the metric components, 6 for $K_{ij}$, 4 for the $z^{\mu}$ four vector, 3 for $A_i$, 9 for $B_i^{\,\,j}$, 18 for $D_{ijk}$, 1 for $K_0$ and 3 for $b^i$.},
${\bf F}(\u)=(\f_1(\u), \f_2(\u), \f_3(\u))$ is the flux tensor for the conservative (hydrodynamic) part of the PDE system,
while ${\B}(\u)=\left( \B_1(\u),\B_2(\u),\B_3(\u) \right)$ represents the non-conservative part of the system, essentially all of the Einstein sector.
Finally, $\bf S(\u)$ is the source term, which contains  algebraic terms only.
When written in pure quasilinear form, the system (\ref{eqn.pde.mat.preview}) becomes
\begin{equation}	
\label{Csyst}
\frac{\partial \u}{\partial t}+ \A_i(\u) \frac{\partial \u}{\partial x_i}  = \textbf{S}(\u)\,,
\end{equation}
where the matrix ${\bf A}_i(\u)=\partial {\bf f}_i(\u)/\partial \u+{\bf B}_i(\u)$ contains both the conservative and the non-conservative contributions.  
Sect.~\ref{sec.numerical} below describes the numerical methods adopted to solve such a system of equations.

\subsection{Hyperbolicity of the first order Z4 system}
Even before the Z4 formalism was introduced, in~\cite{Bona97a} the hyperbolic nature of the first--order conservative  formulation of the Einstein field equations was highlighted. It was subsequently confirmed after the introduction of the Z4 approach~\cite{Bona:2003fj}.  
However, our analysis differs from theirs, since our system~\eqref{eqn.rho}--\eqref{eqn.D} is written 
in non--conservative form.
In the context of the CCZ4 formulation
\cite{Dumbser2018conformal,Dumbser2017strongly}, we have already emphasized that the hyperbolicity of a system like
\eqref{Csyst} is favoured if one makes the \textit{maximum possible use} of the auxiliary
variables defined in Eq.~\eqref{eq:Auxiliary}. 
In other words, our first-order Z4
system does \textit{not} contain \textit{any} spatial derivatives of
$\alpha$, $\beta^i$, ${\gamma}_{ij}$, which  have been moved to the purely algebraic source term
$\boldsymbol{S}(\u)$ precisely by using the auxiliary quantities defined in~\eqref{eq:Auxiliary}. We have verified the hyperbolicity of the 
subsystem~\eqref{eqn.gamma}--\eqref{eqn.D} governing the space-time evolution by computing the eigenvalues and the corresponding eigenvectors through
the symbolic mathematical software Maple\footnote{See \url{https://maplesoft.com/}}.
The results for a general metric are reported in ~\ref{sec:eigenappendix}.

\section{The numerical scheme}
\label{sec.numerical}


\subsection{A well-balanced ADER-DG scheme for non conservative systems}
\label{sec.WBADERDG}

For problems where a stationary equilibrium solution needs to be maintained in time, the well-balancing properties of a numerical scheme 
can play a major difference. Such techniques were first introduced for the shallow water equations in~\cite{Bermudez1994,leveque1998balancing,GNVC00,gosse2001well,audusse2004fast,Castro2006,Castro2008,Pares2006} and further developed over the years
with a number of significant contributions, see~\cite{castro2020well} and references therein. Later, the concept of well-balancing was also extended to the Newtonian Euler equations with gravity, see e.g.~\cite{BottaKlein,Kapelli2014,KM15_630,Klingenberg2015,bermudez2016numerical, Gaburro2018MNRAS,Klingenberg2015,desveaux2016well,klingenberg2019arbitrary,Thomann2020,grosheintz2019high}.  
The resulting numerical schemes are able to remove the discretization errors from the equilibrium solution, while focusing on the development of real physical perturbations
that may act on a system.
A well-balanced scheme for the numerical solution of the Einstein equations was
first proposed by~\cite{Gaburro2021WBGR1D}, who showed that, if an initial perturbation is introduced in a stationary solution, only the well-balanced 
algorithm is able to recover the shape of the equilibrium over long timescales. On the contrary, the solution obtained through a not well-balanced scheme
will be significantly deteriorated. 

Unfortunately, the extension to three space dimensions and to adaptive mesh refinement (AMR) with time-accurate local time stepping (LTS) of the well-balanced scheme presented by~\cite{Gaburro2021WBGR1D} is quite cumbersome, as the scheme essentially relies on the incorporation of well-balanced reconstruction operators. Therefore,
we propose here an alternative approach which is conceptually much simpler, yet extremely effective. 
{\color{black}
The obtained method, presented here below, is exactly well-balanced for \textit{any} equilibrium solution that is known \textit{a priori}, exactly or in a discrete way. 
Thus, the equilibrium can be given in a closed analytical form, but it may also be just a numerical equilibrium, 
as it is for example in the TOV star test case, presented in Sect.~\ref{sec:tov}, 
where the equilibrium solution has been obtained by solving an ODE system in radial direction with a high order accurate numerical method.  
From the point of view of preserved equilibria this is for example the same context of \cite{Gaburro2021WBGR1D} and \cite{berberich2021high}, the latter being similar also for the structure of the proposed well-balanced methodology. 
}

In the following we use $\u^e=\u^e(\x)$ to denote a general stationary equilibrium solution, for 
which we know that 
\begin{equation}
	\partial_t \u^e=0.
	\label{eqn.stationary}
\end{equation}	
Hence, as a consequence, the equilibrium solution $\u^e$ must satisfy the stationary PDE system
\begin{equation}
	\label{eqn.equi}
	 \frac{\partial \f_i(\u^e)}{\partial x_i} +
	\B_i(\u^e) \frac{\partial \u}{\partial x_i} 
	= \S(\u^e),
	\qquad
	\textnormal{or, equivalently,}
	\qquad 
	\A_i(\u^e) \frac{\partial \u}{\partial x_i} 
	= \S(\u^e).
\end{equation}
Since we can always subtract~\eqref{eqn.equi} from the governing PDE~\eqref{eqn.pde.mat.preview} we obtain 
\begin{equation}	
\label{Csyst-wb}
\frac{\partial \u }{\partial t} + \frac{\partial \f_i(\u)}{\partial x_i} 
- 
\frac{\partial \f_i(\u^e)}{\partial x_i}
+
\B_i(\u) \frac{\partial \u}{\partial x_i} 
-
\B_i(\u^e) \frac{\partial \u^e}{\partial x_i} 
= \S(\u) - \S(\u^e). 
\end{equation}
Having done that, we create an extended vector of quantities $\tilde\u=[\u,\u^e]^T$ to be evolved in time, essentially doubling the number of variables. In practice,
the vector $\u^e$ is slightly smaller than $\u$, since we do not need to consider the equilibrium values of the cleaning four vector $z^\mu$, neither of the scalar $K_0$,  nor of the
three vector $b^i$ related to the {\emph{gamma--driver}}. Eventually, the full vector $\tilde\u$ contains $59+51=110$ variables, and with the above property
$\partial_t \u^e=0$ of a stationary equilibrium the system~\eqref{Csyst-wb} translates into
\begin{equation}	
\label{Csyst-wb-compact}
\frac{\partial \tilde{\u} }{\partial t} + \frac{\partial \tilde{\f}_i(\u)}{\partial x_i} +
\tilde{\B}_i(\u) \frac{\partial \tilde{\u}}{\partial x_i} 
= \tilde{\S}(\tilde{\u}),
\qquad \textnormal{or, equivalently,} \qquad 
\frac{\partial \tilde{\u} }{\partial t} + \nabla \cdot \tilde{\F}(\tilde{\u}) +
\tilde{\B}(\tilde{\u}) \cdot \nabla \tilde{\u}  
= \tilde{\S}(\tilde{\u}), \end{equation}
with $\tilde{\F}(\tilde{\u}) = 
\left( 
\tilde{\f}_1(\tilde{\u}),
\tilde{\f}_2(\tilde{\u}),
\tilde{\f}_3(\tilde{\u})
\right)$, 
\begin{equation}  
	{\tilde\S}(\tilde \u)=\left(  
	\begin{array}{c }  
		\S(\u) - \S(\u^e)  \\  
		$0$   
	\end{array}  
	\right),
	\qquad 
	{\tilde\f}_i(\tilde \u)=\left(  
	\begin{array}{c }  
		\f_i(\u) - \f_i(\u^e)  \\  
		$0$   
	\end{array}  
	\right)  	
\end{equation} 
and 
\begin{equation}  
{\tilde\A_i}=\left(  
\begin{array}{c c c}  
\A_i(\u)  & \vline & -\A_i(\u^e)\\  
\hline   
$0$ &\vline &   0
\end{array}  
\right), 
\qquad 
{\tilde\B_i}=\left( 
\begin{array}{c c c}  
	\B_i(\u)  & \vline & -\B_i(\u^e)\\  
	\hline   
	$0$ &\vline &   0
\end{array}  
\right),  
\quad 
\end{equation}
thus obtaining that
the equilibrium sector $\u^e$ contained in the second part of 
$\tilde\u$ remains frozen, while 
the equilibrium solution is subtracted from the first part of the
vector $\tilde\u$, as dictated by Eq.~\eqref{Csyst-wb}. \textcolor{black}{We
note that the approach expressed by Eq.~\eqref{Csyst-wb} closely follows the seminal ideas 
introduced in \cite{Ghosh2016,berberich2021high} for the well-balancing of \textit{completely general} multi-dimensional hyperbolic PDE systems. While the applications presented in  \cite{Ghosh2016,berberich2021high} were related to the Newtonian Euler and MHD equations, the method is general enough so that in this paper it can now for the first time also be applied to a first order reformulation of the Einstein-Euler system that describes the coupled dynamics of matter and spacetime in full general relativity, see \eqref{eqn.rho}-\eqref{eqn.D}. } 

It is obvious that when inserting the extended equilibrium solution $\tilde{\u}^e = (\u^e, \u^e)^T$ into~\eqref{Csyst-wb-compact}
one has $\partial_t \tilde{\u}^e = 0$, i.e. the augmented equation is trivially satisfied since by construction the following fundamental properties hold: 
\begin{equation}
\tilde{\F}(\tilde{\u}^e) = 0, 
\qquad 
\tilde{\B}(\tilde{\u}^e) \cdot \nabla \tilde{\u}^e = 0, 
\qquad   
\tilde{\S}(\tilde{\u}^e) = 0. 
\label{eqn.wb.fundamental.property}
\end{equation} 
 
In the computation of the numerical fluxes via Riemann solvers, 
which is typical for discontinuous Galerkin and finite volume schemes, special care has to be taken in the structure of the numerical viscosity, which must not destroy the well-balancing of the numerical scheme. For this purpose, we will later need a 
\textit{modified identity matrix} or \textit{well-balanced identity matrix}, which acts on the extended state vector
$\tilde\u$ and has the following block structure: 
\begin{equation}  
\tilde{\mathbf{I}} = \left(  
\begin{array}{c c c c}  
\mathbf{I} & \vline & -\mathbf{I} \\  
\hline   
0 &\vline & 0  
\end{array}  
\right).
\label{eqn.wb.id}
\end{equation}
The main property of the above well-balanced identity matrix is that its product with the extended equilibrium state $\tilde{\u}^e = (\u^e, \u^e)^T$ is zero, i.e. $\tilde{\mathbf{I}} \, \tilde{\u}^e = 0$.

In the practical implementation of the numerical scheme solving Eq.~\eqref{Csyst-wb-compact}, we have
allowed for the possibility to switch the well-balancing on or off, according to the problem under consideration.
For equilibrium, or close--to--equilibrium problems, well-balancing is of course important and it is activated. For rather
dynamical problems, on the contrary, well-balancing is abandoned, and only the first $59$ equations are considered with no need to subtract the equilibrium solution. 

The DG and FV discretization is based on the weak form of the PDE~\eqref{Csyst-wb-compact}, which, upon integration
over the spacetime control volume $\Omega_i \times [t^n,t^{n+1}]$, provides
\begin{equation}
\label{eqn.pde.nc.gw1}
\int \limits_{t^n}^{t^{n+1}} \int \limits_{\Omega_i}
\Phi_k \, \frac{\partial \tilde\u}{\partial t} \, d \x \, d t
+\int \limits_{t^n}^{t^{n+1}} \int \limits_{\Omega_i}
\Phi_k \, \left( \nabla \cdot \tilde{\F}(\tilde{\u}) + \tilde\B(\tilde\u) \cdot \nabla \tilde\u  \right) \, d \x \, d t
= \int \limits_{t^n}^{t^{n+1}} \int \limits_{\Omega_i}   \Phi_k
\, \tilde\S(\tilde\u) \, d \x \, d t\,.
\end{equation}
The most important difference between the new scheme presented in this paper and the one used in~\cite{Gaburro2021WBGR1D} is that here we use a \textit{discrete} version of the equilibrium $\u_h^e(\x,t^n)$ by simply setting the nodal degrees of freedom as $\tilde{\u}^n_{i,\ell} = (\u^e(\x_{\ell}), \u^e(\x_{\ell}) )^T$, i.e. the discrete equilibrium is the $L^2$ projection of the exact equilibrium into the space of piecewise polynomials of degree $N$. Instead, in~\cite{Gaburro2021WBGR1D} the discrete solution was the sum of the exact analytical (non-polynomial) equilibrium $\u^e(\x)$ plus a piecewise polynomial perturbation.  

In the following, we will focus on a few relevant aspects calling for attention when integrating Eq.~\eqref{eqn.pde.nc.gw1}, each of which deserves a bit of discussion.

\subsubsection{The DG discretization in space}
\label{sec.DGdiscretization}

We tackle the solution of  the Z4 system  by considering a
computational domain $\Omega$ in dimension $d=2$ or $d=3$ that is given by the union
of a set of non-overlapping
Cartesian tensor-product elements, namely  $\Omega = \bigcup \Omega_i = \bigcup [x_i - \halb \Delta x_i,
x_i + \halb \Delta x_i] \times [y_i - \halb \Delta y_i, y_i + \halb
\Delta y_i] \times [z_i - \halb \Delta z_i, z_i + \halb \Delta z_i] $,
where $\boldsymbol{x}_i = (x_i, y_i, z_i)$ indicates the barycenter of
cell $\Omega_i$ and $\Delta \boldsymbol{x}_i = (\Delta x_i,\Delta
y_i,\Delta z_i)$ defines the size of $\Omega_i$ in each spatial
coordinate direction. 
According to the DG finite-element approach, the discrete solution at time $t^n$
is written in terms of prescribed spatial
basis functions $\Phi_\ell(\boldsymbol{x})$  as
\begin{equation}
	\tilde{\u}_h(\boldsymbol{x},t^n) = \sum \limits_\ell \tilde{\boldsymbol{u}}^n_{i,\ell}
	\Phi_\ell(\boldsymbol{x}) := \tilde{\boldsymbol{u}}^n_{i,\ell} \Phi_\ell(\boldsymbol{x})\,.
	\label{eqn.ansatz.uh}
\end{equation}
Here $\ell:=(\ell_1,\ell_2,\ell_3)$ is a multi-index while the expansion coefficients $\hat{\boldsymbol{u}}_{i,\ell}^n$ are 
the so-called \emph{degrees of freedom}.
The spatial basis functions $\Phi_\ell(\boldsymbol{x}) = \varphi_{\ell_1}(\xi) \varphi_{\ell_2}(\eta)
\varphi_{\ell_3}(\zeta)$ are chosen as 
tensor products of one-dimensional nodal basis functions 
defined on the reference element $[0,1]$. 
In one spatial dimension, the basis functions $\varphi_{\ell_i}(\xi)$ are  
the Lagrange interpolation polynomials, up to degree $N$, 
which pass through the $(N+1)$ Gauss-Legendre 
quadrature points. This is particularly convenient when performing numerical integrals
of the discrete solution, due to the nodal property that 
$\varphi_k(\xi_j) = \delta_{kj}$, with $\xi_j$ being the coordinates of the nodal points\footnote{
	The mapping from physical coordinates $\mathbf{x} \in
	\Omega_i$ to reference coordinates $\boldsymbol{\xi}=\left(
	\xi,\eta,\zeta \right) \in [0,1]^3$ is simply given by $\mathbf{x} =
	\mathbf{x}_i - \halb \Delta \mathbf{x}_i + (\xi \Delta x_i, \eta \Delta
	y_i, \zeta \Delta z_i)^T$.}.

\subsubsection{The spacetime predictor}
\label{sec.predictor}

A crucial aspect has to do with time integration. A common option to integrate Eq.~\eqref{eqn.pde.nc.gw1} in time would be to resort to
Runge--Kutta schemes, thus obtaining RKDG schemes~\cite{cbs0,cbs1}. However, as a valid alternative introduced by~\cite{dumbser_jsc,DumbserEnauxToro,QiuDumbserShu}
and adopted preferentially within our group, we have followed the ADER approach, according to which a high order accurate (both in space and in time)
solution can be obtained through a single time integration step,
provided an approximate \textit{predictor} state $\tilde{\q}_h$ is available 
at any intermediate time between $t^n$ and $t^{n+1}$. Note that unlike in previous publications on ADER schemes in this paper $\tilde{\q}_h$ is an approximation of the \textit{extended} state vector $\tilde{\u}=[\u,\u^e]^T$. 
Furthermore, while in the original ADER version of ADER by
Toro and Titarev~\cite{Titarev2002,Titarev2005,Toro2006} 
the computation of the predictor was obtained through the
Cauchy-Kovalewski procedure, we follow here the more recent approach introduced in~\cite{dumbser2008unified},
which is more suitable for complex systems of equations like the Einstein-Euler equations of general relativity.
The predictor $\tilde{\q}_h$ is thus expanded into a local spacetime basis
\begin{equation}
\label{eqn.spacetime}
\tilde{\q}_h(\boldsymbol{x},t) = \sum \limits_\ell \theta_\ell(\boldsymbol{x},t)
\tilde{\boldsymbol{q}}_{i,\ell} := \theta_\ell(\boldsymbol{x},t) \tilde{\boldsymbol{q}}_{i,\ell}\,,
\end{equation}
with the multi-index $\ell=(\ell_0,\ell_1,\ell_2,\ell_3)$ and where the spacetime basis
functions 
$$\theta_\ell(\boldsymbol{x},t) = \varphi_{l_0}(\tau)
\varphi_{\ell_1}(\xi) \varphi_{\ell_2}(\eta) \varphi_{\ell_3}(\zeta)$$ 
are again
generated from the same one-dimensional nodal basis functions
$\varphi_{k}(\xi)$ as before, namely using the Lagrange interpolation polynomials
up to degree $N$ passing through $N+1$ Gauss--Legendre quadrature nodes. 
The coordinate time is mapped to the
reference time $\tau \in [0,1]$ via $t = t^n + \tau \Delta
t$. Multiplication of the PDE system~\eqref{Csyst-wb-compact} with a
test function $\theta_k$ and integration over the spacetime  control volume $\Omega_i \times [t^n,t^{n+1}]$ 
yields 
\begin{equation}
\label{eqn.pde.st1}
\int \limits_{t^n}^{t^{n+1}} \, \, \int \limits_{\Omega_i}
\theta_k \frac{\partial \tilde{\q}_h}{\partial t} d \x \, d t
+ \int \limits_{t^n}^{t^{n+1}} \, \, \int \limits_{\Omega_i}
\theta_k \left( \nabla \cdot \tilde{\F}(\tilde{\q}_h) +  \tilde{\B}(\q_h) \cdot \nabla \tilde{\q}_h  \right) d \x \, d t
= \int \limits_{t^n}^{t^{n+1}} \, \, \int \limits_{\Omega_i} 
\theta_k \tilde{\S}(\tilde{\q}_h) \, d \x \, d t.
\end{equation}
Since the calculation is performed locally for each cell, no special treatment of the jumps at the element boundaries is needed at this stage, and Riemann solvers are not involved. Rather, Eq.~\eqref{eqn.pde.st1} is integrated by parts in time, providing
\begin{eqnarray}
\label{eqn.pde.st2}
\int \limits_{\Omega_i}   \theta_k(\x,t^{n+1}) \tilde{\q}_h(\x,t^{n+1}) d \x  -
\int \limits_{\Omega_i}   \theta_k(\x,t^{n}) \tilde{\u}_h(\x,t^{n}) d \x
- \int \limits_{0}^{1} \, \, \int \limits_{T_E}
\frac{\partial \theta_k(\x,t)}{\partial t}  \tilde{\q}_h(\x,t)
d \x \, d t = && \nonumber \\
\int \limits_{t^n}^{t^{n+1}} \, \, \int \limits_{\Omega_i} 
\theta_k \, \left( \tilde{\S}(\tilde{\q}_h) - \nabla \cdot \tilde{\F}(\q_h) +  \tilde{\B}(\tilde{\q}_h) \cdot \nabla \tilde{\q}_h  \right) d \x \, d t. &&
\end{eqnarray}
Eq.~\eqref{eqn.pde.st2} generates a nonlinear system for the
unknown degrees of freedom $\tilde{\boldsymbol{q}}_{i,\ell}$ of the spacetime
polynomials $\tilde{\q}_h$. The solution of~\eqref{eqn.pde.st2} is obtained
via a simple fixed-point iteration, the convergence of which was proven in~\cite{BCDGP20}.  

\paragraph{Well-balanced property of the predictor}
When the discrete solution $\tilde{\u}_h(\x,t^n)$ at time $t^n$ coincides with the discrete equilibrium, i.e. when 
$\tilde{\u}_h(\x,t^n) = (\u_h^e, \u_h^e)^T$ with nodal degrees of freedom $\tilde{\u}^n_{i,\ell} = (\u^e(\x_{\ell}), \u^e(\x_{\ell}) )^T$, then it is obvious that $\tilde{\q}_h = \tilde{\q}_h^e = (\u_h^e, \u_h^e)^T$
is a solution of~\eqref{eqn.pde.st2} since $\tilde{\S}(\tilde{\q}_h^e) - \nabla \cdot \tilde{\F}(\tilde{\q}_h^e) +  \tilde{\B}(\tilde{\q}_h^e) \cdot \nabla \q_h^e =0$ due to the fundamental properties~\eqref{eqn.wb.fundamental.property}.
Hence, the predictor is by construction well-balanced.

\subsubsection{ADER-DG schemes for non-conservative systems}
\label{sec.ADER-DG}
Another aspect related to the solution of Eq.~\eqref{eqn.pde.nc.gw1}
has to do with the presence of non--conservative terms, 
indeed the vast majority in the Einstein--Euler system that we are considering. 
Our strategy is based on the so-called \emph{path-conservative approach} of
\cite{Castro2006, Pares2006}, which 
was first applied to DG schemes by~\cite{Dumbser2009a, Dumbser2010}
and subsequently considered in the context of  the first--order formulation of the CCZ4 
Einstein system by~\cite{Dumbser2017strongly}. In practice, 
after integration by parts of the flux divergence and the introduction of a Riemann solver that accounts for the jumps at the element boundaries, the fully discrete one-step ADER-DG scheme resulting from~\eqref{eqn.pde.nc.gw1} reads 
\begin{eqnarray}
\label{eqn.pde.nc.gw2a}
&&\left( \int \limits_{\Omega_i}  \Phi_k \Phi_\ell \, d \x \right)
\left( \tilde{\boldsymbol{u}}^{n+1}_{i,\ell} - \tilde{\boldsymbol{u}}^{n}_{i,\ell}  \right)
+ \int \limits_{t^n}^{t^{n+1}} \, \, \int \limits_{\Omega_i^\circ}
\Phi_k \, \tilde{\B}(\tilde{\q}_h) \cdot \nabla \tilde{\q}_h   d \x \, d t
-
 \int \limits_{t^n}^{t^{n+1}} \, \, \int \limits_{\Omega_i^\circ}
\nabla \Phi_k  \cdot \tilde{\F}(\tilde{\q}_h)  \, d \x \, d t
 \nonumber \\
&& + \int \limits_{t^n}^{t^{n+1}} \, \, \int \limits_{\partial \Omega_i}
 \Phi_k \, \mathcal{D}\left( \tilde{\q}_h^-, \tilde{\q}_h^+ \right) \cdot \boldsymbol{n} \,d S d t
 +
  \int \limits_{t^n}^{t^{n+1}} \, \, \int \limits_{\partial \Omega_i}
 \Phi_k \, \mathcal{F}\left( \tilde{\q}_h^-, \tilde{\q}_h^+ \right) \cdot \boldsymbol{n} \,d S d t
 = 
\int \limits_{t^n}^{t^{n+1}} \, \, \int \limits_{\Omega_i}
\Phi_k \, \tilde{\S}(\tilde{\q}_h)  d \csi \, d \tau\,,
\end{eqnarray}
where the boundary integrals in~\eqref{eqn.pde.nc.gw2a} become relevant only when 
the boundary extrapolated states at the left $\tilde{\q}_h^-$ and at the right
$\tilde{\q}_h^+$ of the interface are different, 
$\tilde{\q}_h^-\neq \tilde{\q}_h^+$, namely when there is a true jump.
According to a now well-established procedure, 
developed in~\cite{Pares2006,Castro2006,Dumbser2011}, the jump terms in the non-conservative product are computed through a path-integral in phase space as
\begin{equation}
\label{eqn.pc.scheme}
\mathcal{D}\left( \tilde{\q}_h^-, \tilde{\q}_h^+ \right) \cdot \boldsymbol{n} =
\frac{1}{2} \left( \, \int \limits_{0}^1
\tilde{\B}(\boldsymbol{\psi}) \cdot \boldsymbol{n} \, d s \right)
\left( \tilde{\q}_h^+ - \tilde{\q}_h^- \right)\,,
\end{equation}
which we have solved via a Gaussian quadrature formula composed of three points.
For simpler systems of equations, one might even think about using the Riemann invariants of the PDE system 
as optimal paths along which to perform the integration~\cite{MuellerToro2013a}, but for the Einstein equations such an option is absolutely impracticable,
thus we have used a simple segment path 
\begin{equation}
\boldsymbol{\psi} = \boldsymbol{\psi}(\tilde{\q}_h^-, \tilde{\q}_h^+, s) =
\tilde{\q}_h^- + s \left( \tilde{\q}_h^+ - \tilde{\q}_h^- \right),
\qquad 0 \leq s \leq 1\,.
\end{equation}
The simplest possible numerical flux for the conservative part of the equations, i.e. for the Euler subsystem, is a Rusanov-type flux given by 
\begin{equation}
	\mathcal{F}\left( \tilde{\q}_h^-, \tilde{\q}_h^+ \right) \cdot \boldsymbol{n} = 
	\halb \left( 
	\tilde{\F}(\tilde{\q}_h^-) + \tilde{\F}(\tilde{\q}_h^+) \right) \cdot
	 \boldsymbol{n}  
	- \frac{1}{2} s_{\max}
	\tilde{\mathbf{I}} \, \left(
	\tilde{\q}_h^+ - \tilde{\q}_h^- \right). 
	\label{eqn.rusanov} 
\end{equation}
The last term in Eq.~\eqref{eqn.rusanov} contains the numerical viscosity, which employs the well-balanced identity matrix $\tilde{\mathbf{I}}$. 
For a Rusanov-type flux the numerical viscosity is provided by the knowledge of a single 
characteristic speed, $s_{\max}$, which denotes the maximum of the absolute values of the 
characteristic velocities 
$\left| \Lambda(\tilde{\q}_h^-) \right|$, $\left| \Lambda(\tilde{\q}_h^+)
\right|$ at the interface 
\begin{equation}
 \qquad s_{\max} =
\max \left( \left| \Lambda(\tilde{\q}_h^-) \right|, \left| \Lambda(\tilde{\q}_h^+)
\right| \right). 
\end{equation}
A more sophisticated HLL-type flux, which also employs the use of the well-balanced identity matrix $\tilde{\mathbf{I}}$ reads 
\begin{equation}
	\mathcal{F}\left( \tilde{\q}_h^-, \tilde{\q}_h^+ \right) \cdot \boldsymbol{n} = 
 \frac{ 
s_R \tilde{\F}(\tilde{\q}_h^-) - s_L \tilde{\F}(\tilde{\q}_h^+) }{s_R - s_L} \cdot
\boldsymbol{n}  
+ \frac{s_R s_L}{s_R - s_L}
\tilde{\mathbf{I}} \, \left(
\tilde{\q}_h^+ - \tilde{\q}_h^- \right), 
	\label{eqn.hll} 
\end{equation}
with the left and right signal speeds $s_L \leq 0$ and $s_R \geq 0$ computed, e.g., according to~\cite{Einfeldt88,munz91}. 

\paragraph{Well-balanced property of the final ADER-DG scheme}

We now assume that the discrete solution coincides with the discrete equilibrium, i.e. $\tilde{\u}_h=(\u_h^e,\u_h^e)^T$. Since the predictor is well-balanced, the resulting predictor
solution is $\tilde{\q}_h=(\u_h^e,\u_h^e)^T$.   
Due to the fundamental property~\eqref{eqn.wb.fundamental.property} it is obvious that all terms
in~\eqref{eqn.pde.nc.gw2a} cancel by construction.  
However, at this point we emphasize again that in order to preserve the well-balancing property of the numerical scheme, in the numerical fluxes one must make use of the well-balanced identity matrix $\tilde{\mathbf{I}}$ introduced
in~\eqref{eqn.wb.id}, since $\tilde{\mathbf{I}} \left( \tilde{\q}_h^{e,+} - \tilde{\q}_h^{e,-} \right) =0$ for two arbitrary discrete equilibrium states $\tilde{\q}_h^{e,\pm} = (\u_h^{e,\pm}, \u_h^{e,\pm})$. 

\textcolor{black}{Finally, since it is quite often a crucial quantity in a numerical simulation, it is worth providing some information about the total memory consumption produced by our numerical scheme. Let us first quantify the memory load of the spacetime predictor of Sect.~\ref{sec.predictor} for a single variable and a single numerical cell, i.e.
\begin{equation}
	\label{load1}
	\textrm{MemLoad}_P=(N+1)^4\cdot[\underbrace{1}_{\tilde{\q}_h}+\underbrace{3}_{\tilde{\F}}+\underbrace{3}_{\nabla \tilde{\q}}+\underbrace{1}_{\tilde{\S}}]=8\cdot (N+1)^4  \,,
\end{equation}
where $N$ is the degree of the DG polynomial, the exponent 4 refers to the number of spacetime dimensions, while the terms in square brackets correspond to the contribution of the variable itself, the fluxes, the gradients and the source, respectively. Secondly, the memory load produced by the true DG scheme of Sect.~\ref{sec.ADER-DG} is given by
\begin{equation}
	\label{load2}
	\textrm{MemLoad}_{DG}=2\cdot(N+1)^3\cdot[\underbrace{1}_{\tilde{\q}_h}+\underbrace{3}_{\tilde{\F}}+\underbrace{1}_{\tilde{\S}-\tilde{\mathbf{B}}\cdot\nabla \tilde{\q}}]=10\cdot(N+1)^3  \,,
\end{equation}
where the multiplication factor $2$ is required to account for the two time levels at $t^n$ and $t^{n+1}$, the exponent 3 refers to the number of space dimensions, while the terms in square brackets correspond to the contribution of the variable itself, the fluxes, and the compactified term $\tilde{\S}-\tilde{\mathbf{B}}\cdot\nabla \tilde{\q}$, the latter one being an optimization feature of our implementation. Summing  Eq.~\eqref{load1} and Eq.~\eqref{load2}, and multiplying by the 110 variables of the fully well-balanced scheme, we obtain the total memory load per numerical cell
\begin{equation}
	\label{loadtot}
	\textrm{MemLoad}=110\cdot(\textrm{MemLoad}_P + \textrm{MemLoad}_{DG})= 220\cdot(N+1)^3\cdot(4N+9)  \,.
\end{equation}
Table~\ref{tab.memload} shows the total memory load for a few values of the polynomial degree $N$ according to Eq.~\eqref{loadtot}.
 \begin{table}[!t] 
 	\begin{center}
	\renewcommand{\arraystretch}{1.1}
	\caption{Memory load per cell of the fully well-balanced Z4 ADER-DG scheme according to Eq.~\eqref{loadtot}.}  
	\begin{tabular}{|l||l|l|}
		\hline
		$N$ & \textrm{MemLoad} & \textrm{MemLoad} (Byte, double precision)  \\
		\hline 
		\hline
     	1      & 22880   &  183040  $\sim$ 0.2\,\textrm{MB} \\ 
     	2      & 100980  &  807840  $\sim$ 0.8\,\textrm{MB} \\ 
		3      & 295680  & 2365440  $\sim$ 2.2\,\textrm{MB} \\ 
		4      & 687500  & 5500000  $\sim$ 5.2\,\textrm{MB}\\
		5      & 1378080 & 11024640 $\sim$ 10.5\,\textrm{MB}\\
		\hline
	\end{tabular}
	\label{tab.memload}
\end{center}	
\end{table}
}
\subsection{\textit{A posteriori} sub-cell finite volume limiter}
\label{ssec.limiter} 

At this point, we need to point out that DG schemes are \textit{linear} in the sense of Godunov~\cite{godunov}. This means that, while the solution is represented within each cell by higher order polynomials, the update rule is \textit{linear} when applied to a linear PDE.  
Hence, they represent a highly accurate method to describe the smooth features of the metric variables, 
but, as proven by the Godunov theorem~\cite{godunov}, starting from second order, they will inevitably oscillate in presence of discontinuities or strong gradients. 
Thus, we need to endow our DG scheme with a technique able to strengthen its robustness, 
maintaining at the same time its desirable high order of accuracy. 

Among the different strategies proposed over the years  
(see for example~\cite{cockburn1990runge,cockburn2000development,krivodonova2004shock,persson2006sub,qiu2004hermite,qiu2005runge} for some seminal introductory papers), 
we select the so-called \textit{a posteriori} sub-cell finite volume limiter, 
which has proved its capabilities in previous works both from the 
authors themselves~\cite{DGLimiter1,DGLimiter2,DGLimiter3,Zanotti2015d,SolidBodies2020,Gaburro2020Arepo,Gaburro2021PNPMLimiter} and also from other research groups~\cite{SonntagDG,Sonntag2,DeLaRosaMunzDGMHD,hajduk2019new,markert2021sub,rueda2022subcell,popov2023space}. 
While referring to the aforementioned references for a detailed description, 
in particular to Section 3.4 of~\cite{DGLimiter2} and Section 4 of~\cite{Zanotti2015d} where the sub-cell finite volume limiter has been also outlined on adaptive Cartesian meshes (AMR), 
here we only briefly recall the key concepts. 

First, our limiter acts in general in an \textit{a posteriori}
fashion: 
indeed, at the beginning of each timestep we apply our \textit{unlimited} DG scheme, 
everywhere on the domain, in order to obtain a candidate  solution $\tilde{\u}_h^{n+1,*} = \tilde{\u}_h^{n+1}$.
Then, the candidate solution is checked against physical and numerical admissibility criteria to verify that it does not presents 
nonphysical values (as negative densities, negative pressures or superluminal velocities) or spurious oscillations (according to a relaxed discrete maximum principle).
The cells where one of these criteria is not respected are marked as \textit{troubled}
and, only in those cells, we completely recompute the solution by employing a more robust scheme; 
in particular, in this work we rely either on a second order Total Variation Diminishing (TVD) finite volume scheme or on a third order ADER-WENO~\cite{balsara2009efficient} FV method. 

Furthermore, we emphasize that the key point for maintaining the resolution capabilities of the DG scheme, when using instead a less accurate FV scheme, consists in applying it on a locally \textit{refined mesh}. 
So, we subdivide each original troubled cell $\Omega_i$ in $(2N+1)^d$ \textit{sub-cells} $\omega_\alpha$.
Then, we perform an $L^2$ projection of the DG solution $\tilde{\u}_h^{n}$ on the space of constant polynomials 
obtaining the sub-cell averages values $\hat{u}^{n}|_{\omega_\alpha}$ with $\alpha=[1,(2N+1)^d]$, and we evolve these sub-cell values with the FV scheme. 
In this way, we obtain the updated sub-cell averages information $\hat{u}^{n+1}|_{\omega_\alpha}$ 
from which we reconstruct back a high order polynomial $\tilde{\u}_h^{n+1}$ with a least square operator coupled with a conservation constraint on the main cell $\Omega_i$.
We also notice that this reconstruction technique might still lead to an oscillatory solution, 
being an unlimited linear procedure. 
In this case, the oscillatory cell will be marked again as troubled at the next timestep $t^{n+2}$, so we will apply again the FV scheme there but using as sub-cell averages directly the oscillation-free $\hat{u}^{n+1}|_{\omega_\alpha}$ obtained at the previous timestep without passing through the reconstruction-projection step. 

Finally, we remark that FV schemes have a less restrictive CFL stability condition than that imposed by Eq.~\eqref{eqn.cfl} for DG schemes. In particular, the choice of the $\Delta t$ is not affected at all by the requested order of accuracy, 
thus the factor $(2N+1)$ is not appearing in the finite volume CFL formula. 
This justifies the stability of our FV limiter scheme which can be safely applied to the cells $\omega_\alpha$ whose mesh size is exactly a factor $(2N+1)$ smaller than the original $\Omega_i$ cell size, thus leading exactly to the same CFL constraint of the original unlimited DG scheme. Further details can be found in~\cite{DGLimiter1}.

Concerning the well-balancing property, the subcell FV limiter is also by construction well-balanced for discrete equilibria due to the fundamental properties~\eqref{eqn.wb.fundamental.property}.

%
\subsection{The choice of coordinates}
\label{sec.coord}
In general relativity the choice of coordinates is completely arbitrary, in the sense that, since  the original equations are covariant,
the mathematical form of the equations is always the same, irrespective of the coordinates  chosen. However, this does not mean that all coordinate systems behave
equally well, especially when performing numerical simulations. In this paper we have adopted
the following systems of coordinates: 
\begin{enumerate}
\item {Spherical coordinates $(t,r,\theta,\phi)$}, which can be used either in flat spacetime or in the presence of a central (non--rotating) mass, 
as for the case described in Sect.~\ref{sec:tov}. The corresponding
metric is
\begin{eqnarray}
\label{eq:symm_metric}
ds^2 = -e^{2\phi} dt^2 + e^{2\psi} dr^2 + r^2 d\theta^2  + r^2 \sin^2\theta d\phi^2 \,,
\label{tov5}
\end{eqnarray}
where $\phi$ and $\psi$ are functions of $r$ only.
\item  {Kerr--Schild spheroidal  coordinates $(t,r,\theta,\phi)$.} These are special coordinates\footnote{In view of the coordinate transformation 
\eqref{cart1}--\eqref{cart4},
we are not allowed to interpret the
Kerr--Schild coordinates as standard spherical coordinates. For an extended discussion about different coordinate systems in the Kerr spacetime see~\cite{Visser2007}.} 
that are very convenient to describe the stationary spacetime
of either non-rotating (Schwarzschild, with $a=0$) 
or  rotating (Kerr, with $0<a<1$) black holes, since they do not show any singularity at the  event horizon. In terms of such coordinates the metric can be written as~\cite{Kerr63,Komissarov04b}
\begin{align}
\label{eq:kerr1}
ds^2 =& (z-1)\,dt^{2} - 
2 z a \sin^2\theta\, dt \, d\phi +
2 z dt  \ dr - 2 a ( 1+z) \sin^2\theta \, dr \, d\phi
\nonumber \\
&+ ( 1+z )\, dr^2 + \rho^2\, d\theta^2 +
\frac{\Sigma\sin^2\theta}{\rho^2}\, d\phi^{ 2} \,,
\end{align}
where $z = {2Mr}/{\rho^2}$, $\rho^2=r^2+a^2\cos^2\theta$, $\Sigma=(r^2+a^2)^2-a^2\Delta\sin^2\theta$, $\Delta=r^2+a^2-2Mr$. The lapse of the metric is 
$\alpha=1/\sqrt{1+z}$, while there is a 
non--zero shift $\beta^i=\left(z/(1+z),0,0\right)$ even in the absence of black hole rotation. The spatial part of the metric is given by
\begin{equation}
	\gamma_{ij} = \left( \begin{array}{ccc}
	 1                    & 0           & -a\sin^2\theta (1+z) \\
	 0                    & \rho^2      & 0                    \\
	 -a\sin^2\theta (1+z) & 0           & \Sigma\sin^2\theta/\rho^2
	\end{array} \right)\,.
	\label{eqn.kerr.metric}
\end{equation}
The only physical singularity of the Kerr
spacetime, which is also a coordinate singularity, is at \hbox{$\rho^2=0$}, namely, at $r=0$ and $\theta=\pi/2$.

\item  {Kerr--Schild Cartesian coordinates $(t,x,y,z)$.} These coordinates are obtained from the Kerr-Schild spheroidal coordinates through the transformation
\begin{align}
\label{cart1}
x &= \sqrt{r^2+a^2}\sin\theta\cos\left[\phi-
\arctan\left(\frac{a}{r}\right)\right]\,,
\\
\label{cart2}
y &= \sqrt{r^2+a^2}\sin\theta\sin\left[\phi-
\arctan\left(\frac{a}{r}\right)\right]\,,
\\
\label{cart3}
z &= r\cos\theta\,, \\
\label{cart4}
t &= t^\prime \,,
\end{align}
such that the metric can be expressed as a deviation from the flat Minkowski spacetime, namely
\begin{equation}
\label{eq:kerr2}
ds^2=\left(\eta_{\mu\nu} + 2 H l_\mu l_\nu  \right)dx^\mu \,dx^\nu \hspace{2cm}\mu,\nu=1,2,3
\end{equation}
where
\begin{equation}
	 H = \frac{Mr^3}{r^4 + a^2 z^2}, \qquad
	 l_x = \frac{r x + a y}{r^2 + a^2}, \qquad
	 l_y = \frac{r y - a x}{r^2 + a^2}, \qquad
	 l_z = \frac{z}{r},
\end{equation}
and
\begin{equation}
	 r = \sqrt{ (x^2 + y^2 + z^2 - a^2)/2 + \sqrt{((x^2 + y^2 + z^2 - a^2)/2)^2 + z^2 a^2} }.
\end{equation}
Note that the lapse and the shift are given, respectively, \textcolor{black}{by $\alpha=1/\sqrt{G}$ and $\beta^i=\frac{2H}{G}l_i$, where $	 G = 1 + 2 H$.}
%
In these coordinates, the physical singularity, that in spheroidal coordinates is at $r=0$, $\theta=\pi/2$, corresponds to the points
with $x^2+y^2= a^2$ on the $z=0$ plane,  and it is therefore represented by a circle, the so-called ring singularity.

\end{enumerate}
For each of  the numerical tests reported in Sect.~\ref{sec.tests} we will specify which kind of coordinates have been adopted, among those just described.

\subsection{Recovering of the primitive hydrodynamical variables}
\label{sec:cons2prim}

\begin{figure}[!htbp]
\begin{center}
    \includegraphics[width=0.45\textwidth]{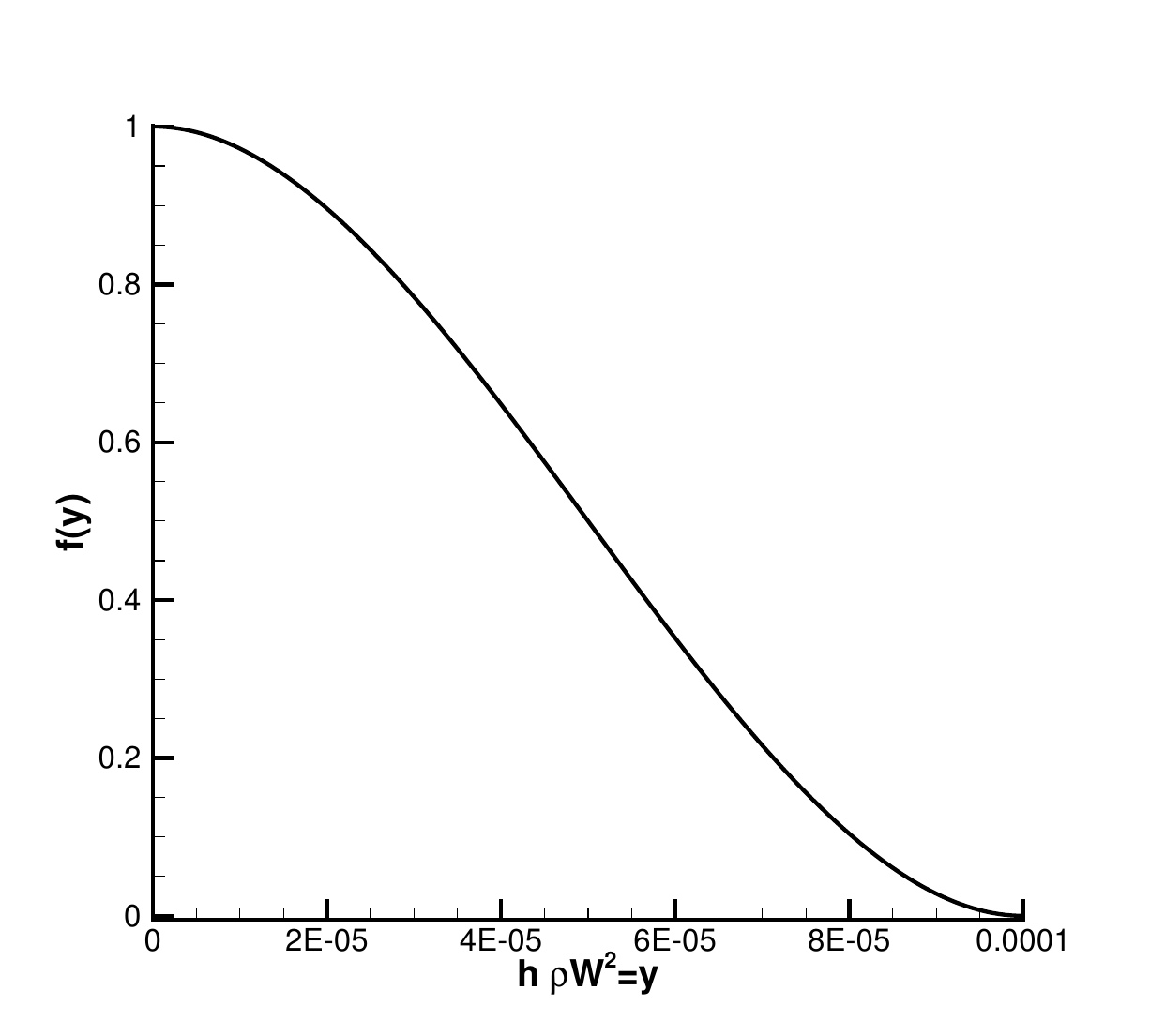}
	\includegraphics[width=0.45\textwidth]{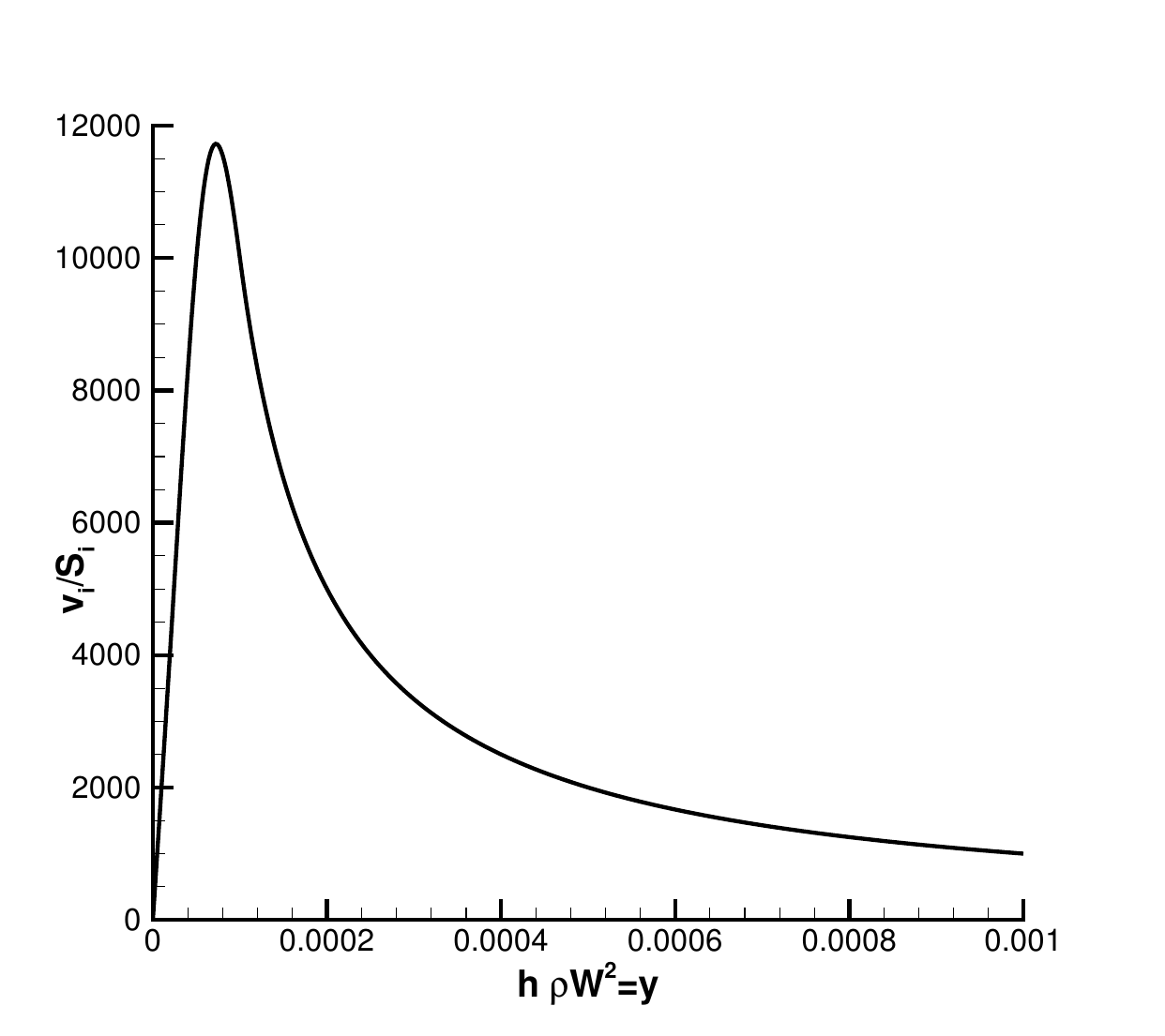}
	\caption{ \textcolor{black}{Lelft panel: plot of the polynomial $f(y)$ built in such a way to have $f(0)=1$, $f(y_0)=0$, $f^\prime(0)=0$, $f^\prime(y_0)=0$.}
	Right panel: plot of the ratio between velocity and momentum $v_i/S_i$ when the filter function is applied. If $y=\rho h W^2 < 10^{-4}$, the filter is activated, and the velocity
	decreases smoothly to zero.}
	\label{fig.filter}
\end{center}	
\end{figure}

Notoriously, in the relativistic framework the recovering of the 
primitive variables $(\rho,v_i,p)$ from  
the conserved variables $(D,S_i,E)$ is not 
analytic, and a numerical root-finding 
approach is necessary. The primitive variables are in fact required for the computation of the numerical fluxes
in the evolution of the matter variables (see equations~\eqref{eqn.rho}--\eqref{eqn.E} above).
Here, following the third method reported in Sect. 3.2 of 
\cite{DelZanna2007}, we solve the system
\begin{eqnarray}
\label{eq:F1}
F_1(x,y)&=& y^2 x - S^2=0, \\
\label{eq:F2}
F_2(x,y)&=&y - p - E=0\,,
\end{eqnarray}
where $x=v^2$, $y=\rho h W^2$, and where the pressure, at least for an ideal gas equation of state \textcolor{black}{considered in this paper}, can be written in terms of $x$ and $y$ as
\begin{equation}
p = \frac{\gamma-1}{\gamma}\left[(1-x)y-D\sqrt{1-x}\,\,\right]\,.
\label{eq:eos2}
\end{equation} 
In practice, we first  derive $y=y(x)$ from Eq.~\eqref{eq:F2} and then we find the root of $F_1[x,y(x)]=0$ via a Newton scheme.
As any other root solver, however, also this one might have troubles when the gas variables become very small, a problem that has been
afflicting numerical relativistic hydrodynamics since its birth. We have found a rather efficient 
strategy to solve this problem in such a way that allows us
to treat even cases when $\rho=0$ exactly. The idea can be split in the following steps:
\begin{enumerate}
\item We first check whether $D$ is smaller than a given tolerance, say $D<10^{-14}$. If that is the case, we set $\rho=max(0,D)$ and $v^i=0$.
This accounts also for the cases when $D$ becomes less or equal than zero, and reflects the idea that where there is no matter, the velocity field also vanishes, hence the associated Lorentz factor is one.
\item If $D>10^{-14}$, then we apply our standard root solver as outlined above. If the root solver fails, then again we set $\rho=p=v^i=0$.
\item If the root solver finds a root, namely a value of $x=v^2$, the following check is performed. If $y>y_0=10^{-4}$, the velocity is computed normally
as
\begin{equation}
	\label{standard-v}
v_i=\frac{S_i}{y}\,.
\end{equation} 
If instead $y<y_0=10^{-4}$, then a filter function is introduced
\begin{equation}
f(y)= 2(y/y_0)^3 -3(y/y_0)^2 + 1
\end{equation}
and the velocity field is computed by a filtered division as 
\begin{equation}
\label{filtered-v}
v_i=S_i\frac{y}{y^2 + f(y)\varepsilon}\,,
\end{equation}
where $\varepsilon=5\times 10^{-9}$. The filter function $f(y)$ in the denominator of~\eqref{filtered-v} is a cubic polynomial chosen in such a way 
to have vanishing first derivatives in $y=0$ and in $y=y_0$, as well as the correct interpolating property in those two points, namely $f(0)=1$, $f(y_0)=0$.

\end{enumerate}
In this way it is possible to solve regions characterized by very low matter densities, including even $\rho=0$, and the potentially harmful division by zero
is controlled by the filter function in the denominator of~\eqref{filtered-v}, which never vanishes. 
\textcolor{black}{
We stress that the value of $\varepsilon$ in Eq.~\eqref{filtered-v}
does not come from a rigorous proof, but it is related to the choice $y_0=10^{-4}$
roughly as $\varepsilon \leq y_0^2$ according to the following arguments:
since $f(0)=1$ and $f(y_0)=0$ (see the left panel of Fig.~\ref{fig.filter}), 
when $y\rightarrow 0$, the product $f(y)\varepsilon \rightarrow \varepsilon$, which is a small but finite quantity,
thus avoiding division by zero in the denominator of  Eq.~\eqref{filtered-v}.
When $y\rightarrow y_0$, on the contrary, 
the product $f(y)\varepsilon \rightarrow 0$ 
and we approach the safe regime of normal division, namely Eq.~\eqref{filtered-v} reduces to Eq.~\eqref{standard-v}.	
The effect of the filter is 
plotted in Fig.~\ref{fig.filter}, showing both the polynomial $f(y)$ (left panel) and the ratio $v_i/S_i$ (right panel), which 
reduces smoothly to zero when $y\rightarrow 0$.}

\textcolor{black}{The method that we have just described is decoupled from the 
well-balanced property of Sect.~\ref{sec.WBADERDG}, in the sense that it can be applied successfully even in a not well-balanced implementation. Of course it will require appropriate adaptations in case of more complicated equations of state. Actually, having $\rho=p=v^i=0$ corresponds to removing the fluid, while preserving the underlying 
equilibrium  solution of the spacetime. Therefore, the algorithm itself adheres perfectly to the well-balanced
approach.
}

\section{Numerical tests}
\label{sec.tests}

In this Section we present a large set of numerical results to show all the capabilities,
in terms of robustness, long-term stability and resolution, 
of our high order finite volume and discontinuous Galerkin schemes 
for the simulation of the proposed 
first--order hyperbolic Einstein-Euler Z4 system. If not stated otherwise, in all numerical tests we use the standard Z4 cleaning speed $e=1$ in our modified Z4 system. 

We also recall that the timestep in DG schemes 
is restricted according to 
\begin{equation}
	\Delta t < \frac{1}{d} \frac{1}{(2N+1)} \frac{h}{|\lambda_{\max}|}\,,
	\label{eqn.cfl} 
\end{equation}
where $h$ and $|\lambda_{\max}|$ are a characteristic mesh size and the maximum signal velocity, respectively. 


\subsection{Linearized gravitational wave test}
\label{sec:linearized-grav-wave}

As a first validation of our approach we consider a simple test,
essentially one-dimensional, taken from~\cite{Alcubierre2004} for which the metric is given as a wave perturbation of the flat Minkowski space time
\begin{equation} 
	\label{eqn.lw.metric}
	d s^2 = - d t^2 + d x^2 + (1+b)\,d y^2 + (1-b)\,d z^2, \quad \textnormal{with} \quad b = \epsilon \sin \left( 2 \pi (x-t) \right),
\end{equation} 
where $\epsilon = 10^{-8}$ is small enough so that the model behaviour is linear and the terms depending on $\epsilon^2$ can be neglected. 
According to~\eqref{eqn.lw.metric} 
$\gamma_{xx}=1$, $\gamma_{yy}=1+b$, $\gamma_{zz}=1-b$; next, we use the \emph{harmonic gauge condition}, while the \emph{gamma--driver} can be turned off, i.e. $s=0$.
Furthermore, the extrinsic curvature is given by 
$K_{ij} = \partial_t  \gamma_{ij} /(2\alpha)$ 
which means that its nonzero components are only
$
K_{yy}  = -1/2\,\partial_t b \text{ and } K_{zz}= 1/2\,\partial_t b.
$
The remaining non zero terms for the problem initialization are
$
D_{xyy}~=~1/2\,\partial_x b \text{ and } D_{xzz} = -1/2\,\partial_x b,
$
with the following setting for the other relevant parameters $\kappa_1=0$, $\kappa_2=0$ and $c=0$.
\begin{figure}[!htbp]
	\includegraphics[width=0.45\textwidth]{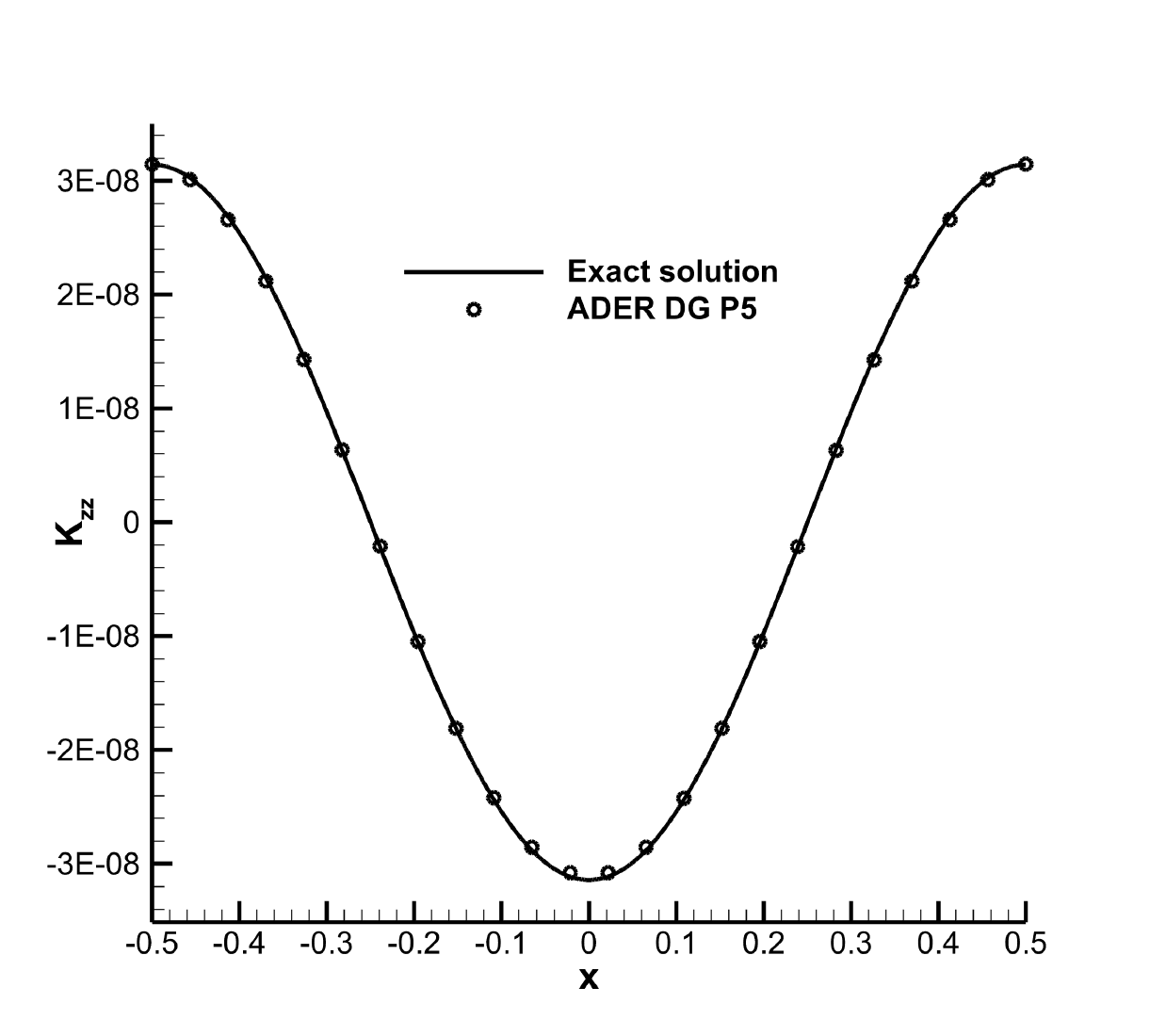}
	\includegraphics[width=0.45\textwidth]{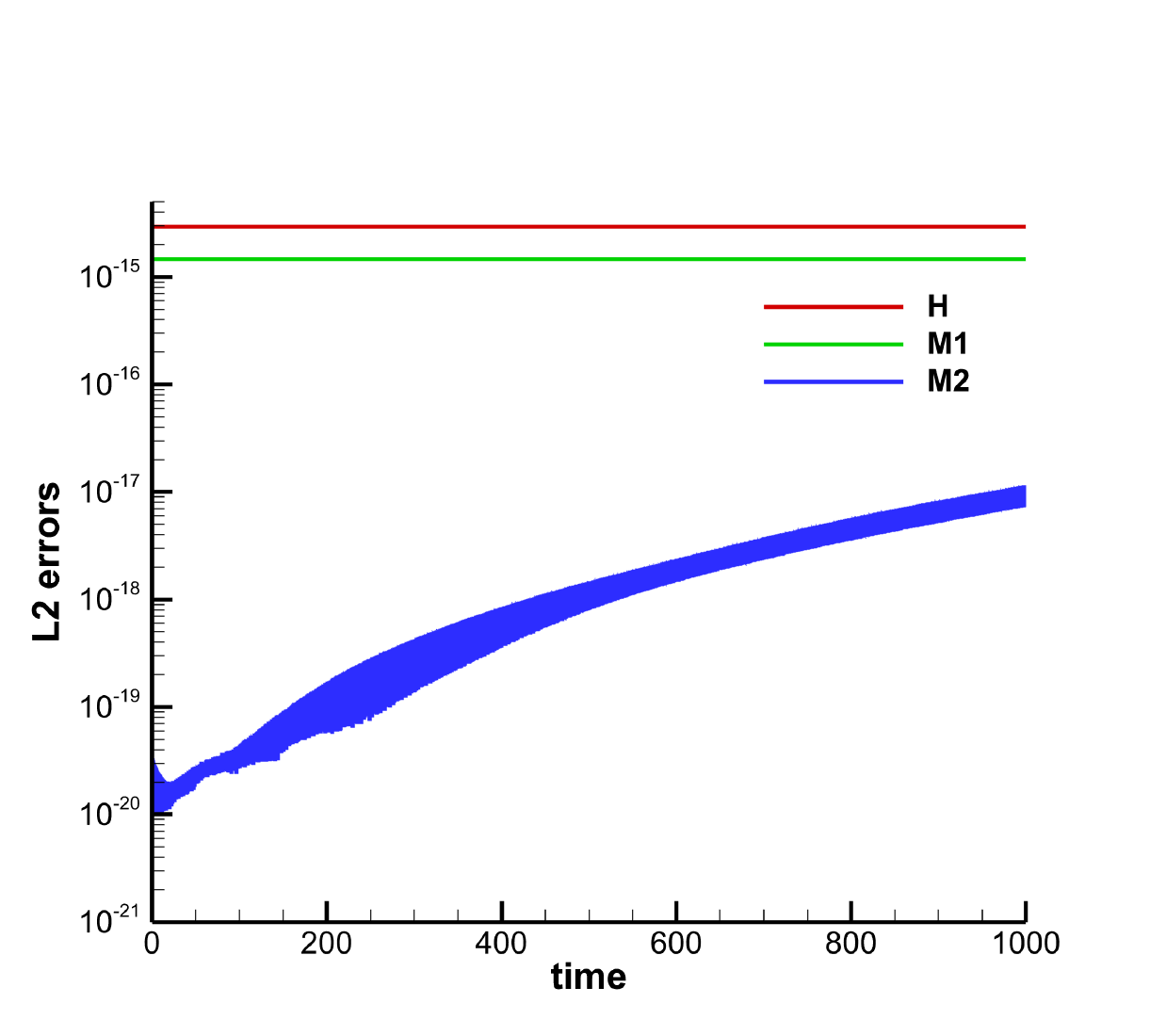}
		\caption{Linearized gravitational wave test solved with an ADER-DG scheme of order 6.
		Left panel: $K_{zz}$ component of the extrinsic curvature at the final time, compared to the exact solution. Right panel: Einstein constraints monitored all along the duration
		of the simulation.}
		\label{fig.linearized-grav-wave}
\end{figure}
Matter is absent in this test.  
To discretize the problem we consider a rectangular domain $[-0.5,0.5]\times[-0.2,0.2]$ with periodic boundary conditions, 
and we employ an unlimited ADER-DG scheme of order 6 on a mesh composed by $4 \times 4$ elements, which corresponds to 24 degrees of freedom in each direction.
We run our simulation until a final time of $t=1000$, 
corresponding to $1000$ crossing times\footnote{
We recall that, for tests in special relativity, having set $c=1$, the unit of time is the time taken by light to cover a
unit distance.}.
Figure~\ref{fig.linearized-grav-wave} shows the results of the calculation. 
In the left panel we present the numerical solution for the $K_{zz}$ component of the extrinsic curvature, 
at the final time, compared with the exact one. Essentially the same perfect matching is exhibited by the other quantities.
In the right panel we display instead the evolution of the Einstein constraints.
As evident, in this simulation the Hamiltonian and momentum constraints are all constant up to machine precision for the entire duration of the simulation.

\subsection{The gauge wave}
\label{sec:gauge-wave}

We continue the benchmarking of our numerical scheme and of the proposed first--order hyperbolic reformulation of the $Z4$ system with the so called \textit{gauge wave test}, 
also taken from~\cite{Alcubierre2004}. 
Here, the metric is given by 
\begin{equation}
	\label{eqn.gw.metric}
	d s^2 = - H(x,t) \, d t^2 + H(x,t) \, d x^2 + d y^2 + d z^2, \quad \text{where} \quad H(x,t) = 
	1-A\,\sin \left( 2\pi(x-t) \right),
\end{equation}
which describes a sinusoidal gauge wave of amplitude $A$ propagating along the $x$-axis.
 This means that the metric variables are set to 
$\gamma_{xx}=H$ and $\gamma_{yy}=\gamma_{zz}=1$ and the shift vector is $\beta^i=0$, 
hence the \emph{gamma--driver} is switched off ($s=0$). For this test the \emph{harmonic gauge condition} is used. 
The extrinsic curvature is again given by 
$K_{ij} = -\partial_t \gamma_{ij}/ (2\alpha)$, i.e.  
\begin{equation}
	K_{yy}=K_{zz}=K_{xy}=K_{xz}=K_{yz}=0 \qquad \text{and} \quad 
	K_{xx} = - \pi A\frac{\cos\left(2\pi(x-t)\right)}{\sqrt{1 - A\sin\left(2\pi(x-t)\right)}}\,.
\end{equation}
All the other quantities follow accordingly, with the lapse function given by $\alpha = \sqrt{H}$. Matter is absent also in this test problem. 
We emphasize that the present test case, even if it can be seen as a nonlinear reparametrization of the flat Minkowski spacetime, is far from trivial: indeed, it is reported that the first and second order formulation of the classical BSSNOK system fail for this
test after a rather short time, see~\cite{Alic:2011a,Brown2012}, and that the original version of the CCZ4 system was stable only in its damped formulation~\cite{Alic:2011a}. The first stable undamped  simulation 
was reported in~\cite{Dumbser2017strongly} for a first--order reformulation of the CCZ4 system.
Also here for this test we use an undamped version of the PDEs with $\kappa_1=0$, $\kappa_2=0$, while we have noticed that 
it is necessary to set $c=1$ in the gauge condition~\eqref{eqn.slicing} chosen with the harmonic version, i.e. $g(\alpha)=1$.

\begin{figure}[!htbp]
	\includegraphics[width=0.45\textwidth]{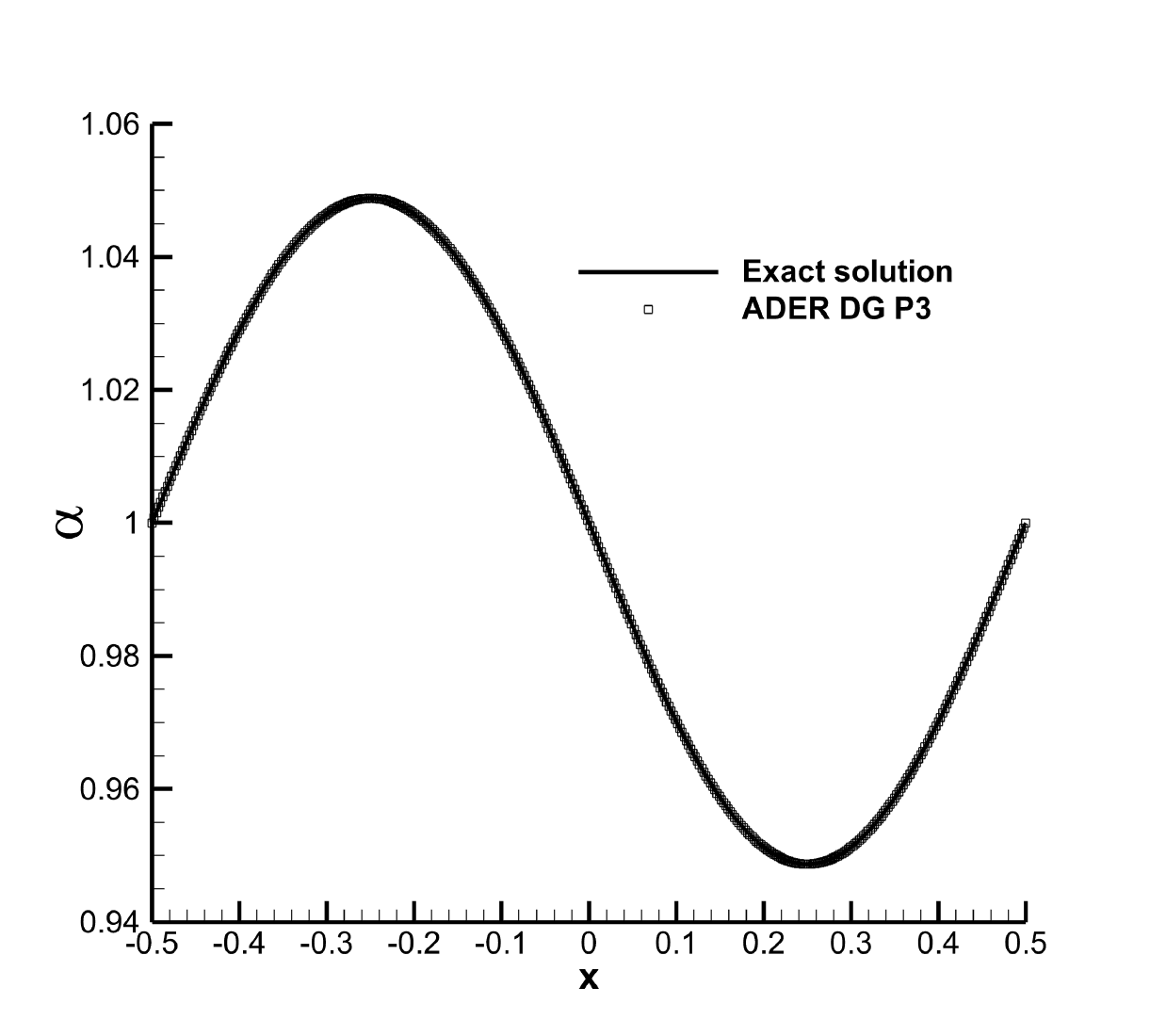}
	\includegraphics[width=0.45\textwidth]{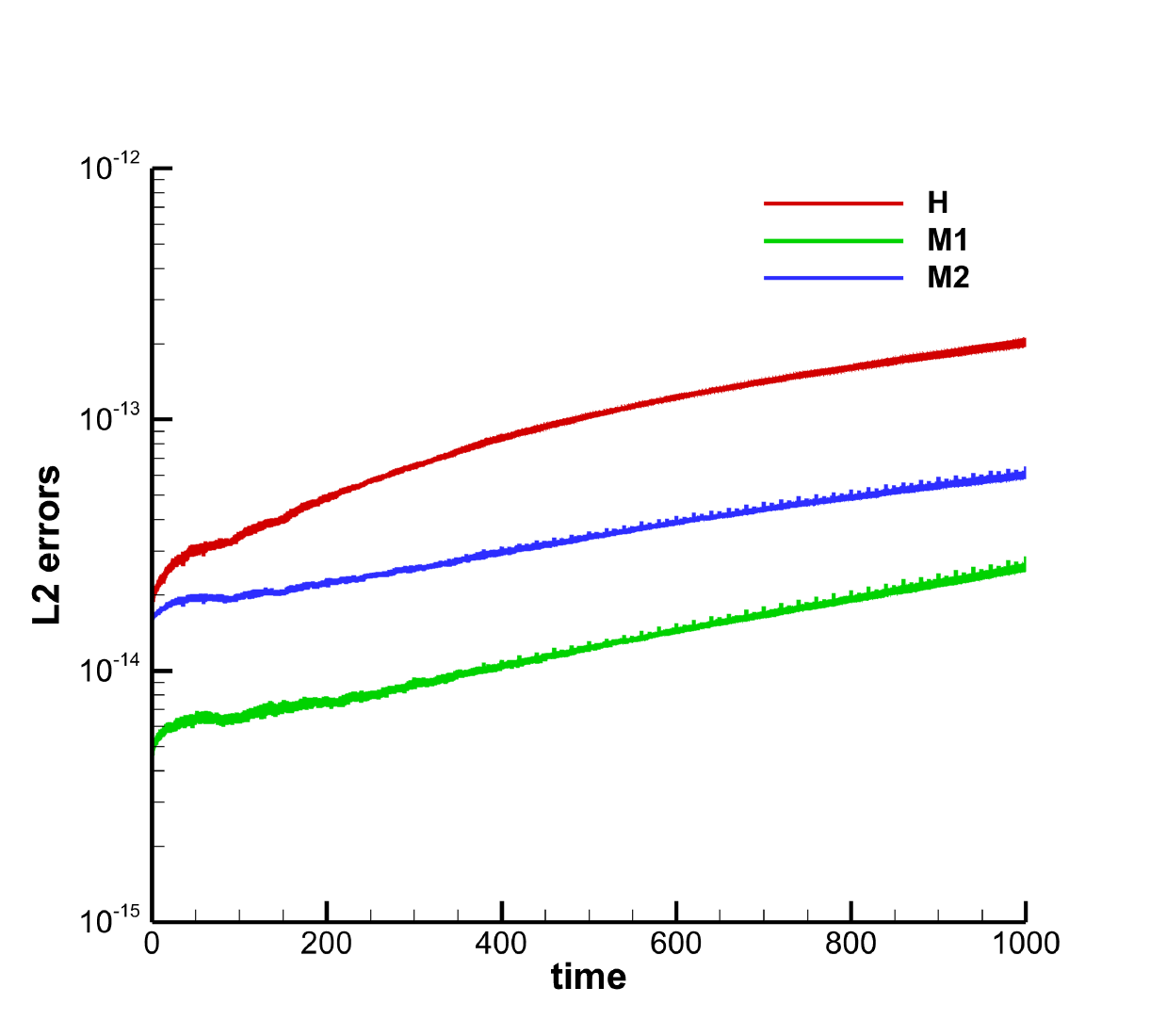}
	\caption{Solution of the gauge wave test at $t=1000$ with $A=0.1$ using an ADER DG scheme of order 4. 
	Left panel: profile of the lapse $\alpha$ compared to the exact solution. 
	Right panel: Evolution of the Einstein constraints. 
	}
	\label{fig.gaugewavetest-smallA}
\end{figure}

We have first run a test case  with a small wave amplitude $A=0.1$ over
a rectangular domain of size  $[-0.5,0.5] \times [-0.02, 0.02]$ with periodic boundary conditions.
We have used an ADER-DG P3 numerical scheme with a uniform grid composed of $100\times 4$ elements, evolving the system until $t=1000$.
Hence in the left panel of Figure~\ref{fig.gaugewavetest-smallA} we show  the profile of the lapse function $\alpha$ as a representative quantity, 
showing a perfect matching with the exact solution at the final time. In the right panel, on the other hand, we monitor as usual the Einstein constraints, which
manifest a moderate  linear growth all along the evolution.

Then, we have considered a large amplitude perturbation with $A = 0.9$, to the extent of performing a numerical convergence analysis of our scheme. The computational domain in this case 
is given by $[-0.5,0.5] \times [-0.05, 0.05]$.
The results, extracted from data at time $t=10$, are reported in Table~\ref{tab.conv-gaugewave} and confirm that the scheme reaches the nominal order of convergence.

 \begin{table}[!htbp] 
 \centering
 \numerikNine
 \begin{tabular}{|l|l||ccc|ccc|c|}
   \hline
   \multicolumn{9}{|c|}{\textbf{Gauge wave --- ADER-DG-$\mathbb{P}_N$}} \\
   \hline
   \hline
    & $N_x\times N_y$ & $L^1$ error & $L^2$ error & $L^\infty$ error & $L^1$ order & $L^2$ order & $L^\infty$ order &
   Theor. \\
   \hline
   \hline
   \multirow{4}{*}{\rotatebox{90}{{DG-$\mathbb{P}_2$}}}
& $40\times4$   & 1.2838E-03	& 4.8661E-03	& 2.5095E-02	&---	&---	&--- & \multirow{4}{*}{3}\\
& $60\times6$  	& 2.6423E-04	& 9.8619E-04	& 4.9053E-03	& 3.90	& 3.94	& 4.03 &\\
& $80\times8$  	& 8.2440E-05	& 3.0322E-04	& 1.5083E-03	& 4.05	& 4.10	& 4.10 &\\
& $100\times10$	& 3.3280E-05	& 1.2108E-04	& 6.0413E-04	& 4.07	& 4.11	& 4.10 &\\
   \cline{2-8}
   \hline
   \multirow{4}{*}{\rotatebox{90}{{DG-$\mathbb{P}_3$}}}
& $40\times4$	  & 5.3398E-05	& 2.0348E-04	& 1.0660E-03	& ---	& ---	& --- & \multirow{4}{*}{4}\\
& $60\times6$  	& 1.2460E-05	& 4.7006E-05	& 2.3760E-04	& 3.59	& 3.61	& 3.70 &\\
& $80\times8$	  & 4.1667E-06	& 1.5621E-05	& 7.7947E-05	& 3.81	& 3.83	& 3.87 &\\
& $100\times10$	& 1.7520E-06	& 6.5436E-06	& 3.2420E-05	& 3.88	& 3.90	& 3.93 &\\
   \cline{2-8}
   \hline
   \multirow{4}{*}{\rotatebox{90}{{DG-$\mathbb{P}_4$}}}
& $40\times4$	  & 1.8236E-06	& 6.7109E-06	& 3.3969E-05	& ---	& ---	& --- & \multirow{4}{*}{5}\\		
& $60\times6$  	& 1.6400E-07	& 5.8994E-07	& 2.8784E-06	& 5.94	& 6.00	& 6.09 &\\
& $80\times8$	  & 2.9500E-08 	& 1.0461E-07	& 4.9922E-07	& 5.96	& 6.01	& 6.09 &\\
& $100\times10$	& 7.7948E-09	& 2.7398E-08	& 1.2988E-07	& 5.96	& 6.00	& 6.03 &\\
   \cline{2-8}
   \hline
	 \multirow{4}{*}{\rotatebox{90}{{DG-$\mathbb{P}_5$}}}
& $40\times4$	  & 5.5287E-08	& 2.0571E-07	& 1.1845E-06	& ---	& ---	& --- & \multirow{4}{*}{6}\\		
& $60\times6$	  & 6.2100E-09 	& 2.2674E-08	& 1.1696E-07	& 5.39	& 5.44	& 5.71 &\\
& $80\times8$  	& 1.2027E-09	& 4.3669E-09	& 2.1883E-08	& 5.71	& 5.73	& 5.83 &\\
& $100\times10$	& 3.3009E-10	& 1.1974E-09	& 5.9321E-09	& 5.79	& 5.80	& 5.85 &\\
   \cline{2-8}
   \hline
   \multirow{4}{*}{\rotatebox{90}{{DG-$\mathbb{P}_6$}}}
& $40\times4$		& 2.8610E-09	& 1.0215E-08	& 5.2758E-08	& ---	& ---	& --- & \multirow{4}{*}{7}\\		
& $50\times5$		& 5.0341E-10	& 1.7825E-09	& 8.9322E-09	& 7.79	& 7.82	& 7.96 &\\
& $60\times6$		& 1.2258E-10	& 4.3434E-10	& 2.5857E-09	& 7.75	& 7.74	& 6.80 &\\
& $70\times7$		& 3.8840E-11	& 1.3929E-10	& 1.0035E-09	& 7.46	& 7.38	& 6.14 &\\
   \cline{2-8}
   \hline
 \end{tabular}
 \caption{Numerical convergence results for the gauge wave test at $t=10$ with a wave amplitude $A=0.9$. In the table we report the $L_1$, $L_2$, $L_{\infty}$
		error norms and the corresponding numerical order of convergence for the lapse $\alpha$.}
		\label{tab.conv-gaugewave}
 \end{table}

\subsection{The robust stability test}
\label{sec:stability-test}

Another important validation for any numerical GR code is represented by the so--called {\emph {robust stability test}} in a flat Minkowski spacetime without matter, already 
treated by~\cite{Alcubierre2004,Dumbser2017strongly}. It consists of a random  perturbation with 
amplitude $\pm 10^{-7}/\varrho^2$ which is 
applied to all quantities of the PDE system in a flat  Minkowski spacetime. The amplitude of the perturbation that we have chosen is three orders of magnitude
higher than that reported in~\cite{Alcubierre2004}.
The computational domain is given by the square $[-0.5;0.5]\times[-0.5;0.5]$,
for which we have considered
four simulations with an unlimited ADER-DG $P_3$ scheme on 
a sequence of refined meshes formed by $10 \varrho \times 10 \varrho$
elements, where $\varrho \in \left\{ 1, 2, 4, 8 \right\}$ is the refinement factor.

\begin{figure}[!htbp]
\begin{center}
    \begin{tabular}{cc}
  \includegraphics[width=0.45\textwidth]{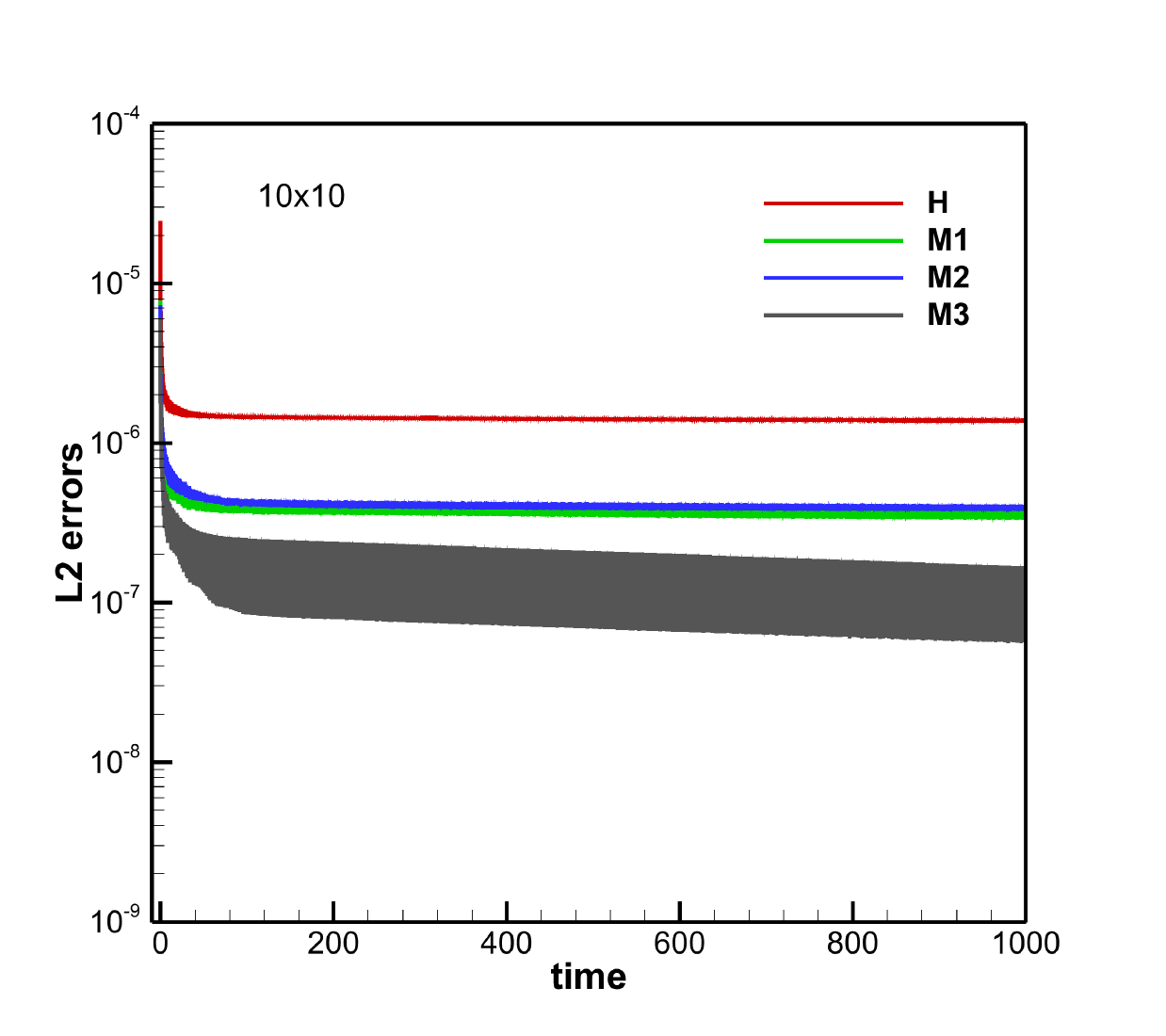} &
  \includegraphics[width=0.45\textwidth]{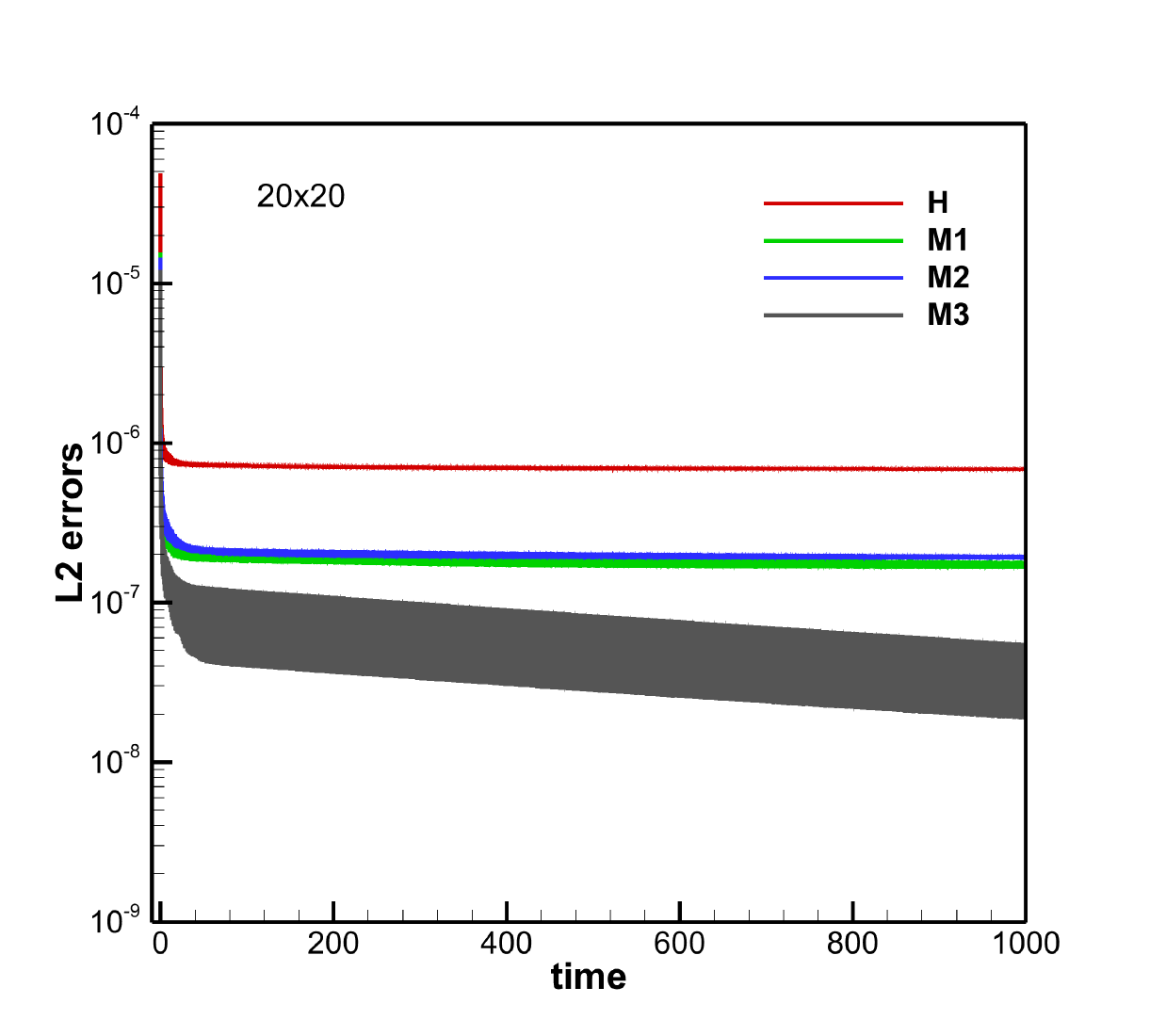} \\
  \includegraphics[width=0.45\textwidth]{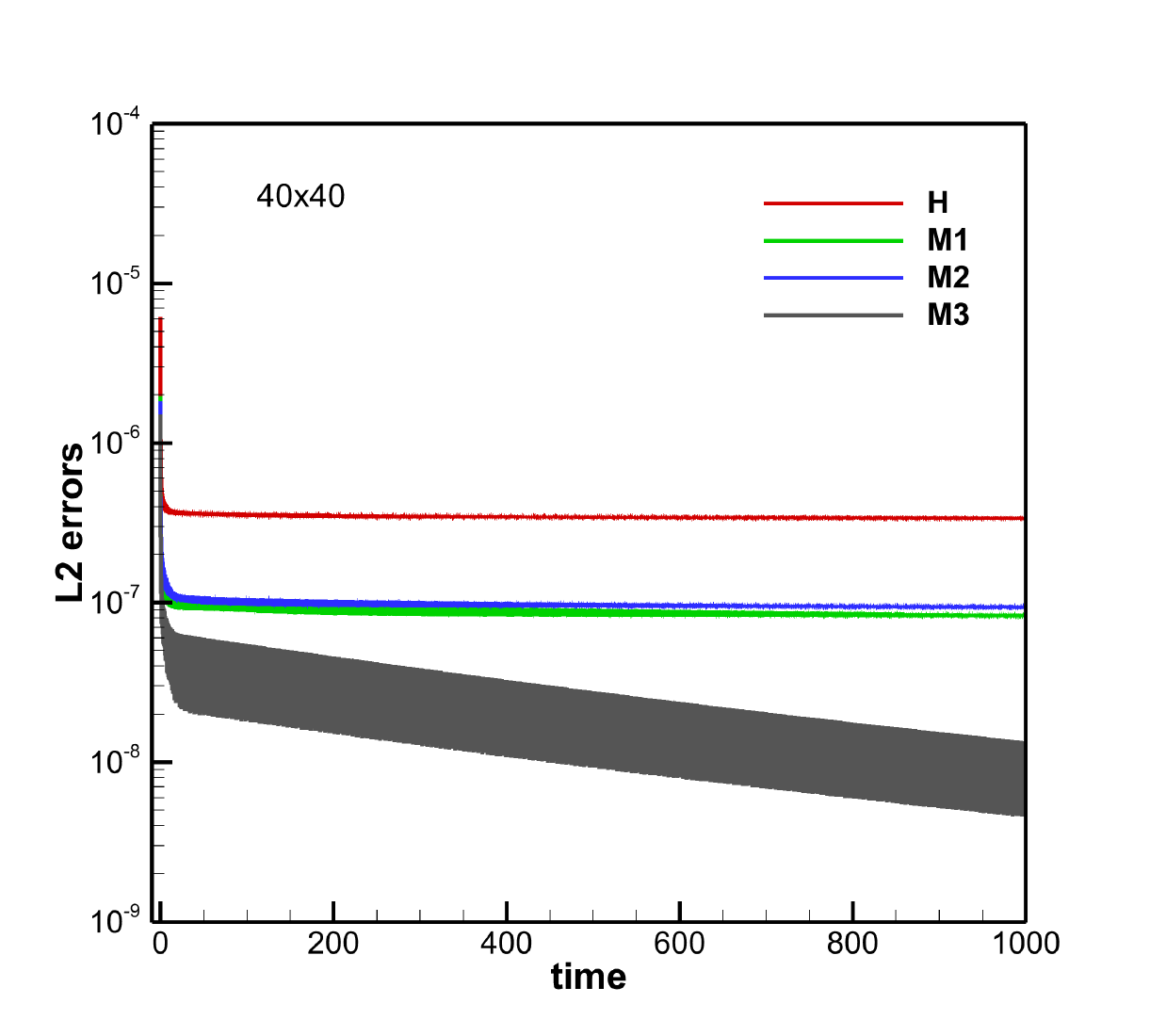} &
  \includegraphics[width=0.45\textwidth]{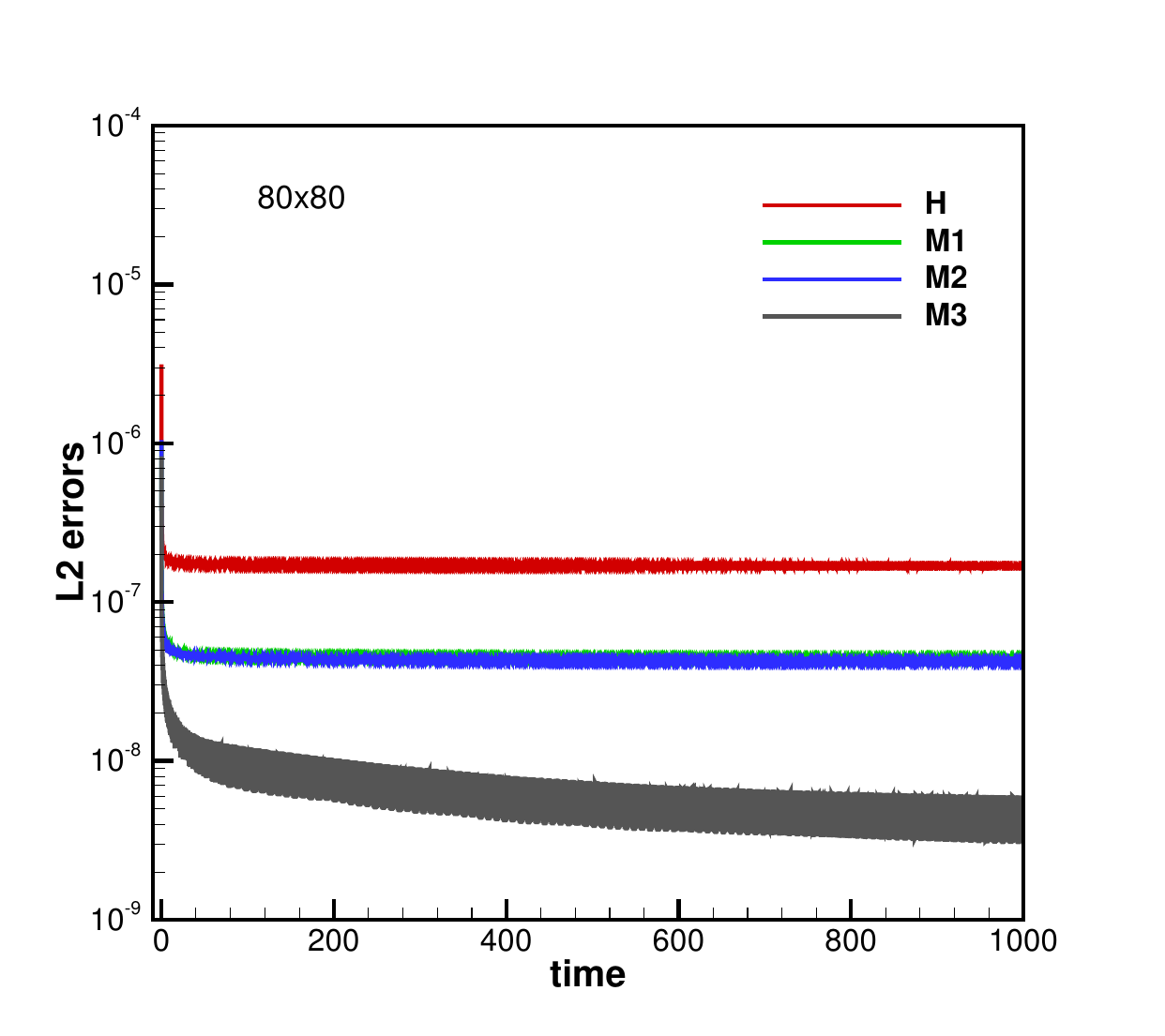} \\
    \end{tabular}
    \caption{Robust stability test case  with a random initial perturbation of amplitude
      $10^{-7}/\rho^2$ in all quantities on a sequence of successively
      refined meshes on the unit square in 2D.
			The {\emph {gamma--driver}} shift condition, $1+\log$ slicing and ADER-DG $P_3$
      scheme have been used. Top left: $10\times10$ elements, corresponding to $40\times40$
      degrees of freedom ($\varrho=1$). Top right: $20\times20$ elements,
      corresponding to $80\times80$ degrees of freedom ($\varrho=2$). Bottom
      left: $40\times40$ elements, corresponding to $160\times160$ degrees of
      freedom ($\varrho=4$). Bottom right: $80\times80$ elements, corresponding
      to $320\times320$ degrees of freedom ($\varrho=8$). }
    \label{fig.robstab}
\end{center}
\end{figure}

This is also a test for the {\emph {gamma--driver}} shift condition, which, in principle, would not be necessary for this kind of problem but is nevertheless activated
to solve the PDE system in its full generality. 
The other relevant parameters have been chosen as $\kappa_1=0$, $\kappa_2=0$, $c=0$, $\mu=0.2$, 
$\eta=0$, see \eqref{g-driver2} and \eqref{eqn.B}. Fig.~\ref{fig.robstab} shows
the results of our calculations, where we have reported the evolution of the four Einstein constraints for a sample of progressively refined meshes.
The unit of time is again the travel time taken by light to cover the edge of the square domain.

\subsection{Spherical Michel accretion}
\label{sec:spherical-accretion}
As a further test, we have evolved the transonic spherical accretion solution of matter onto a Schwarzschild black hole 
obtained by~\cite{michel1972accretion} (see also~\cite{Rezzolla_book:2013} for a modern presentation). 
We recall that this is not a solution of the full Einstein--Euler equations,
but rather just of the Euler equations in the stationary background spacetime of a non--rotating black hole.
However, if the whole mass accretion rate is small enough, we can
neglect the increase of the black hole mass that would in principle be produced by the accreted matter.
Under such circumstances we can consistently evolve the Euler equations while freezing the evolution of the metric,
i.e. assuming what is referred to as the Cowling approximation~\cite{Cowling41}.

The numerical details for obtaining the initial conditions can be found in~\cite{Anton06}. We have performed 
this simulation in spheroidal Kerr--Schild coordinates (see case 1. of Sect.~\ref{sec.coord}) 
over a two dimensional computational domain given by $(r,\theta)\in [0.5;10]\times[0+\epsilon;\pi-\epsilon]$, with $\epsilon=0.005$ and covered by a $50\times32$ uniform grid.
The critical radius, where the flow becomes supersonic, is $r_c=5$ (inside the computational domain). 
We choose the critical density (density at the critical radius) $\rho_c=1.006\times10^{-7}$ such that the mass accretion rate (computed as $4\pi r_c^2 \rho_c u_c^r$) is 
$-1.0\times10^{-5}$, meaning that the total mass accreted onto the black hole from $t=0$ to $t=1000\,M$ is just $1/100$ of  the total mass $M$ of the central black hole,
thus justifying the physical assumption of a stationary spacetime.  We stress that, with these parameters characterized by very low rest mass densities, the test becomes
extremely challenging from the numerical point of view, in spite of the solution being smooth and regular\footnote{For a comparison, the rest mass density chosen in~\cite{DelZanna2007} was much higher, giving a mass accretion rate
$ r_c^2 \rho_c u_c^r=-1$.}.

The equation of state is that of an ideal gas with adiabatic index $\gamma=5/3$.
\begin{figure}[!htbp]
\begin{center}
\begin{tabular}{cc} 
{\includegraphics[angle=0,width=7.3cm,height=7.3cm]{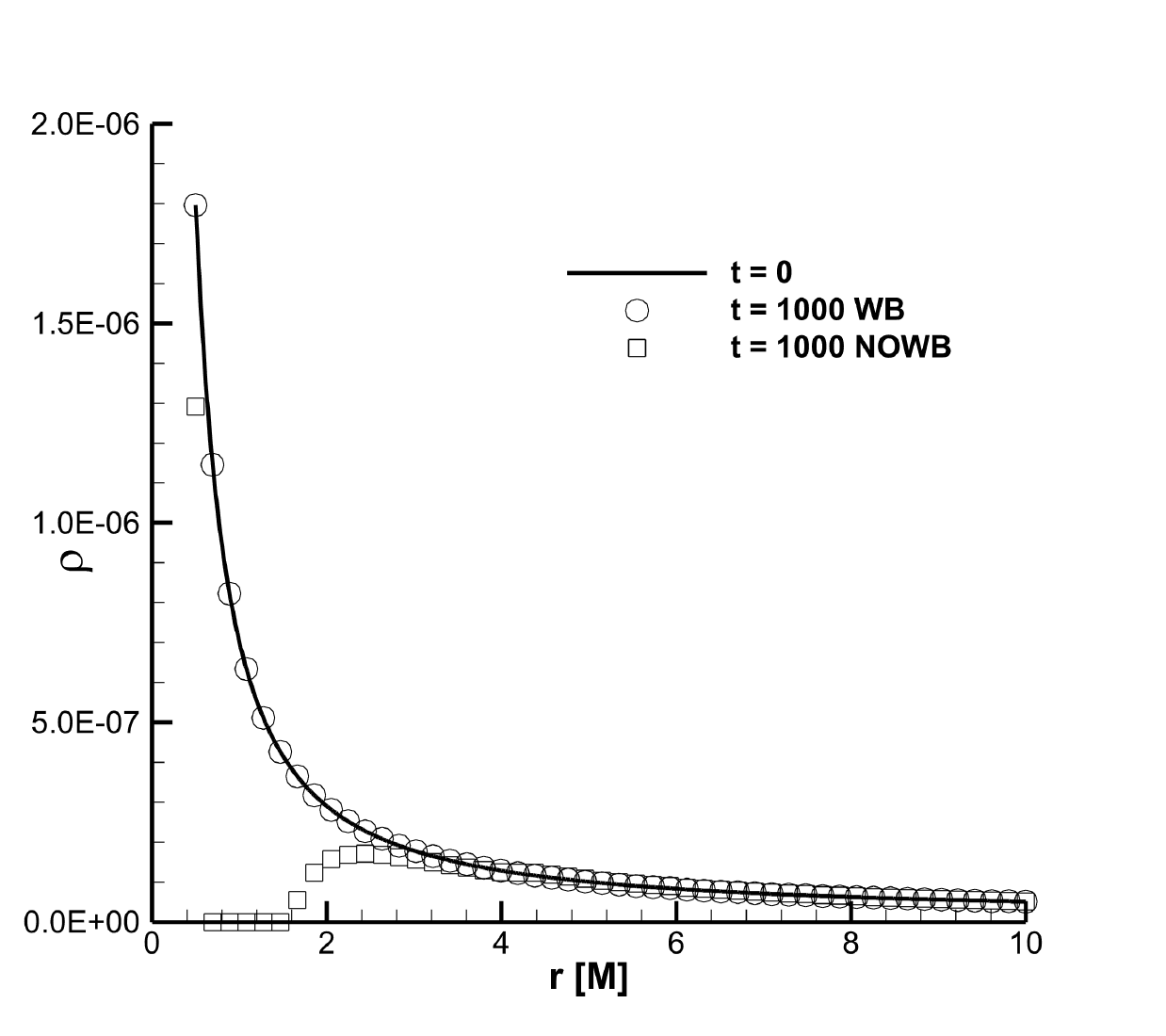}} &
{\includegraphics[angle=0,width=7.3cm,height=7.3cm]{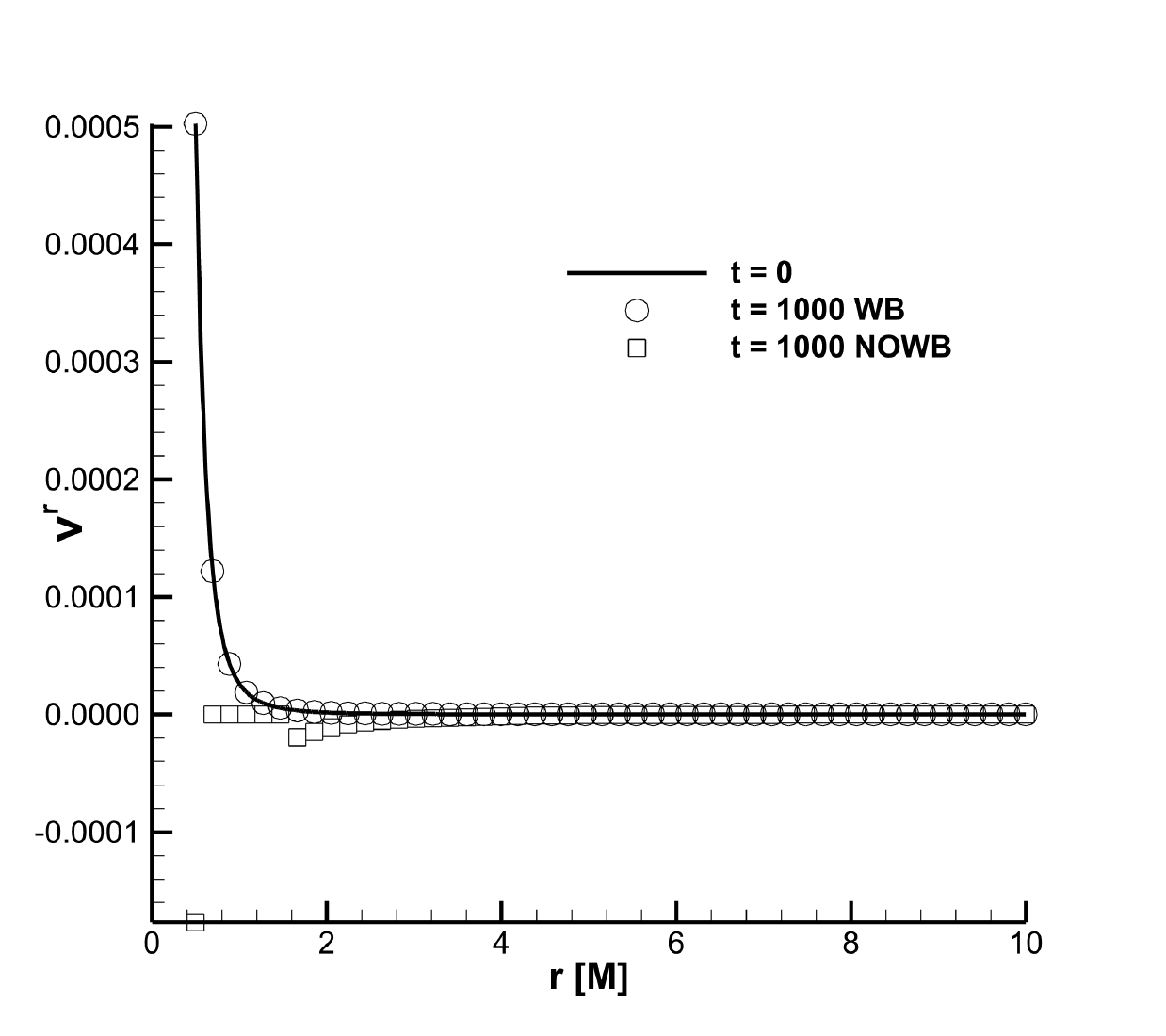}} \\
\end{tabular} 
\caption{Solution of the spherical accretion of matter onto a non-rotating black hole. The final rest mass density (left panel) and the radial velocity (right panel)
at time $t=1000\,M$ are 
compared to their initial profiles. } 
\label{fig:SA-1D}
\end{center}
\end{figure}
At time $t=0$, the rest mass density of the exact solution is perturbed by a Gaussian profile peaked at the critical radius, with an amplitude given by
$\delta\rho = 10^{-3}\rho_c$. 
We have solved this test by considering only the hydrodynamic section of the system~\eqref{eqn.rho}--\eqref{eqn.D}, thus adopting the Cowling approximation.
The numerical scheme is a pure DG scheme at fourth order of accuracy ($N=3$), 
while the other relevant parameters have been chosen as $\kappa_1=0.01$, $\kappa_2=0$, $c=0$, with no {\emph {gamma--driver}}.
We have performed two simulations to the final time $t=1000\,M$, the first one 
with the new well-balancing technique described in Sect.~\ref{sec.numerical}, and a second one 
without it, obtaining rather different results. 
Figure~\ref{fig:SA-1D}  reports the one dimensional profiles of the solution for the rest mass density and for the radial velocity $v^r$ at the final time compared to the exact solution.
If no well-balancing is adopted,
the solution quickly deteriorates,  amounting to a sequence of failures in the recovering of the primitive variables, as can be seen by the zero density values 
reported in the left panel of Figure~\ref{fig:SA-1D}. If the well-balancing is used instead, the 
exact solution is recovered and stationarity is preserved.
We recall that the positive values of the
the radial velocity, which are somewhat counter intuitive given that matter is falling into the black hole with increasing velocity, are
a spurious effect of the Kerr--Schild coordinates, which generate a positive radial shift.     
%
%

We also stress that in these regimes of low density matter, using the filter described in Sect.~\ref{sec:cons2prim} is absolutely crucial, and the simulation encounters
a sequence of catastrophic failures before $t\sim 5M$ if no filter is adopted, irrespective of the well-balancing property being activated, or not.

\subsection{Single stationary black holes in two and three space dimensions}
\label{sec:single-bh}

The Schwarzschild solution, historically the first exact solution that was found for the Einstein field equations, describes the spacetime around a non--rotating black hole and it represents a static solution of the Einstein field equations. 
A generalization to rotating black holes is the stationary Kerr solution. For all simulations reported in this section, the mass of the black hole is $M=1\,M_{\odot}$. In all tests presented here, matter is absent.

\begin{figure}[!htbp]
	\begin{center}
		\begin{tabular}{cc} 
			\includegraphics[trim=10 10 10 10,clip,width=0.45\textwidth]{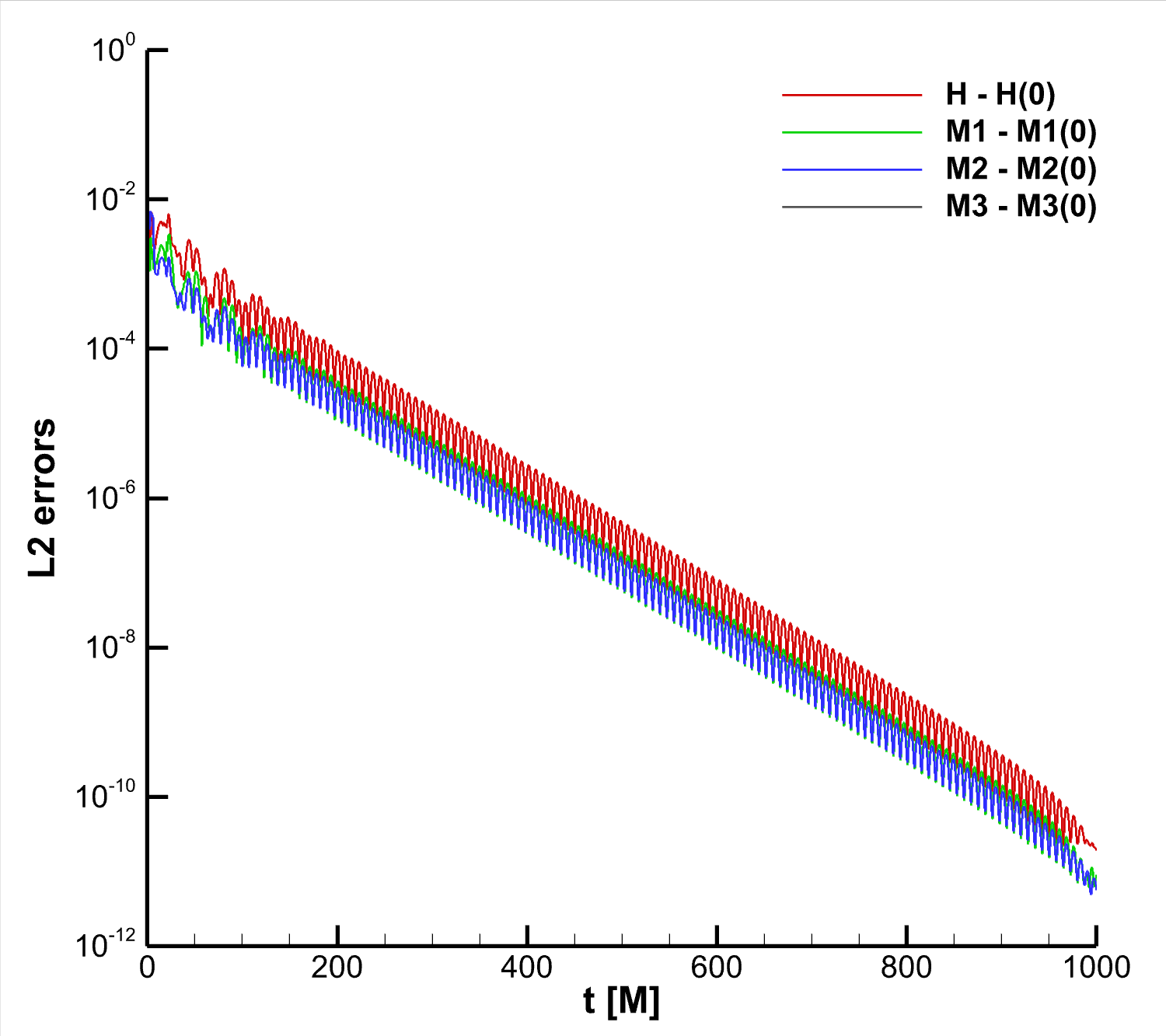} &
			\includegraphics[trim=10 10 10 10,clip,width=0.45\textwidth]{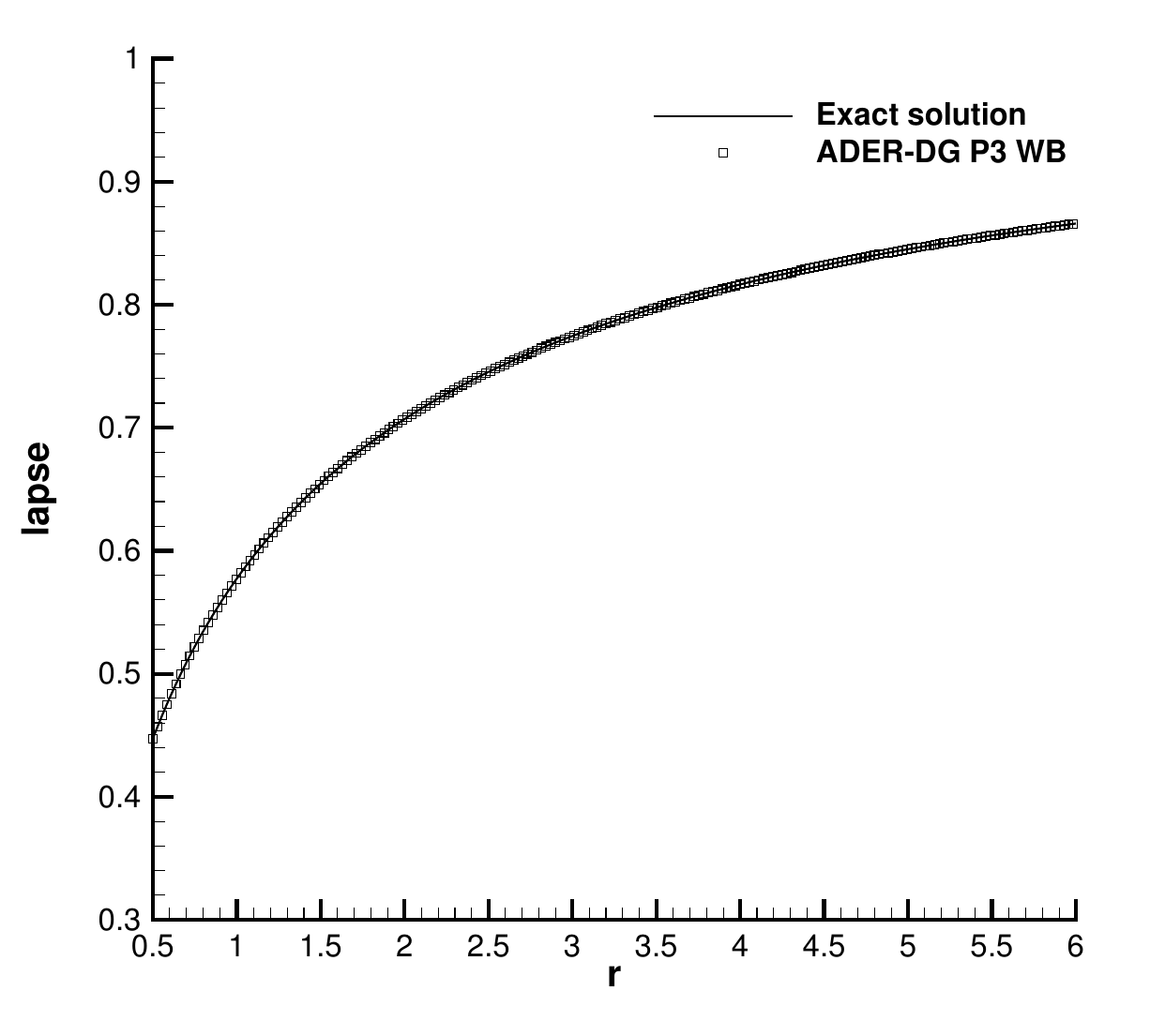} \\  
			\includegraphics[trim=10 10 10 10,clip,width=0.45\textwidth]{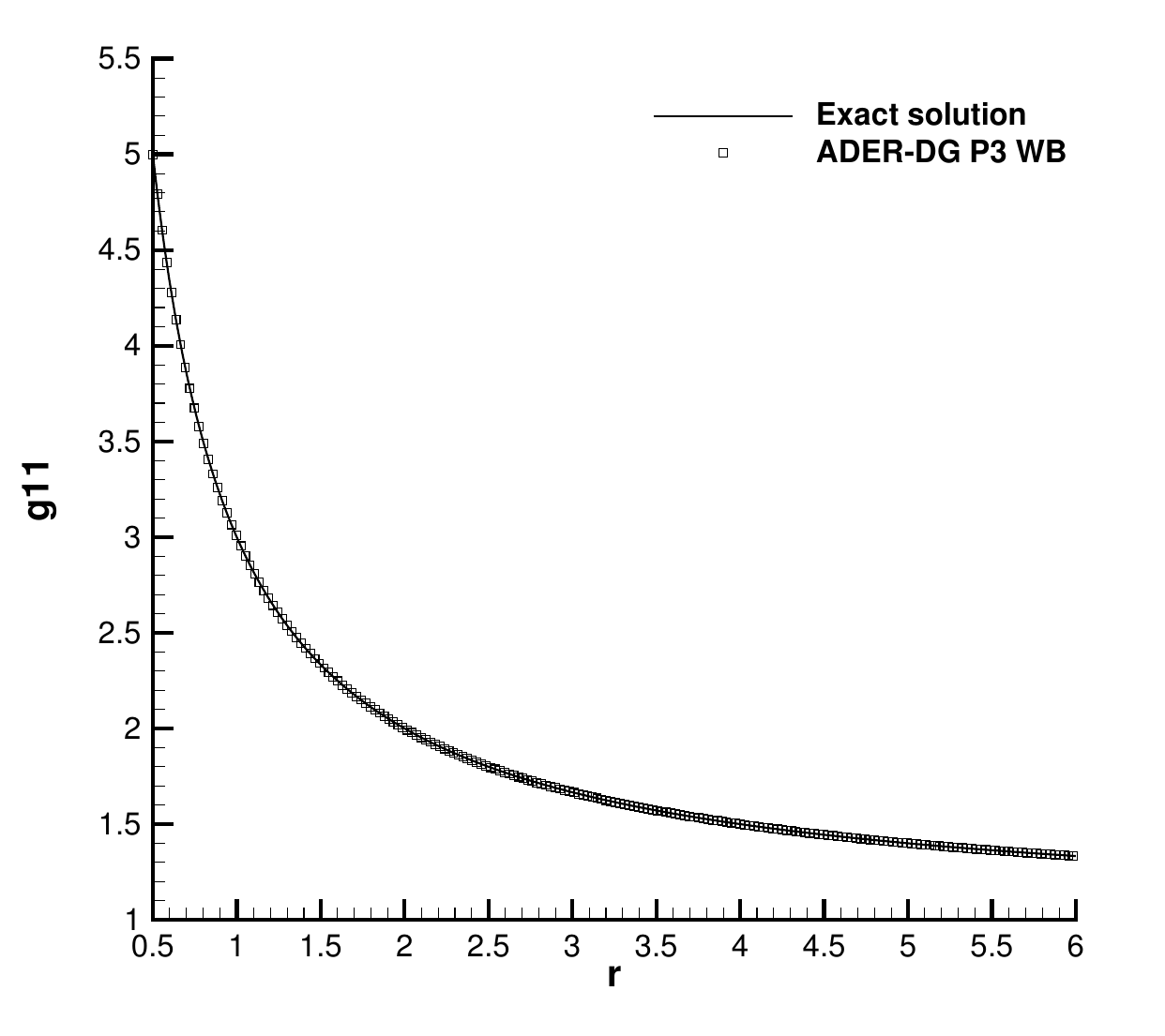} &  
			\includegraphics[trim=10 10 10 10,clip,width=0.45\textwidth]{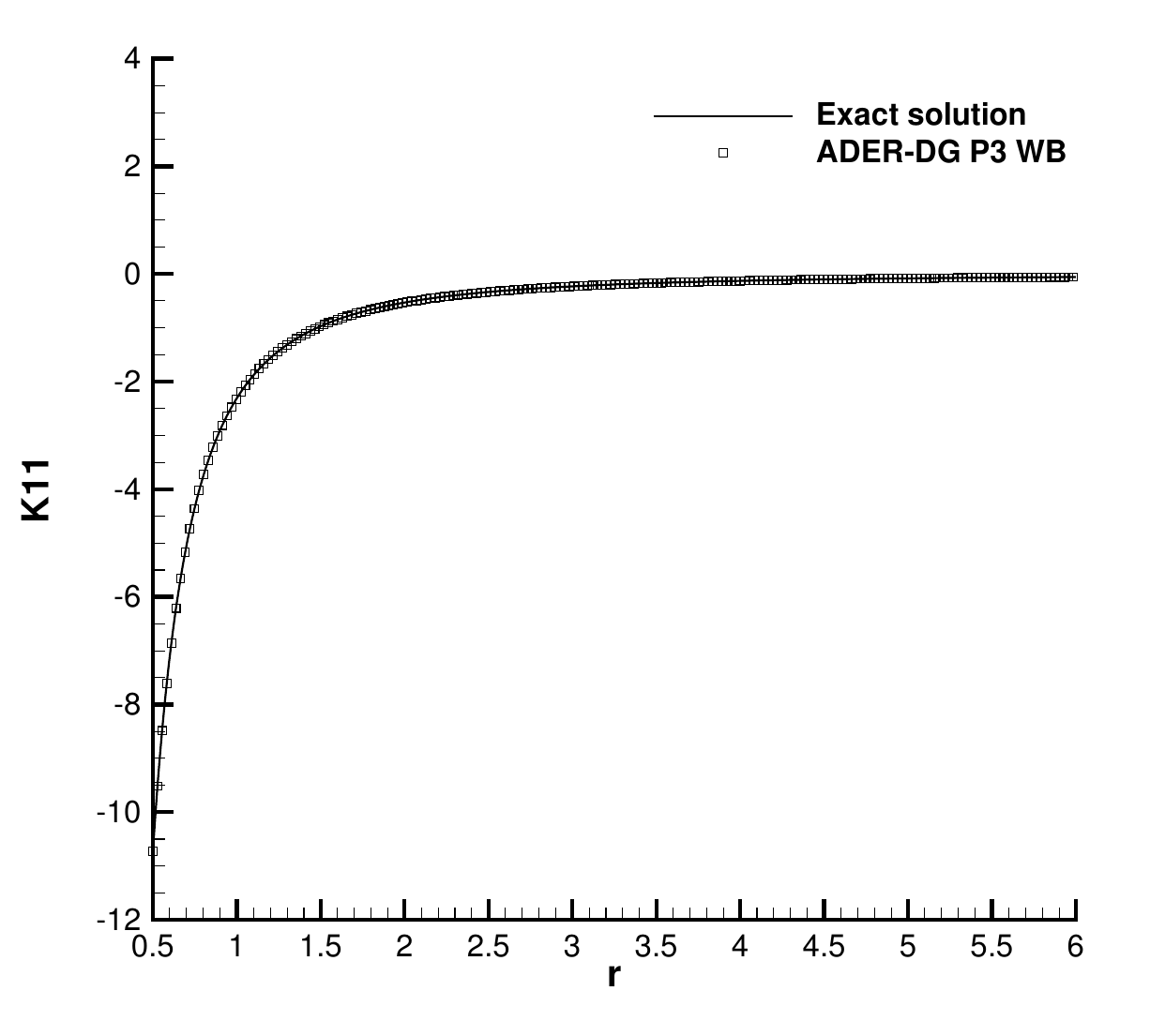}  
		\end{tabular} 
		\caption{2D simulation of an initially perturbed Schwarzschild black hole ($a=0$) in the 2D plane $r-\theta$ using spherical Kerr-Schild coordinates. Top left: time series of the constraint violations until time $t=1000\,M$. It is clearly visible that the initial perturbation decays exponentially in time and that the numerical solution returns to the stationary equilibrium. From top right to bottom right: 1D cuts along the radial direction at $\theta = \pi/2$ for the lapse $\alpha$, the metric tensor component $\gamma_{11}$ and the extrinsic curvature component $K_{11}$ at time $t=1000\,M$ and comparison with the exact solution. }  
		\label{fig:Kerr2Da0}
	\end{center}
\end{figure}

\paragraph{Non-rotating black hole in 2D}
In our first simulation we solve the Z4 equations for a Schwarzschild black hole ($a=0$) in spherical  Kerr--Schild coordinates, see Sect.~\ref{sec.coord}. The two--dimensional computational domain in the $r-\theta$ plane is chosen as 
$\Omega = [0.5,6] \times [\delta,\pi-\delta]$, with $\delta = 0.1415926535$. The domain $\Omega$ is discretized with $80 \times 40$ elements. On all boundaries we prescribe the initial condition as Dirichlet boundary condition for all state variables. We use the fourth order version ($N=3$) of our new exactly well-balanced ADER-DG scheme based on the HLL Riemann solver and without any subcell FV limiter. Concerning the Z4 system we use the \emph{1+log gauge condition}
and set $c=0$, $\kappa_1=1.0$, $\kappa_2=1.0$ and $s=0$, 
i.e. the shift is not evolved in time. 
In order to study the behaviour of the new well-balanced scheme in the presence of a \textit{small perturbation}, the initial condition for the cleaning variable $\Theta$ is chosen as 
\begin{equation}
	\Theta(0,\mathbf{x}) = A_0 \exp \left( -\halb \frac{(X-4)^2 + (Y-0)^2}{\sigma^2} \right)\,,
\end{equation}   
with $(x_1,x_2)=(r,\theta)$, $A_0 = 10^{-3}$, $\sigma=0.2$, $X = r \sin \theta$ and $Y = r \cos \theta$. We expect that during the simulation the perturbation leaves the computational domain and that for large enough times the solution returns back to the exact stationary equilibrium solution. The computational results obtained for this simulation are shown in Figure~\ref{fig:Kerr2Da0}. In the top left panel we plot the $L^2$ norms of the constraint violations $H(t)-H(0)$ and $M_i(t) - M_i(0)$ for the Hamiltonian and the momentum constraints. As expected, the initial perturbation of the order $10^{-3}$ decays exponentially in time and the solution returns back to the exact equilibrium. To the best knowledge of the authors, this is the first long-time simulation ever carried out for the Einstein field equations using a high order exactly well-balanced discontinuous Galerkin finite element scheme and where, after an initial perturbation, the discrete solution returns back to the exact steady equilibrium solution. In the remaining panels of Figure~\ref{fig:Kerr2Da0} we show one dimensional profiles obtained from cuts along the equatorial plane, for various representative quantities  like $\alpha$, $\gamma_{11}$, $K_{11}$. As apparent from the figure, 
perfect agreement with the exact stationary solution is obtained at the final time $t=1000\,M$.

\paragraph{Numerical study of the well-balancing property} 
We now repeat the previous test of the non-rotating black hole in 2D until $t=0.1$ using a fourth order ADER-DG scheme ($N=3$) on $40 \times 20$ elements and employing three different machine precisions, 
namely single, double and quadruple precision.  
We set the perturbation amplitude $A_0$ so that it corresponds to the respective machine precision. The values of $A_0$ as well as the obtained 
$L^{\infty}$ error norms are reported in Table \ref{tab.wb} at time $t=0.1$ for several components of the Z4 system and for all chosen machine precisions. The computational results clearly show that the errors remain of the order of machine precision, hence the new numerical method proposed in this paper 
is well-balanced also in its practical implementation, as expected. 

 \begin{table}[!t] 
 	\renewcommand{\arraystretch}{1.1}
	\caption{Numerical well-balancing test with a fourth order ADER-DG scheme
		using single, double and quadruple precision. 
		$L^{\infty}$ error norms for several quantities of the Z4 system at time $t=0.1$.}  
	\begin{tabular}{|l||ccc|}
		\hline
		Quantity & single precision, {\small $A_0=10^{-8}$} & double precision, {\small $A_0=10^{-16}$}  & quadruple precision, {\small $A_0=10^{-28}$}\\
		\hline 
		\hline
		$\alpha$      & 7.4505806E-06  & 3.2196468E-015 &  4.7282543E-030 \\ 
		$\gamma_{11}$ & 7.8201294E-05  & 3.3306691E-014 &  2.2768344E-030 \\ 
		$K_{11}$      & 8.1777573E-05  & 3.2862602E-014 &  2.8324170E-029 \\ 
		$K_{12}$      & 2.7160518E-06  & 1.9922607E-015 &  1.5911163E-029 \\ 
		$\Theta$      & 2.6383780E-06  & 1.6878889E-015 &  3.4825676E-029 \\ 
		$Z_1$         & 9.8760290E-07  & 1.3437779E-015 &  2.3198075E-029 \\ 
		$A_1$         & 9.4473362E-06  & 4.8849813E-015 &  1.8991764E-029 \\ 
		$D_{111}$     & 3.3855438E-05  & 1.3766766E-014 &  4.5212168E-030 \\ 	 
		\hline
	\end{tabular}
	\label{tab.wb}
\end{table}

\paragraph{Non-rotating black hole in 3D}

\begin{figure}[!htbp]
	\begin{center}
		\begin{tabular}{cc} 
			\includegraphics[trim=10 10 10 10,clip,width=0.45\textwidth]{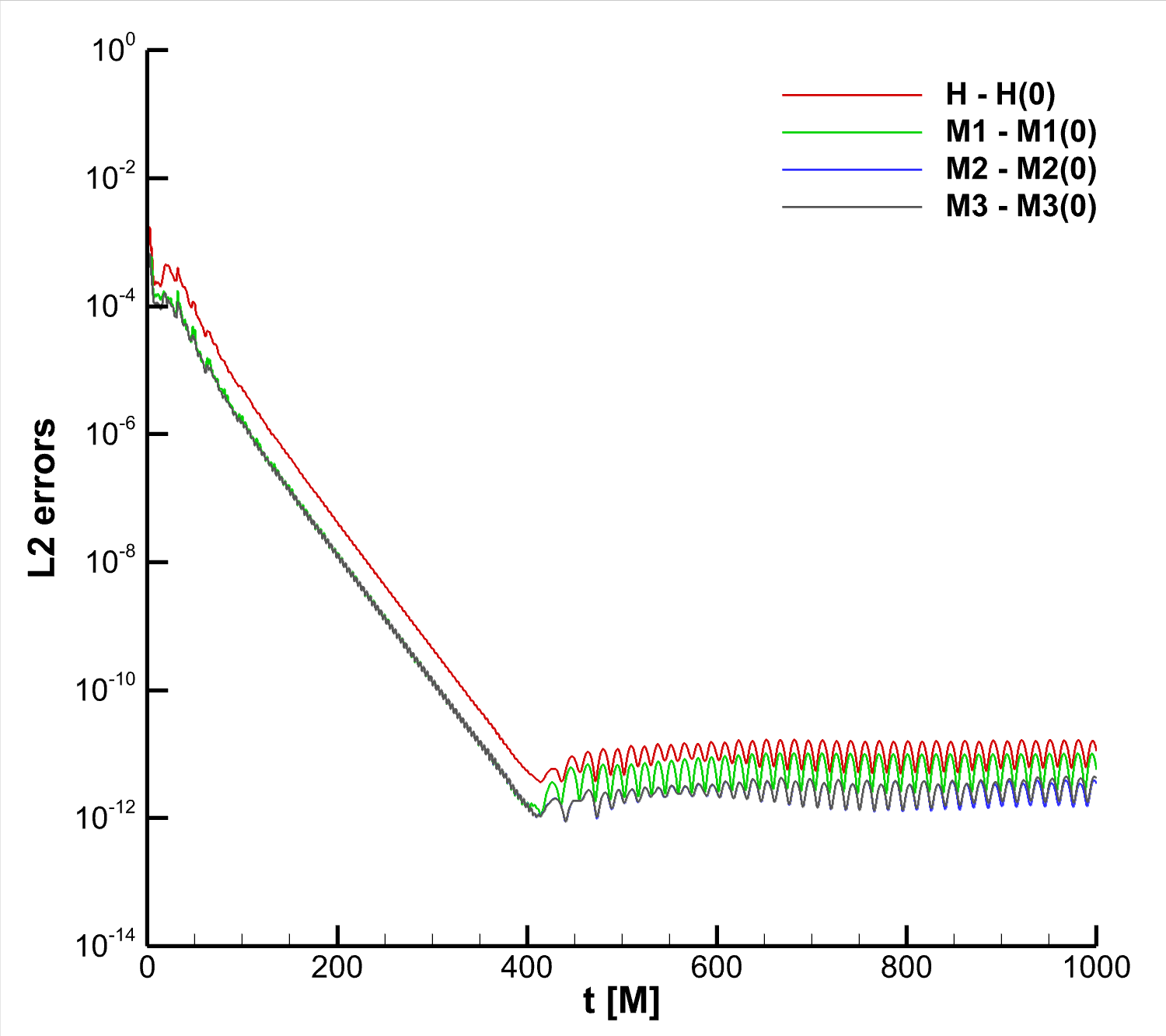} &
			\includegraphics[trim=10 10 10 10,clip,width=0.45\textwidth]{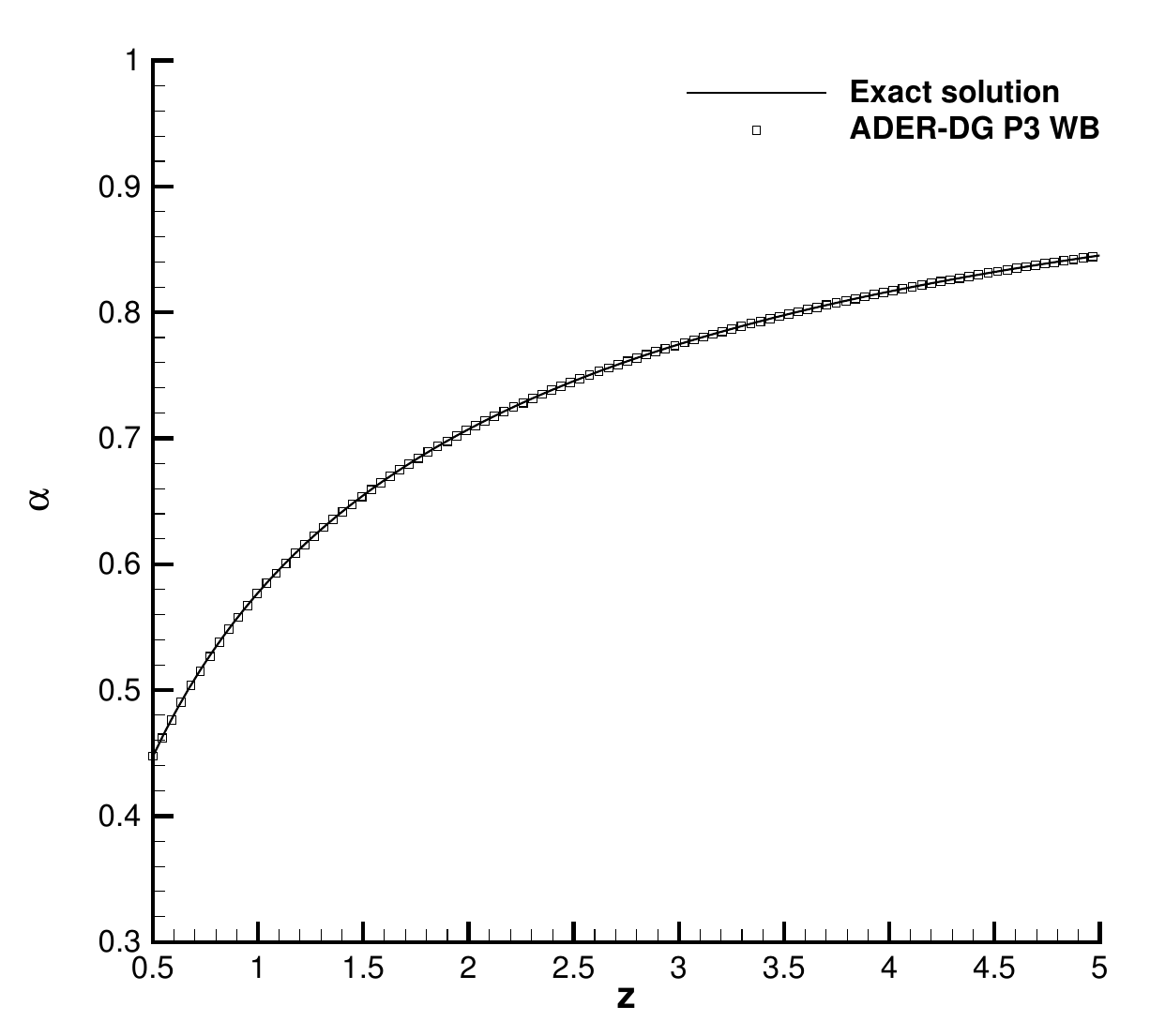} \\  
			\includegraphics[trim=10 10 10 10,clip,width=0.45\textwidth]{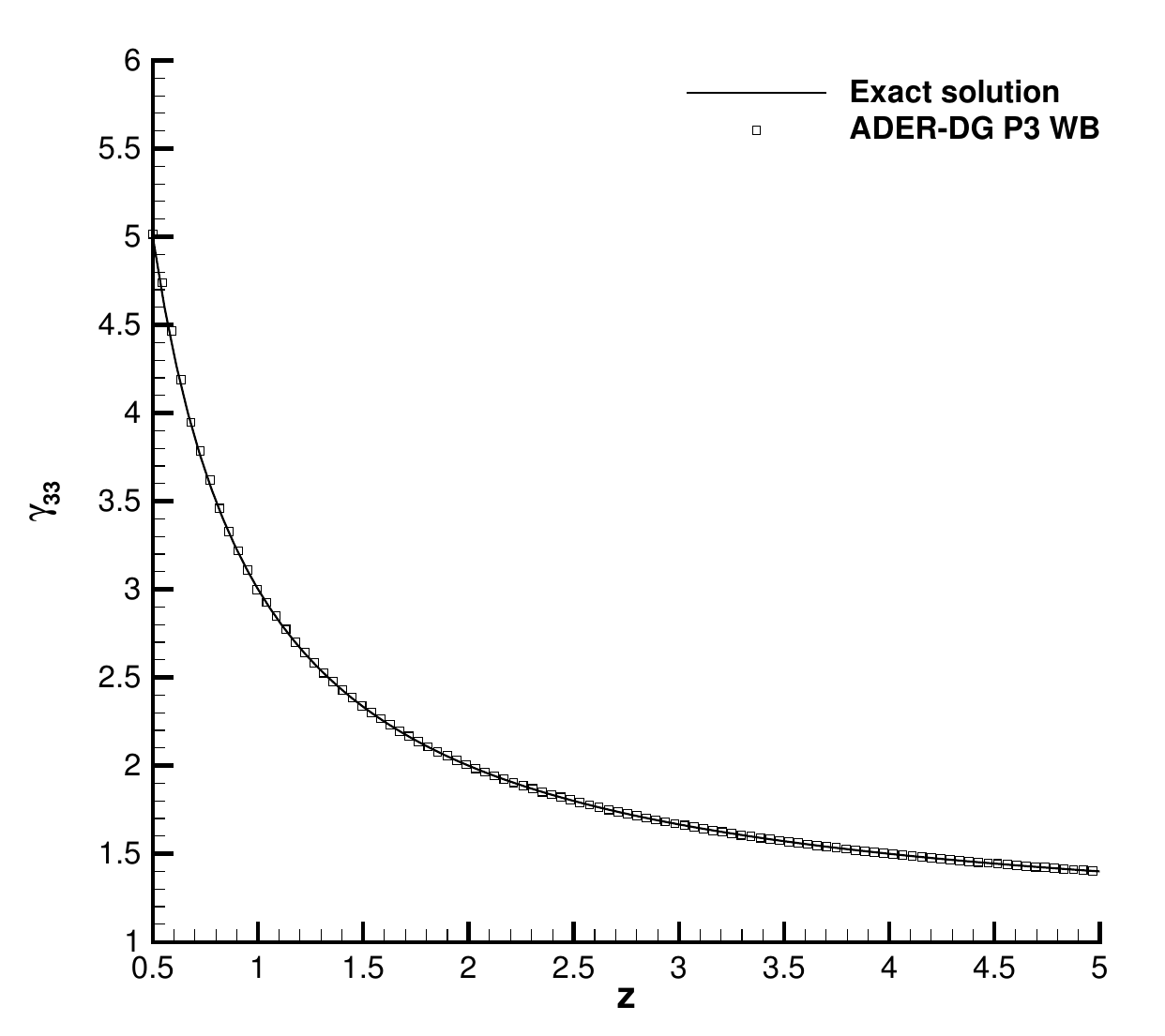} &  
			\includegraphics[trim=10 10 10 10,clip,width=0.45\textwidth]{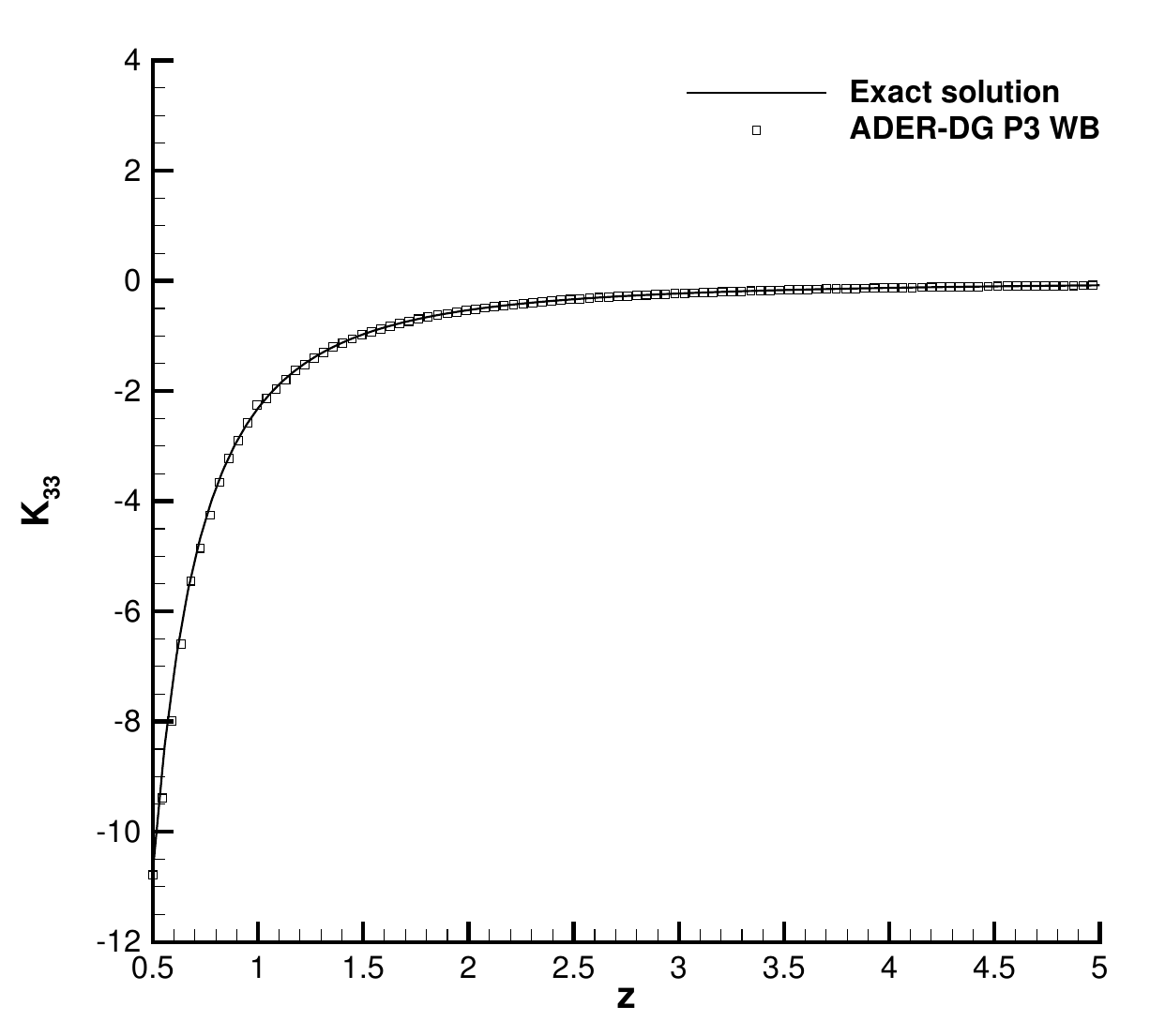}  
		\end{tabular} 
		\caption{3D simulation of an initially perturbed Schwarzschild black hole (spin $a=0$) in 3D Cartesian Kerr-Schild coordinates using a fourth order well-balanced ADER-DG scheme. Top left: time series of the constraint violations until time $t=1000\,M$. It is clearly visible that the initial perturbation decays exponentially in time and that the numerical solution returns to the stationary equilibrium. From top right to bottom right: 1D cuts along the z axis ($x=y=0$) for the lapse $\alpha$, the metric tensor component $\gamma_{11}$ and the extrinsic curvature component $K_{11}$ at time $t=1000\,M$ and comparison with the exact solution. }  
		\label{fig:Kerr3Da00}
	\end{center}
\end{figure}

We have then evolved the same stationary Schwarzschild black hole ($a=0$) in three space dimensions by choosing the 3D Cartesian Kerr--Schild coordinates already discussed in  Sect.~\ref{sec.coord}. 
The computational domain is the box $[-5;5]\times[-5;5]\times[-5;5]$, from which we have
excised  a cubic box with an edge of length $1.0$ centered on the physical singularity at $r=0$.
The  resolution is $20^3$, and similarly to the two-dimensional case, a perturbation is introduced in the variable $\Theta$.
Again with a fourth order well-balanced ADER-DG scheme, we obtain the results that are shown in Fig.~\ref{fig:Kerr3Da00}. The constraint violations decay back 
to the equilibrium at time $t\approx 400\,M$, after which the solution is perfectly stable around machine precision. For this simulation, the 1D cuts are extracted along the $z$ axis.

\paragraph{Rotating black hole in 3D}

\begin{figure}[!htbp]
	\begin{center}
		\begin{tabular}{cc} 
			\includegraphics[trim=10 10 10 10,clip,width=0.45\textwidth]{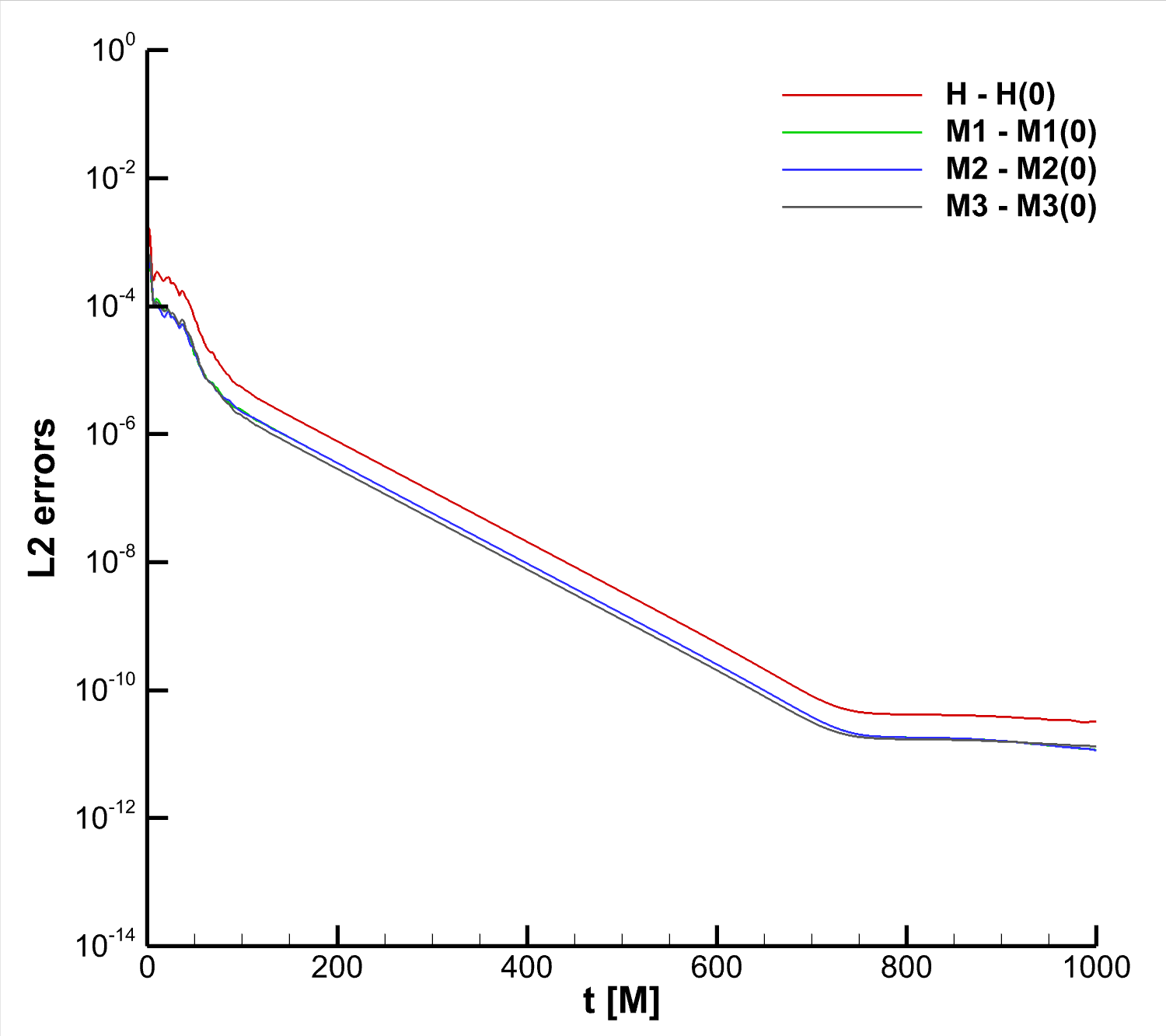} &
			\includegraphics[trim=10 10 10 10,clip,width=0.45\textwidth]{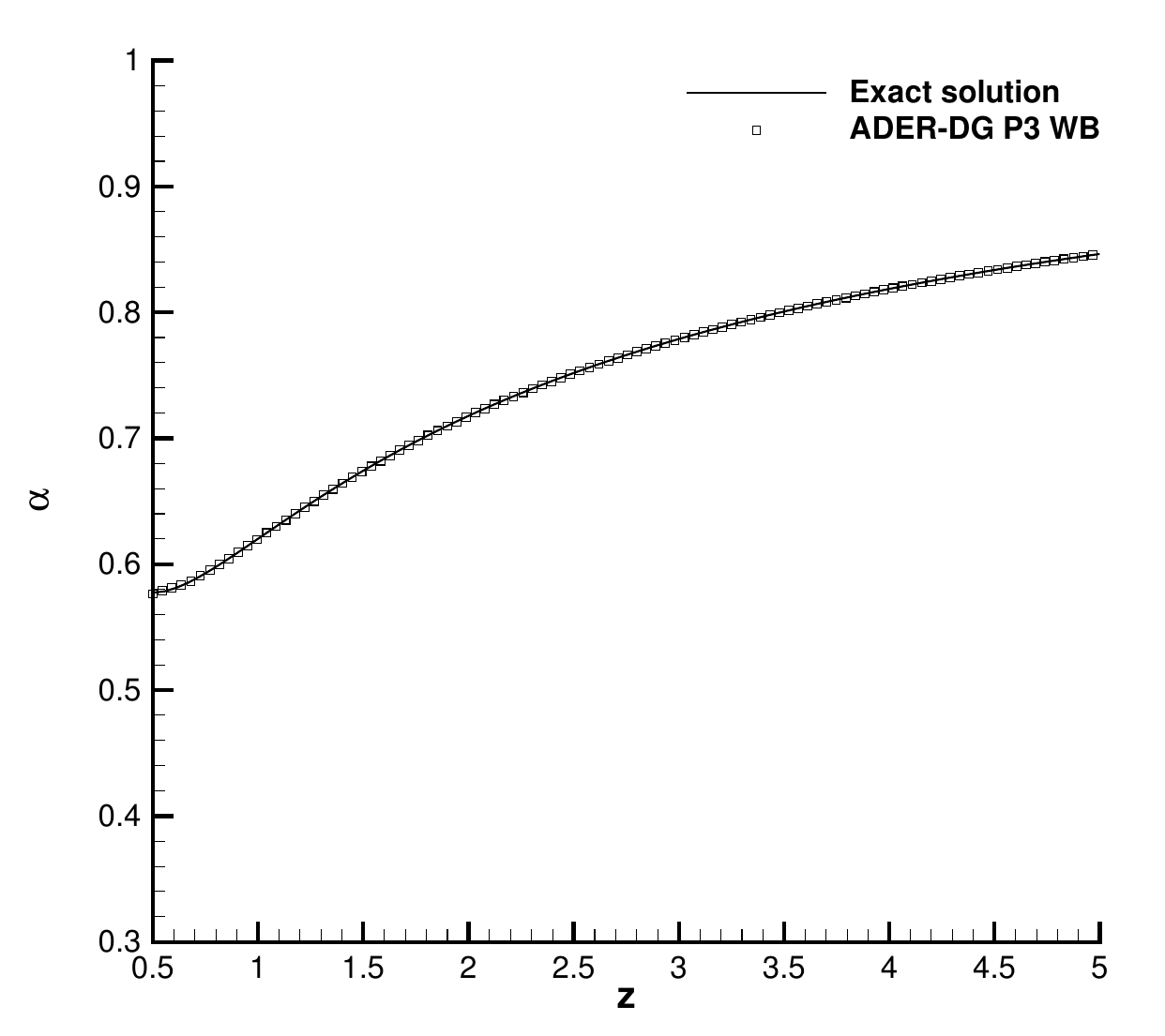} \\  
			\includegraphics[trim=10 10 10 10,clip,width=0.45\textwidth]{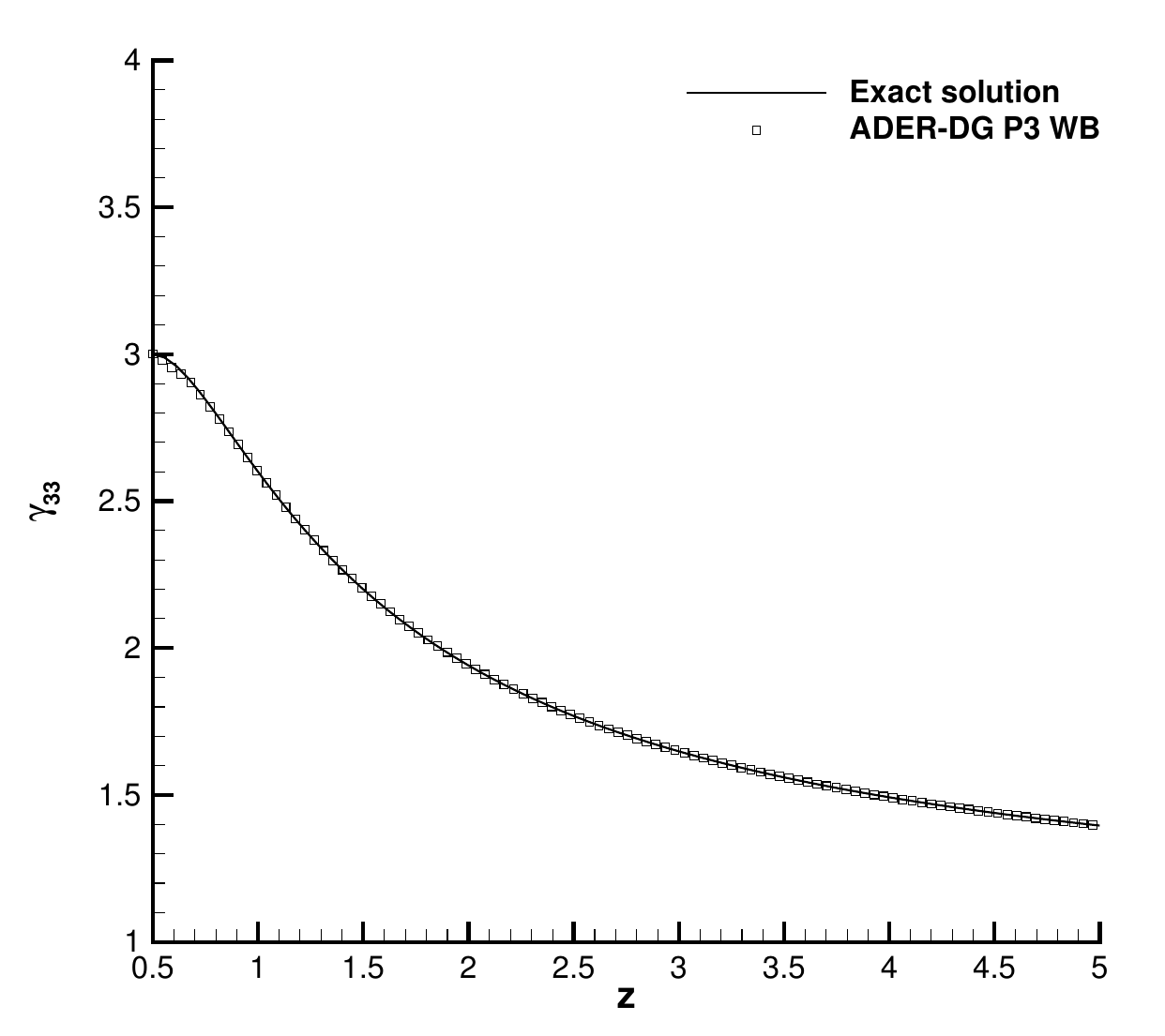} &  
			\includegraphics[trim=10 10 10 10,clip,width=0.45\textwidth]{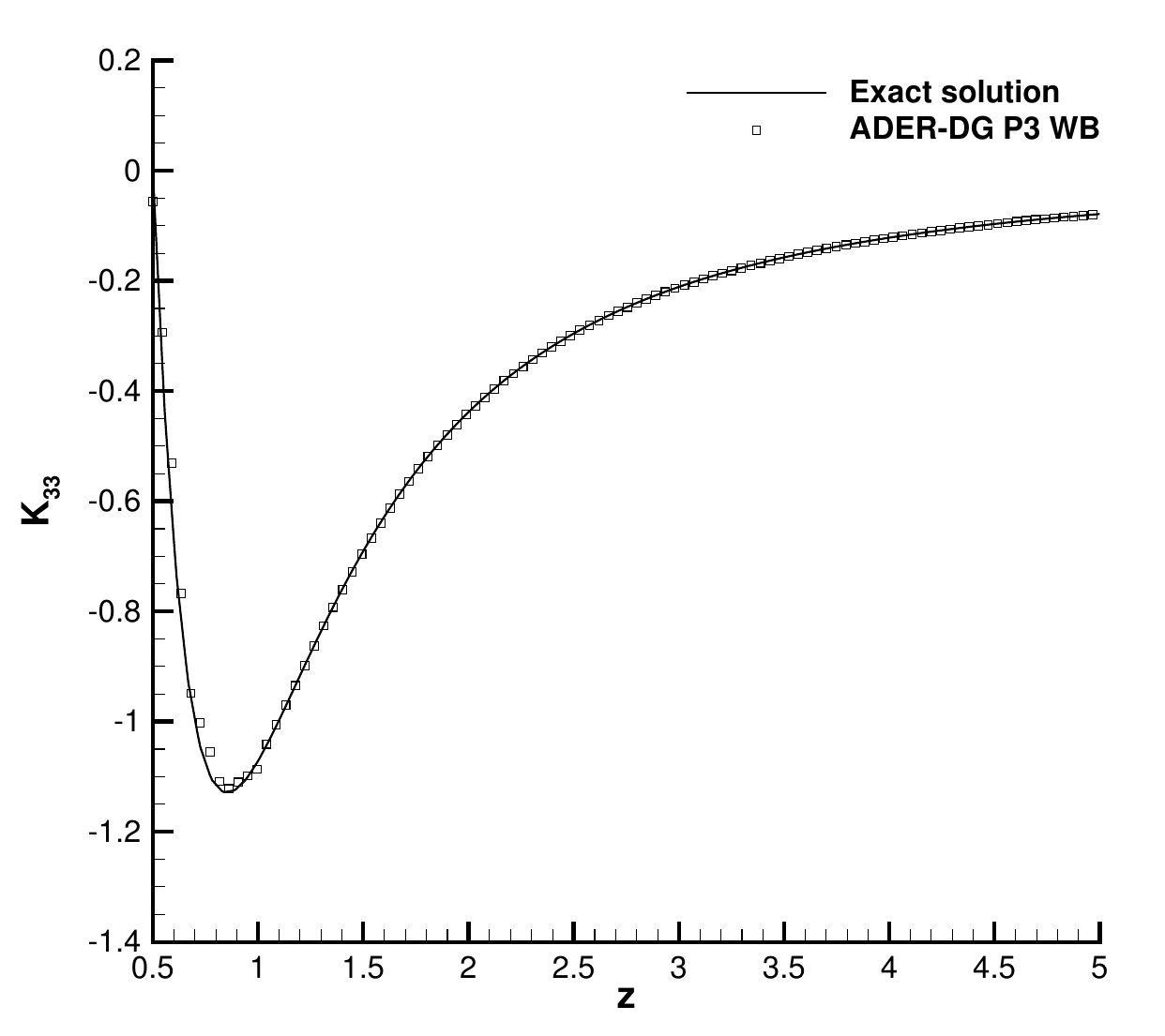}  
		\end{tabular} 
		\caption{3D simulation of an initially perturbed Kerr black hole (spin $a=0.5$) in 3D Cartesian Kerr-Schild coordinates using a fourth order well-balanced ADER-DG scheme. Top left: time series of the constraint violations until time $t=1000\,M$. It is clearly visible that the initial perturbation decays exponentially in time and that the numerical solution returns to the stationary equilibrium. From top right to bottom right: 1D cuts along the z axis ($x=y=0$) for the lapse $\alpha$, the metric tensor component $\gamma_{33}$ and the extrinsic curvature component $K_{33}$ at time $t=1000\,M$ and comparison with the exact solution. }  
		\label{fig:Kerr3Da05}
	\end{center}
\end{figure}

\begin{figure}[h]
	\begin{center}
		\begin{tabular}{cc} 
			\includegraphics[trim=10 10 10 10,clip,width=0.45\textwidth]{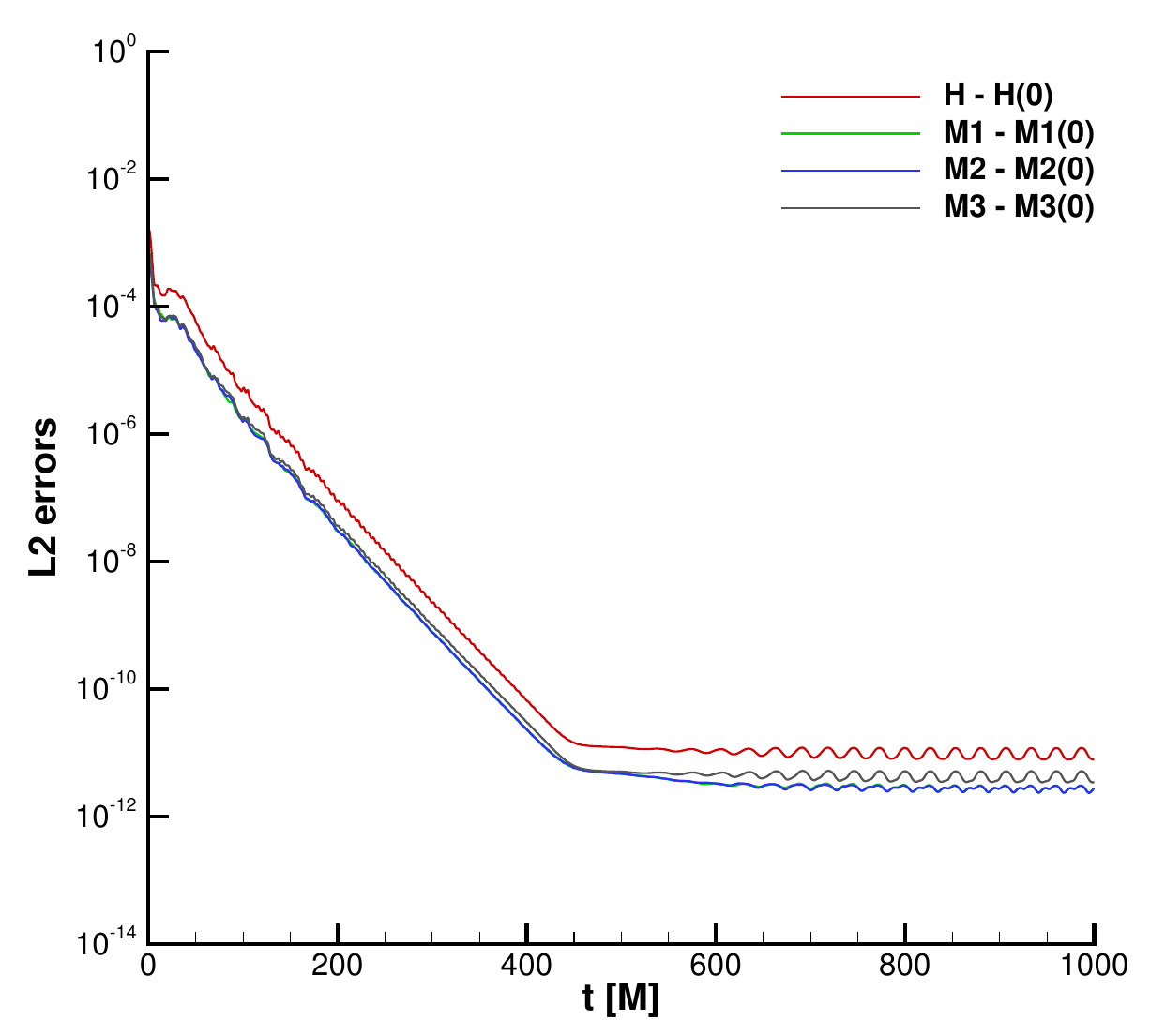} &
			\includegraphics[trim=10 10 10 10,clip,width=0.45\textwidth]{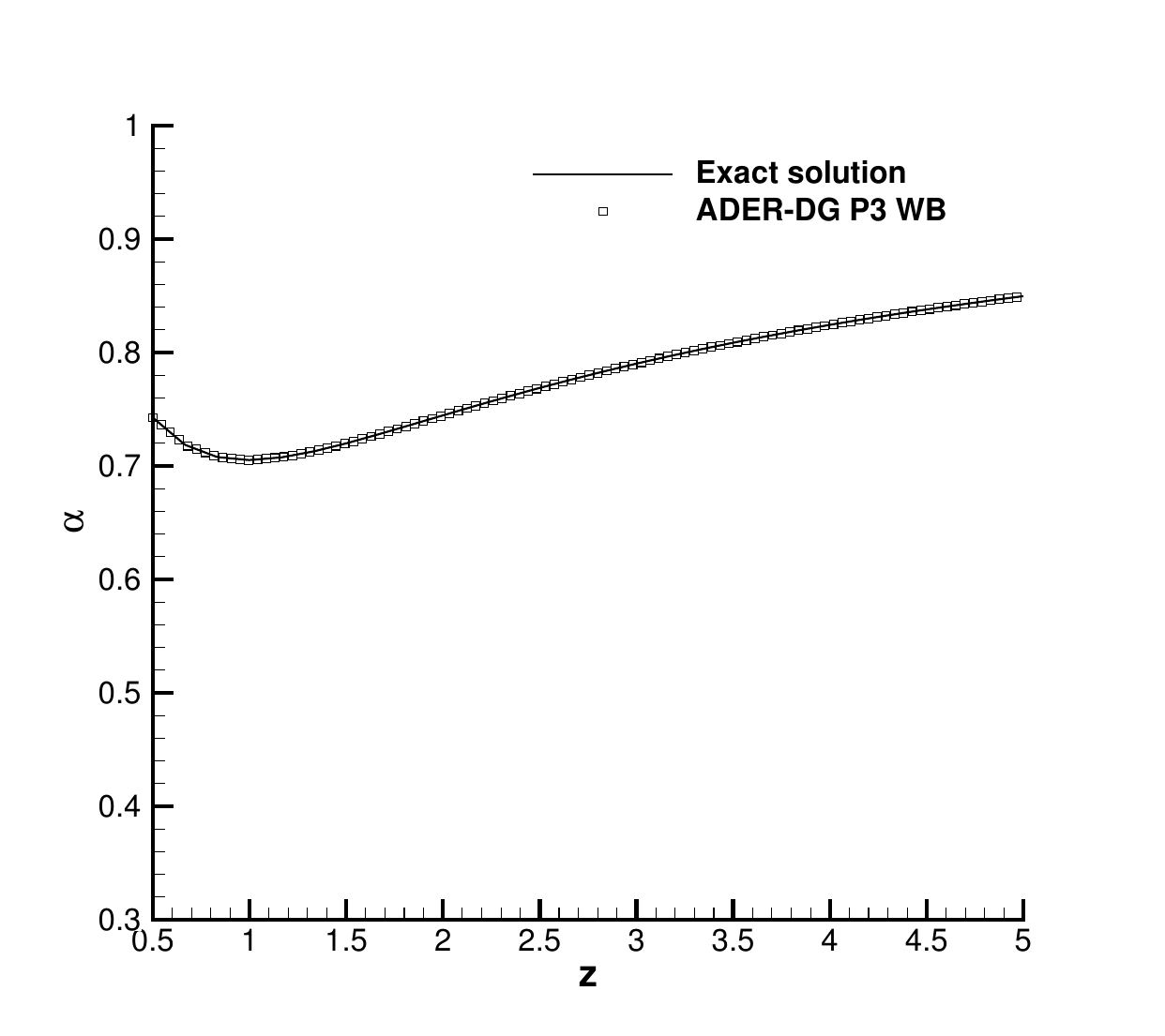} \\  
			\includegraphics[trim=10 10 10 10,clip,width=0.45\textwidth]{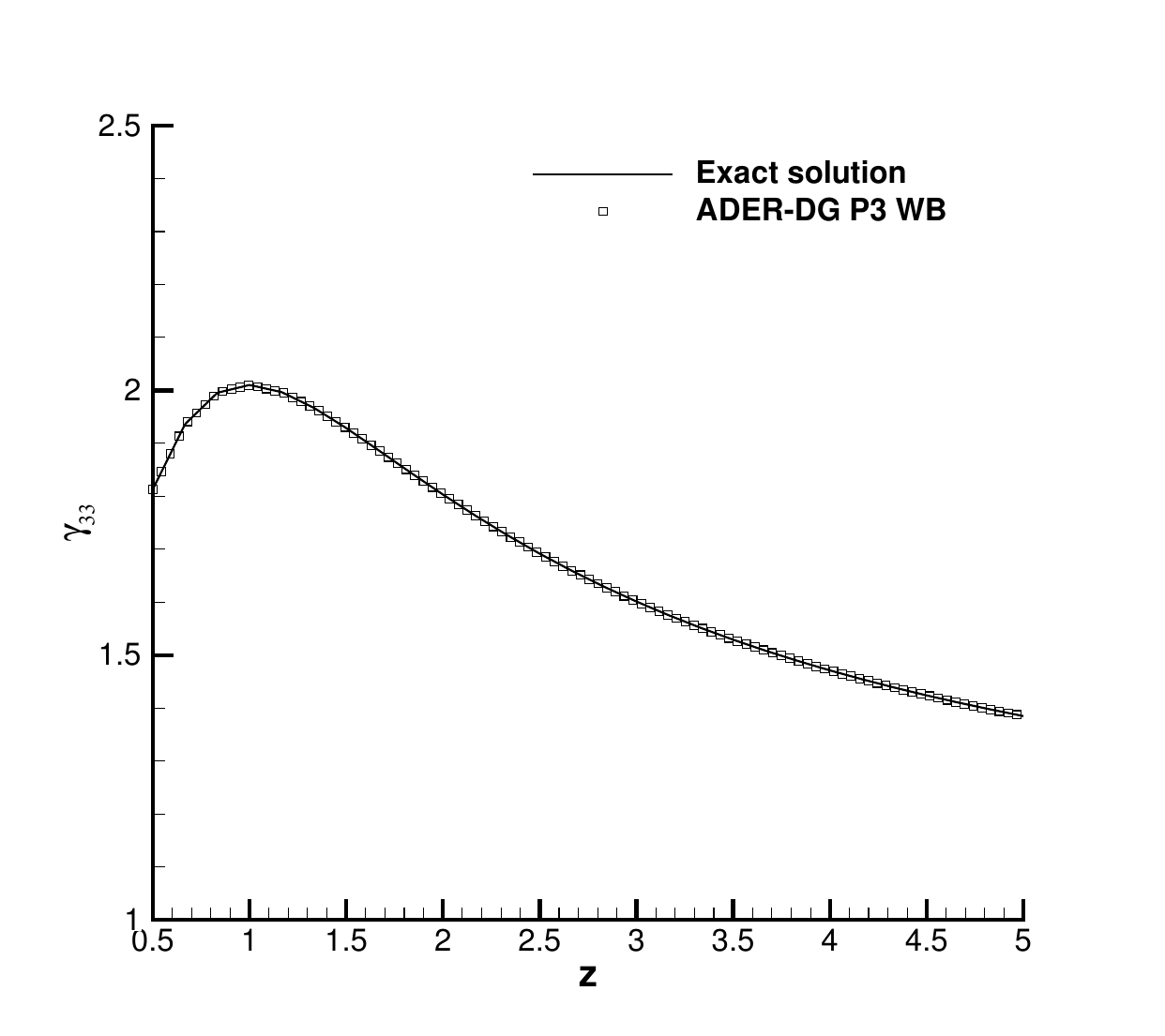} &  
			\includegraphics[trim=10 10 10 10,clip,width=0.45\textwidth]{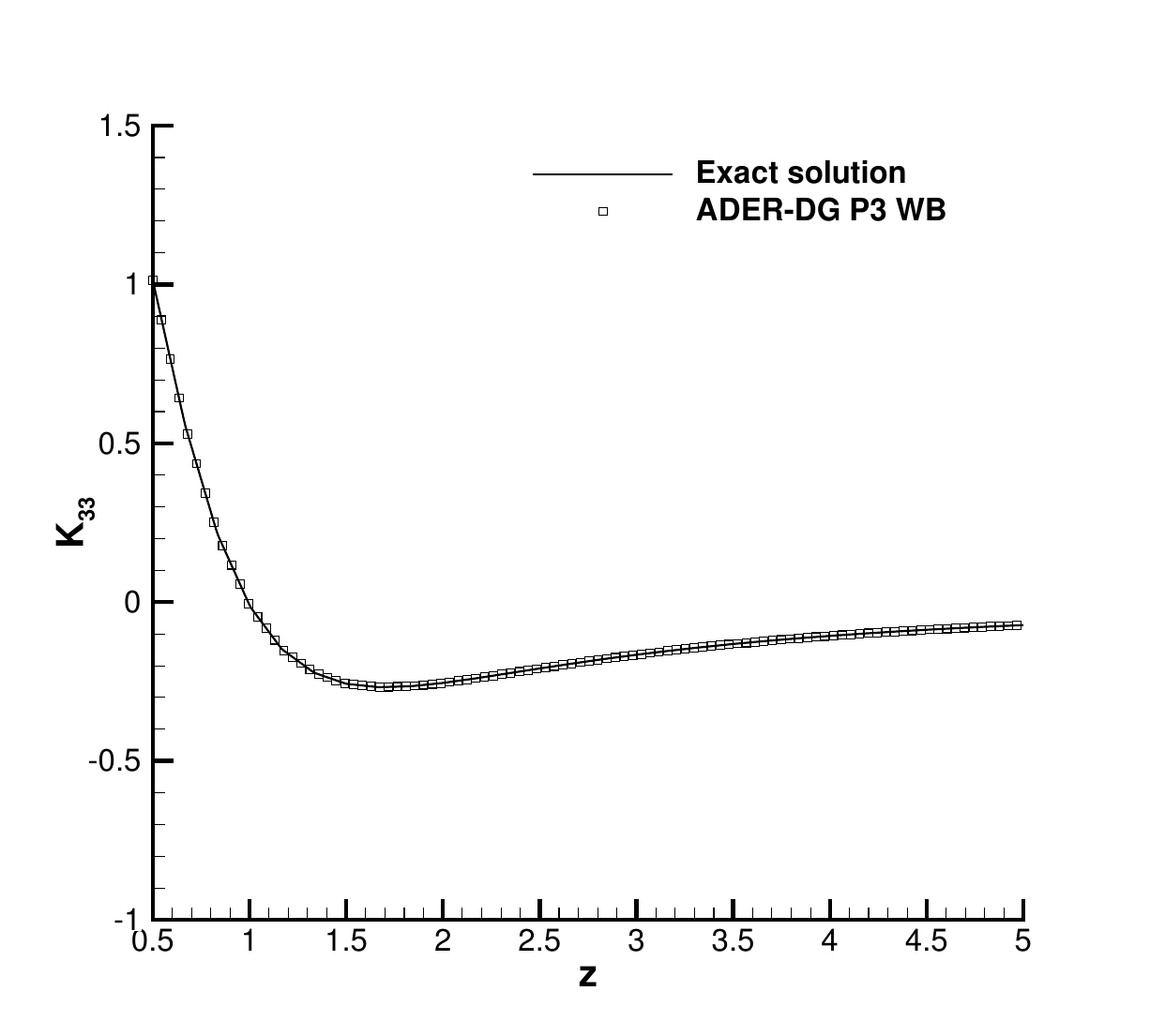}  
		\end{tabular} 
		\caption{3D simulation of an initially perturbed Kerr black hole (spin $a=0.99$) in 3D Cartesian Kerr-Schild coordinates using a fourth order well-balanced ADER-DG scheme. Top left: time series of the constraint violations until time $t=1000\,M$. It is clearly visible that the initial perturbation decays exponentially in time and that the numerical solution returns to the stationary equilibrium. From top right to bottom right: 1D cuts along the z axis ($x=y=0$) for the lapse $\alpha$, the metric tensor component $\gamma_{33}$ and the extrinsic curvature component $K_{33}$ at time $t=1000\,M$ and comparison with the exact solution. }  
		\label{fig:Kerr3Da09}
	\end{center}
\end{figure}

\begin{figure}[!htbp]
	\begin{center}
		\begin{tabular}{cc} 
			\includegraphics[trim=10 10 10 10,clip,width=0.45\textwidth]{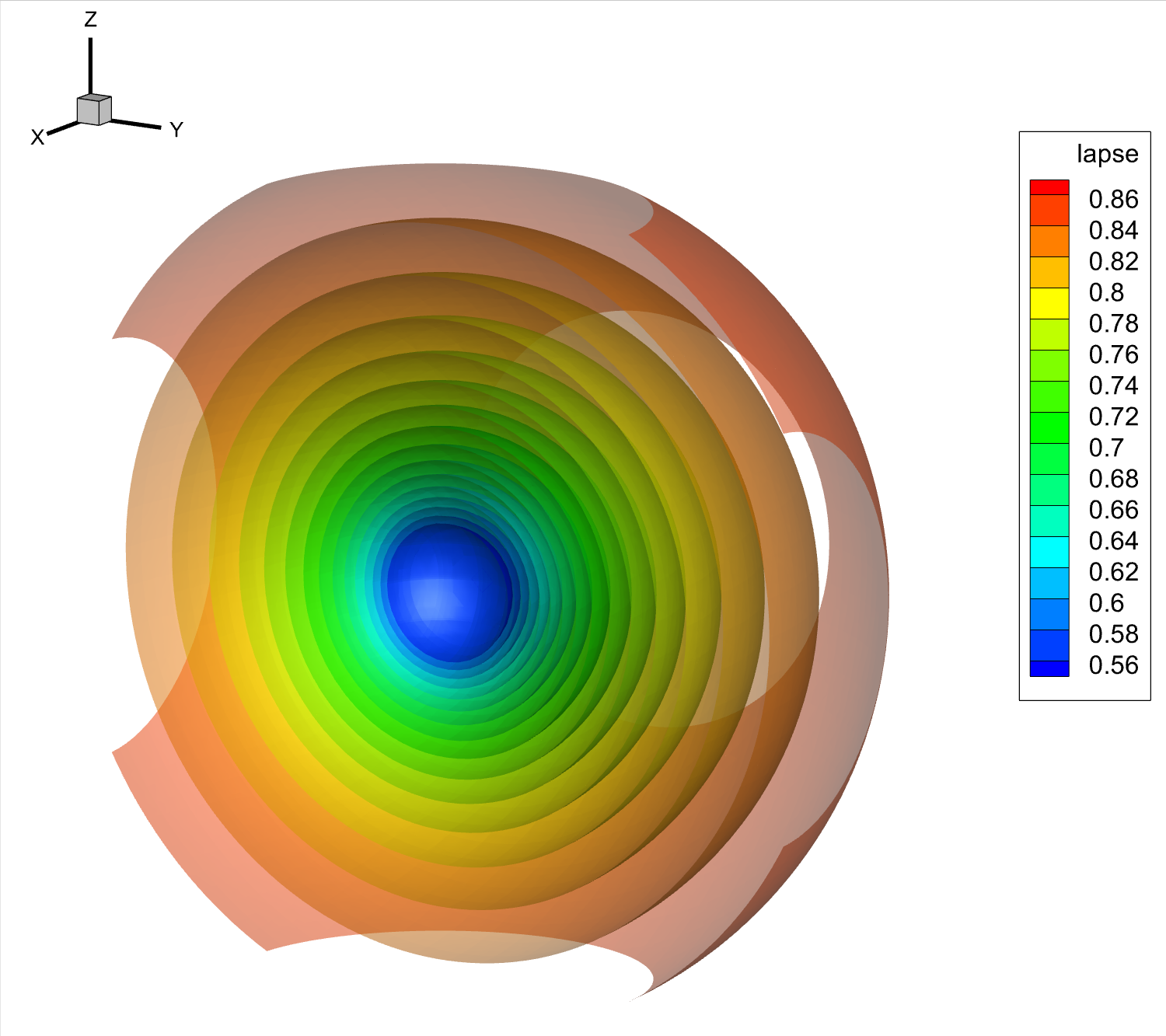} &
			\includegraphics[trim=10 10 10 10,clip,width=0.45\textwidth]{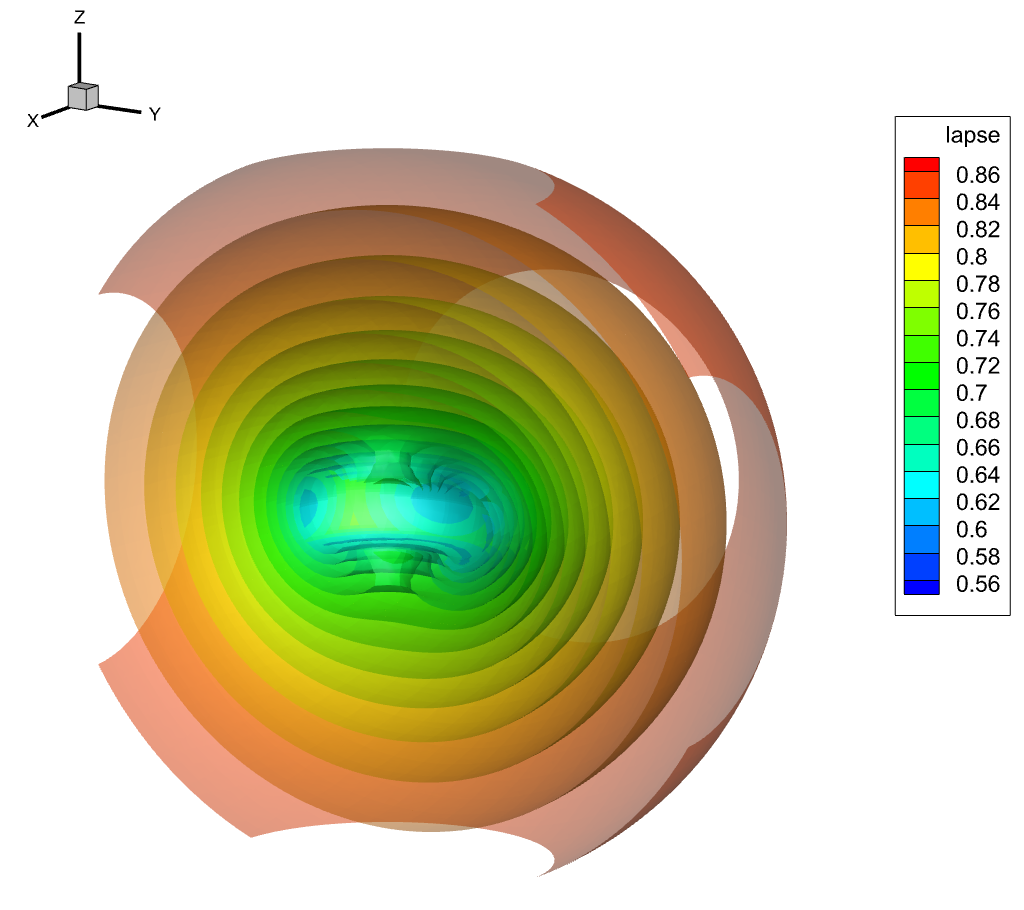} \\  
			\includegraphics[trim=10 10 10 10,clip,width=0.45\textwidth]{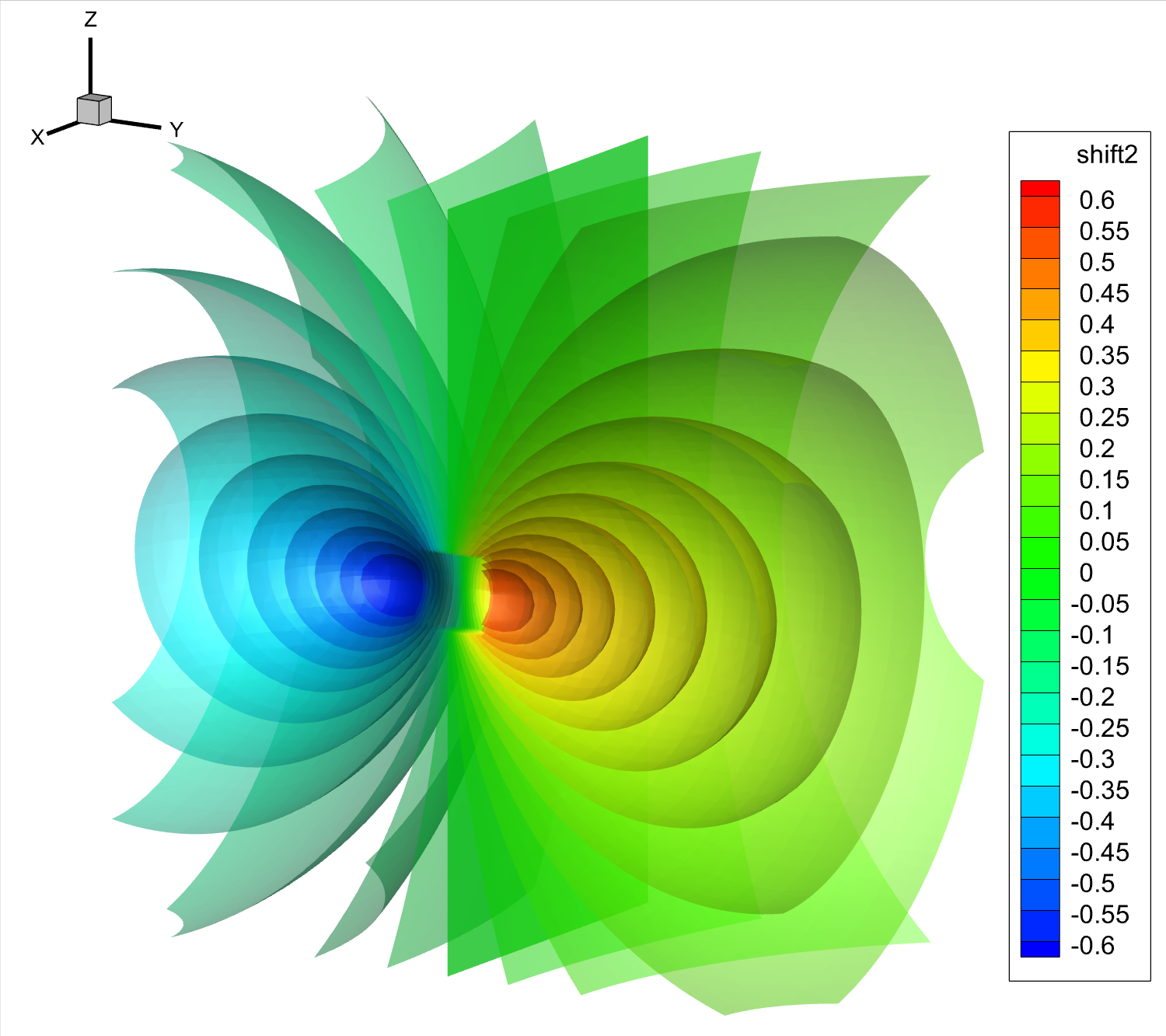} &  
			\includegraphics[trim=10 10 10 10,clip,width=0.45\textwidth]{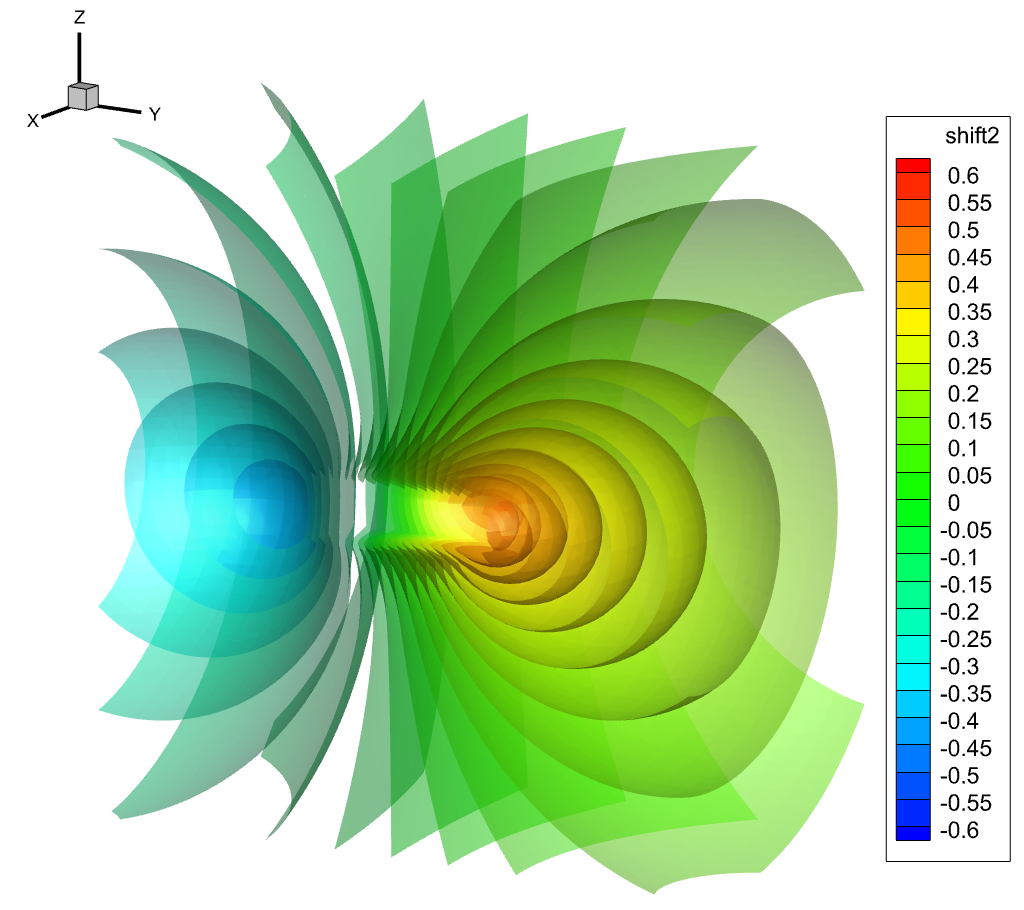} \\  
			\includegraphics[trim=10 10 10 10,clip,width=0.45\textwidth]{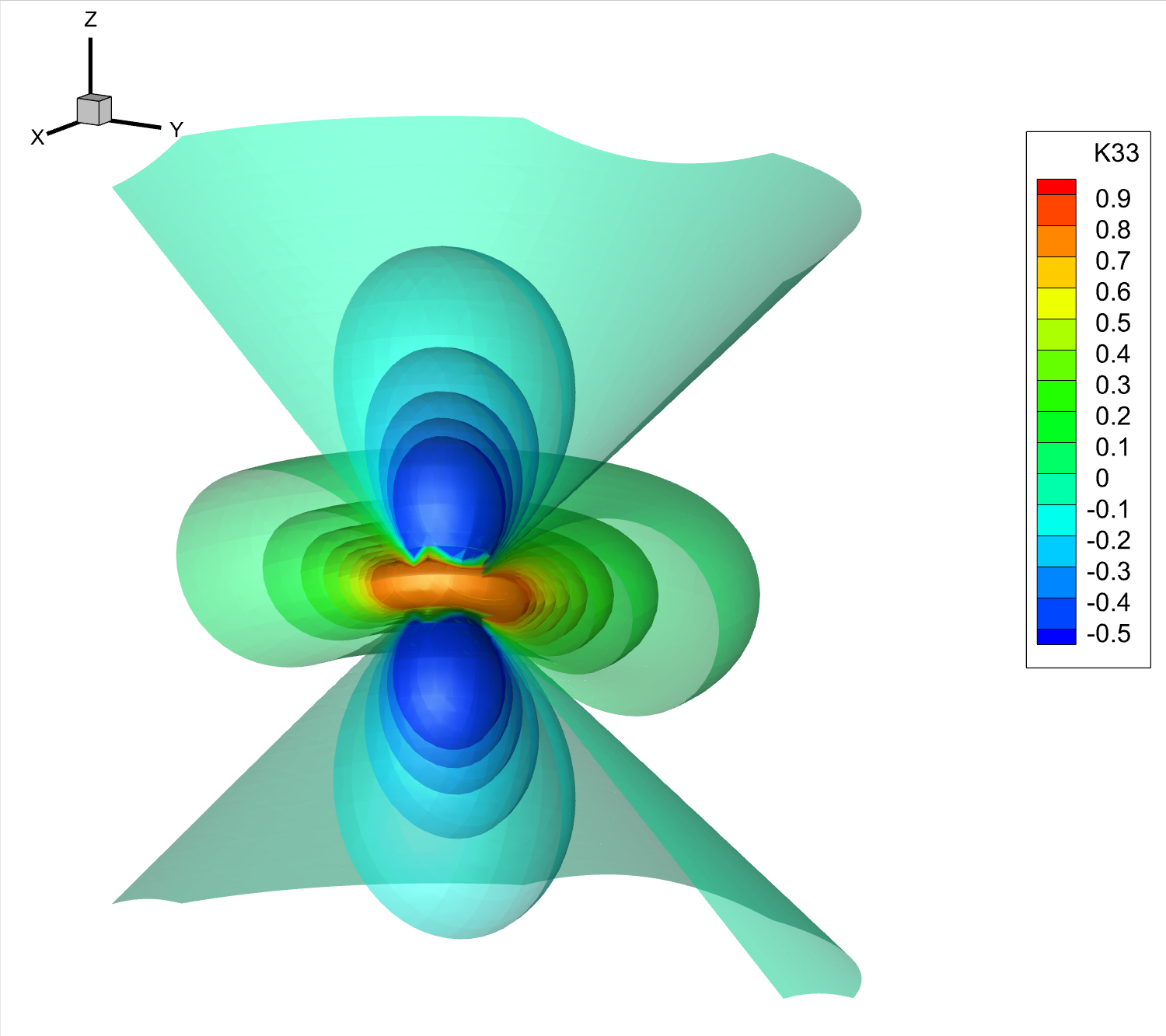} &  
			\includegraphics[trim=10 10 10 10,clip,width=0.45\textwidth]{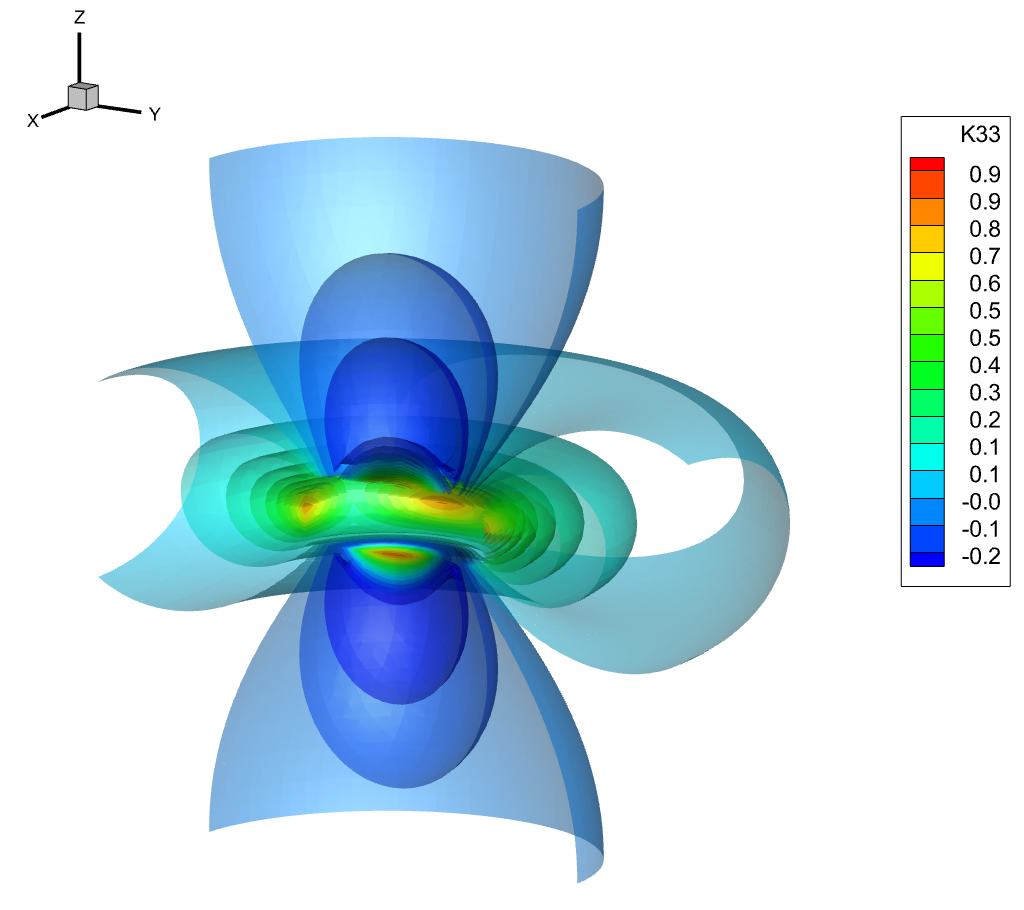}    
		\end{tabular} 
		\caption{Contour surfaces of initially perturbed black hole spacetimes in 3D Cartesian Kerr-Schild coordinates at $t=1000\,M$ using a fourth order well-balanced ADER-DG scheme. Left: Schwarzschild black hole ($a=0$). Right: Kerr black hole ($a=0.99$). From top to bottom: lapse $\alpha$, shift $\beta_2$ and extrinsic curvature component $K_{33}$. }  
		\label{fig:BH3D}
	\end{center}
\end{figure}

Finally, in addition to the previous Schwarzschild black holes with $a=0$, we have also evolved two Kerr black holes in three space dimensions, one  with spin $a=0.5$
and the other one with spin $a=0.99$.
The computational domain is the box $[-5;5]\times[-5;5]\times[-5;5]$, with the same resolution as for the Schwarzschild case, namely $20^3$.
A major difference is given by the fact that the excision box
must enclose the ring singularity on the $z=0$ plane~\cite{deFelice90}, which has an external radius $r_{\rm{ring}}=a$.
Hence, the excision box is effectively a parallelepiped with edges $2\times2\times 1$, and $3.2\times3.2\times 1$, for the two black holes
with spin $a=0.5$ and $a=0.99$, respectively. Keeping the same strategy of perturbing the initial configuration, we obtain results that are
shown in Fig.~\ref{fig:Kerr3Da05} and Fig.~\ref{fig:Kerr3Da09}, and confirming the turning back of the solution to the exact equilibrium. Fig.~\ref{fig:BH3D}, 
on the other hand, shows the contour surfaces of a few representative quantities where the Schwarzschild ($a=0$) and the Kerr ($a=0.99$) black holes are compared.

\subsection{Non--rotating neutron star in equilibrium}
\label{sec:tov}
\begin{figure}[!htbp]
	\begin{center}
		\begin{tabular}{cc} 
			{\includegraphics[width=0.49\textwidth]{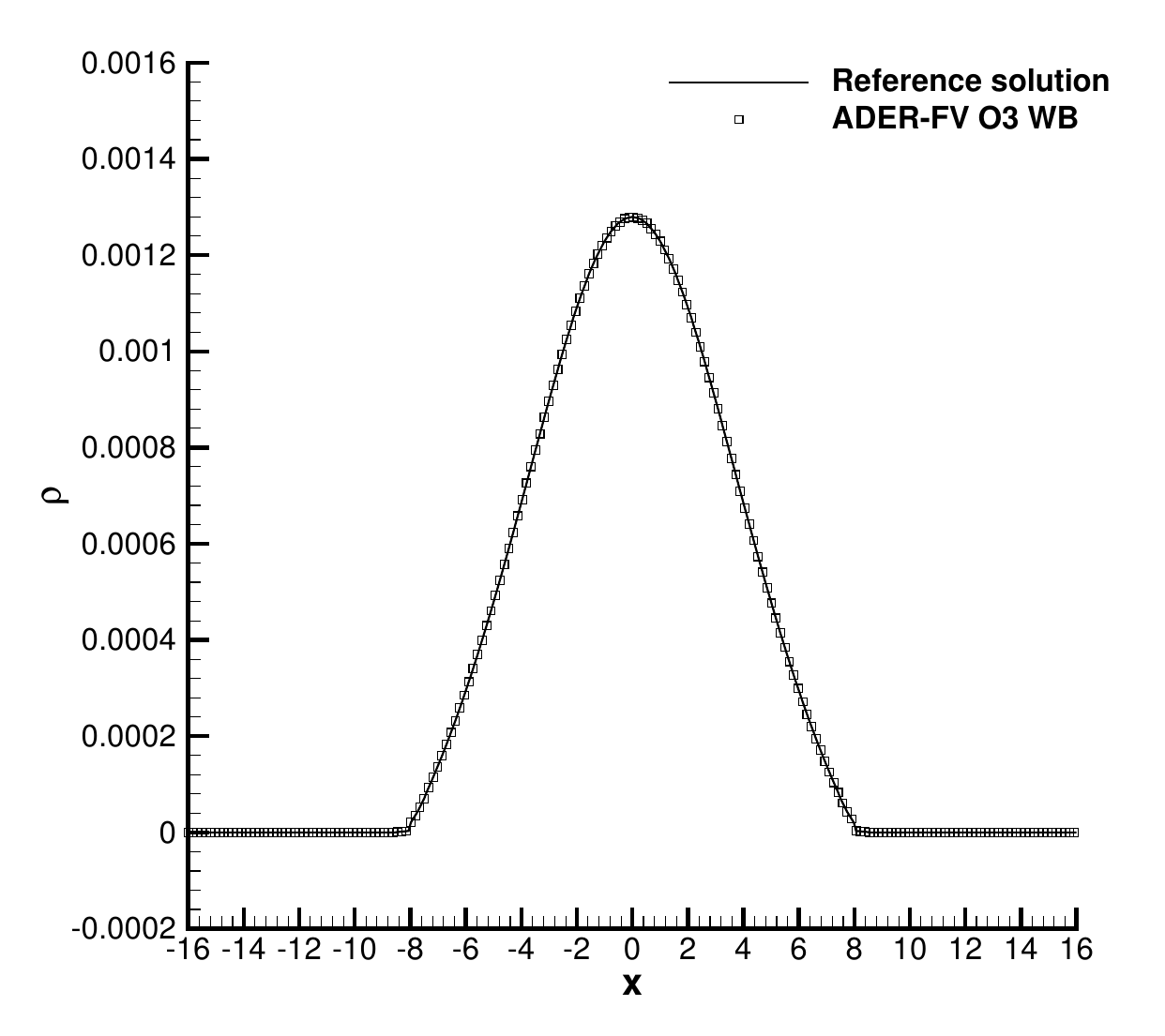}} &
			{\includegraphics[width=0.49\textwidth]{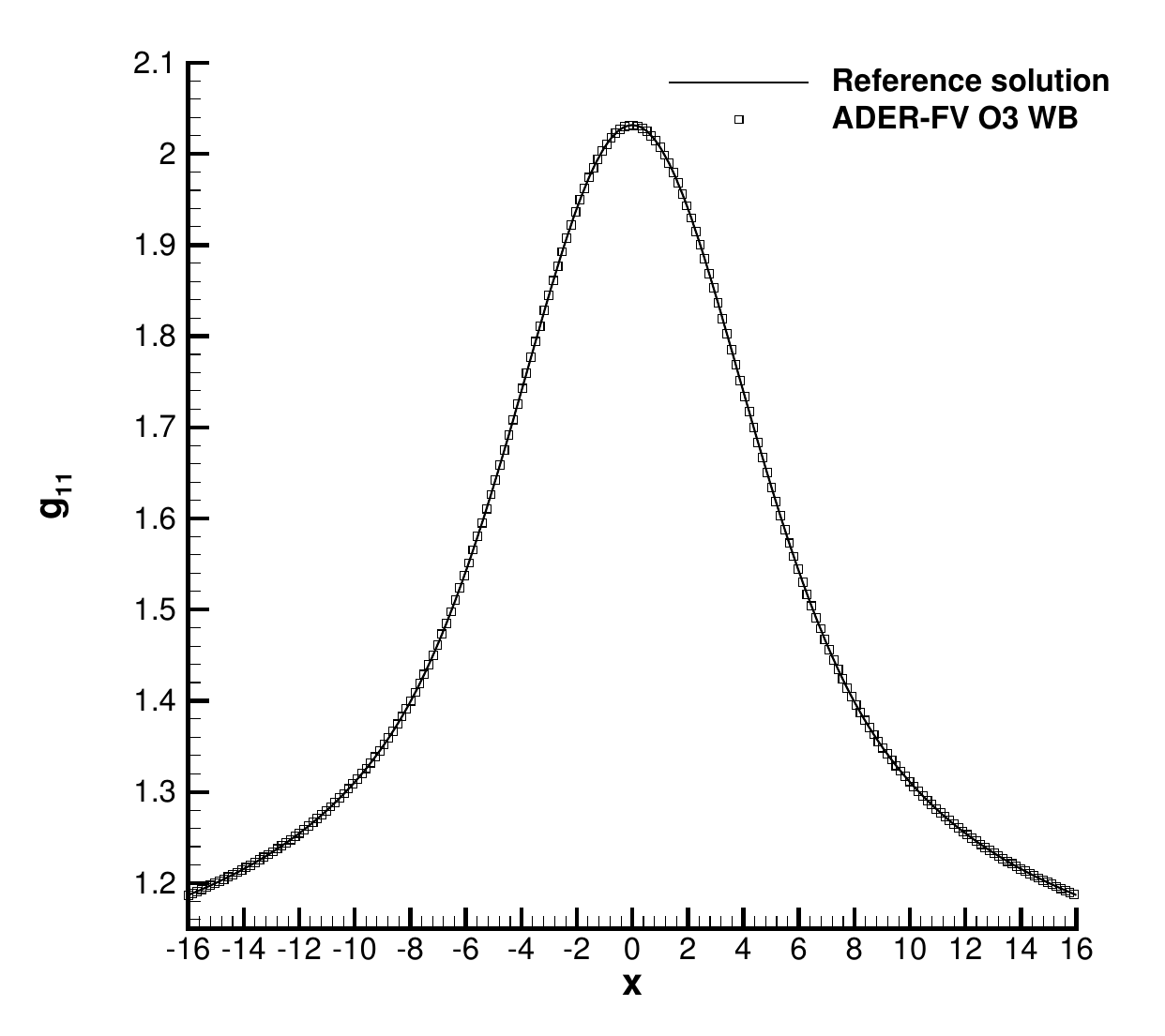}}\\
			{\includegraphics[width=0.49\textwidth]{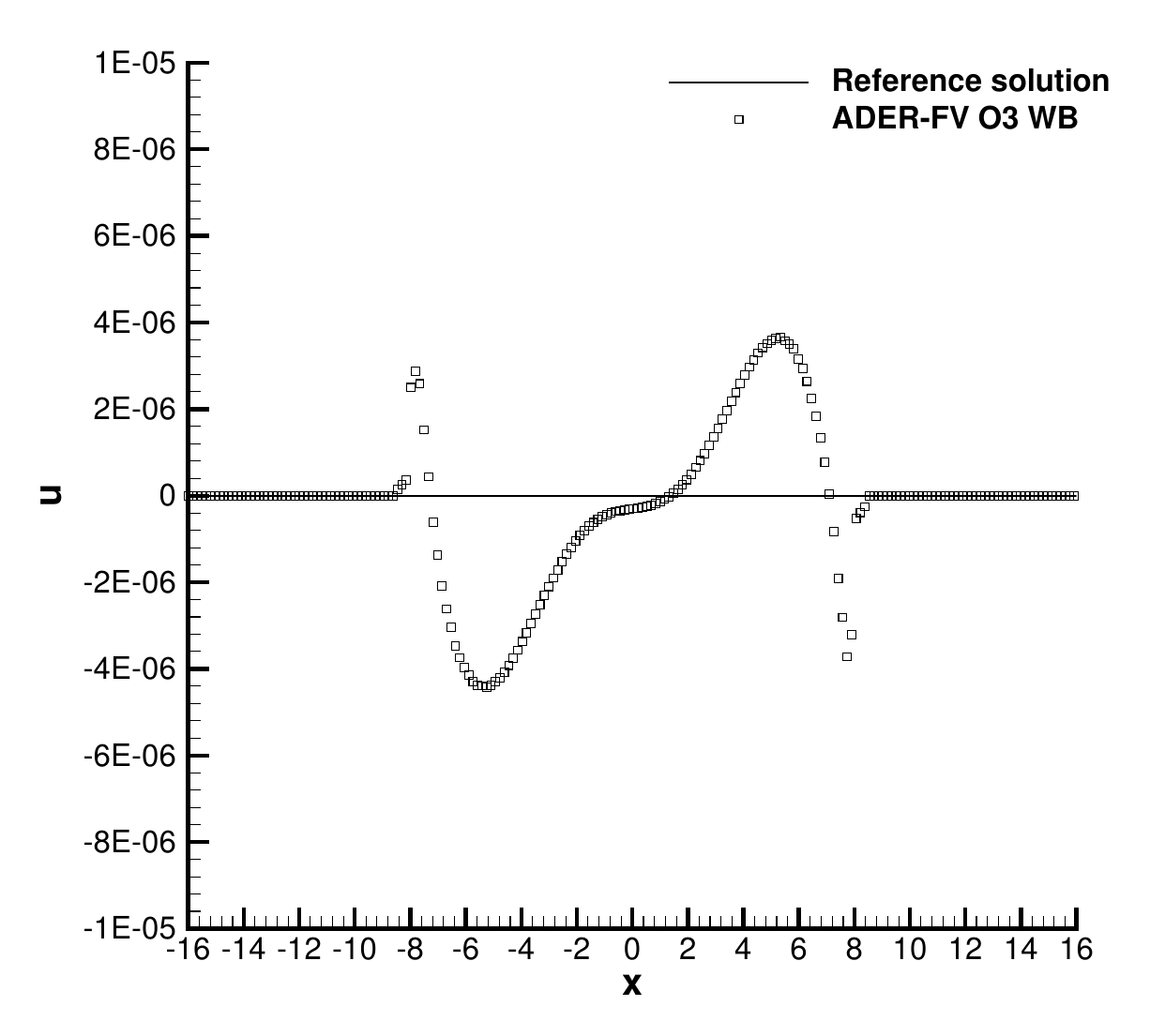}} &
			{\includegraphics[width=0.49\textwidth]{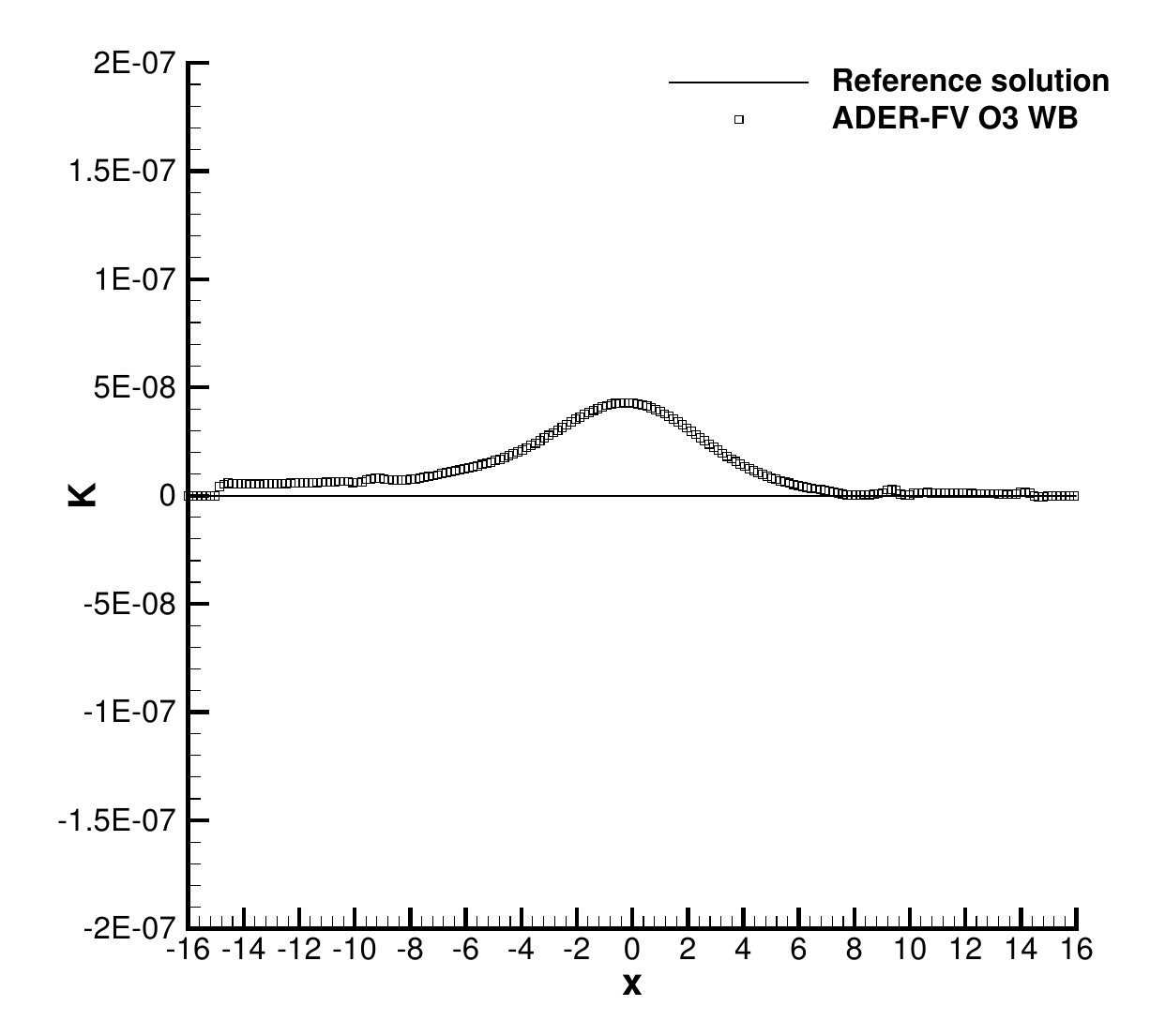}}\\
			{\includegraphics[width=0.49\textwidth]{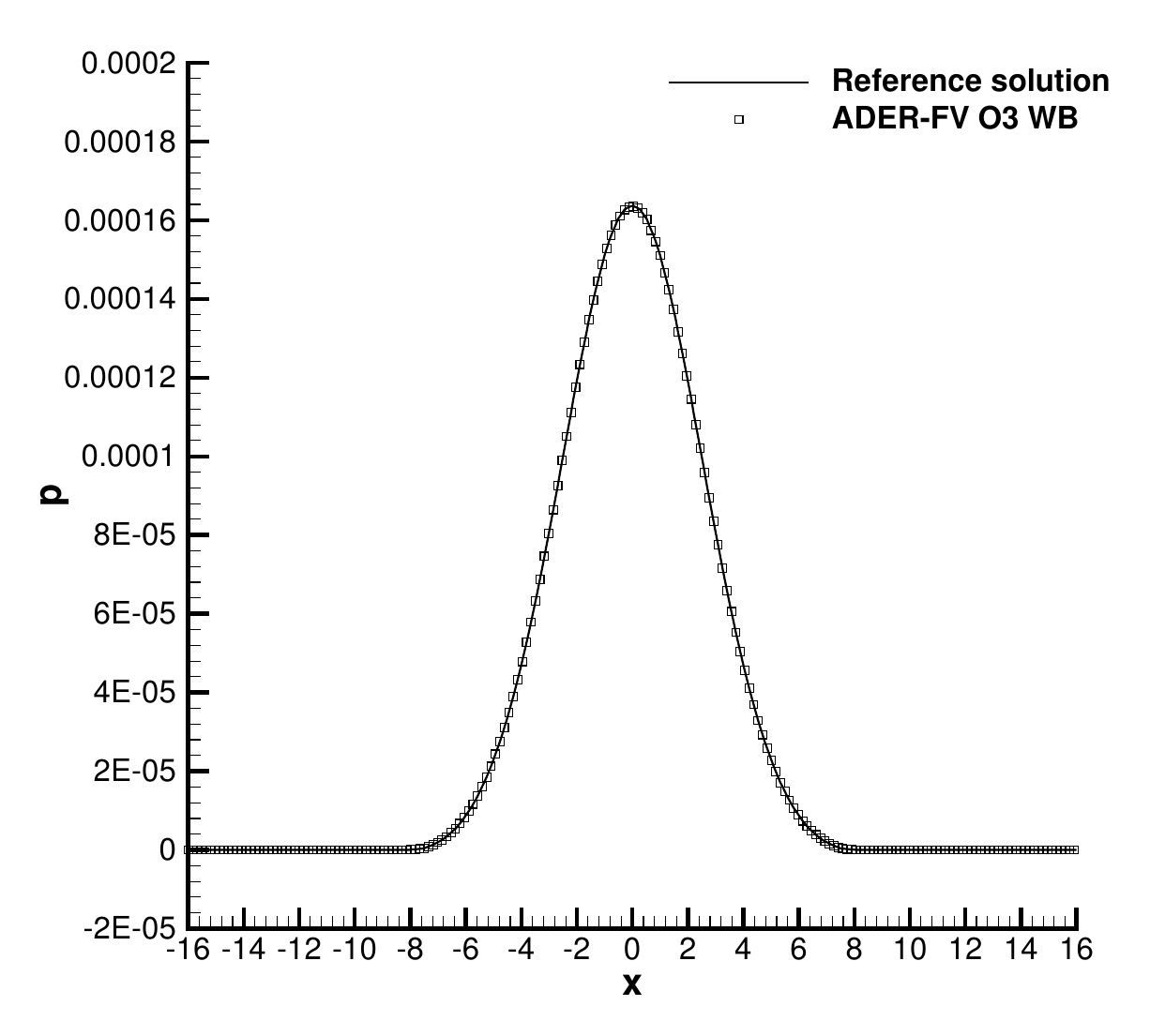}} &
			{\includegraphics[width=0.49\textwidth]{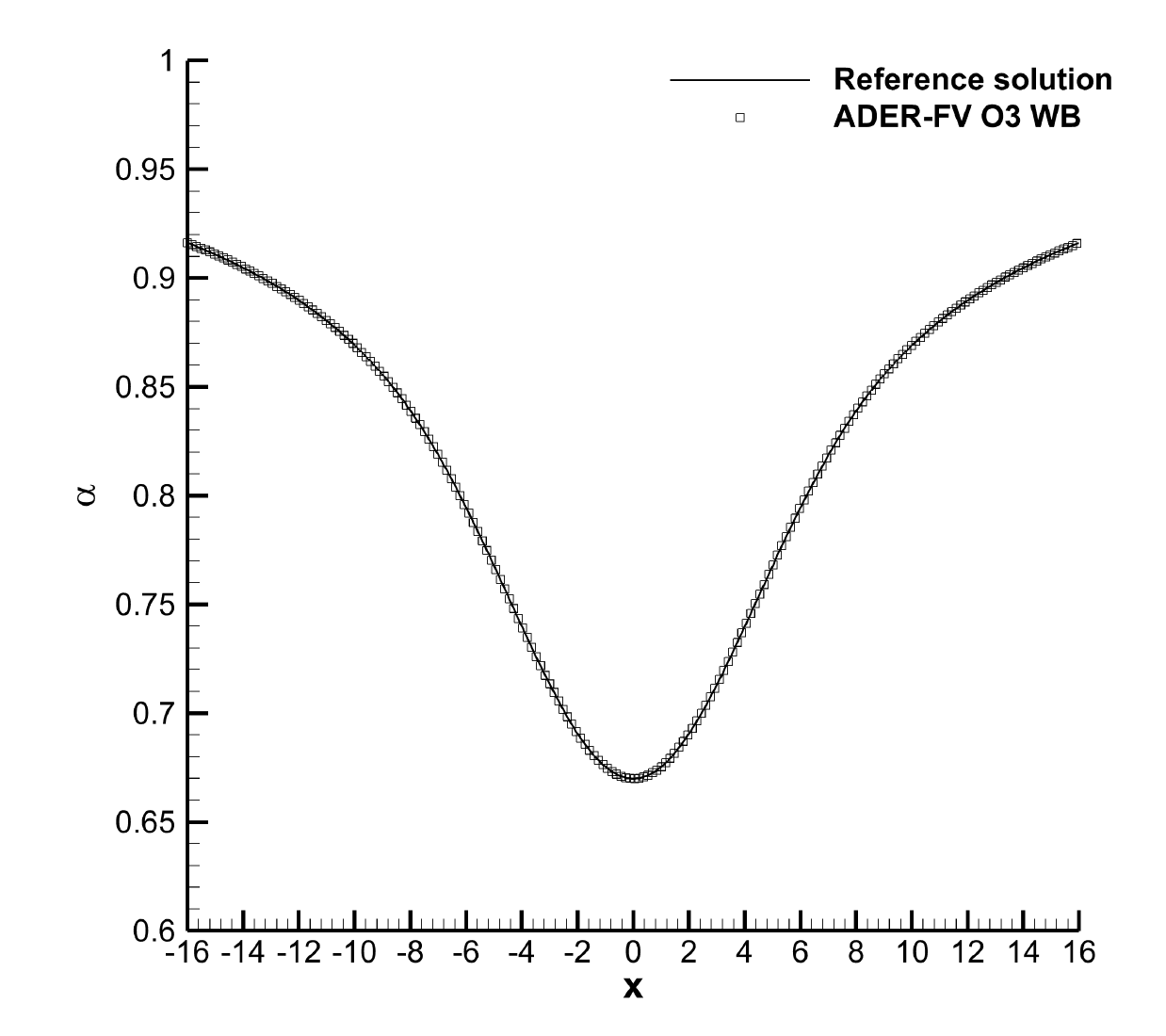}}
		\end{tabular} 
		\caption{1D cuts of some hydrodynamic and metric quantities of the TOV star obtained with the new well-balanced third order ADER-FV scheme at the final time $t=1000\,M$.
		} 
		\label{fig:TOV-profiles}
	\end{center}
\end{figure}
\begin{figure}[!hpb]
	\begin{center}
		\begin{tabular}{cc} 
  		{\includegraphics[angle=0,width=7.3cm,height=7.3cm]{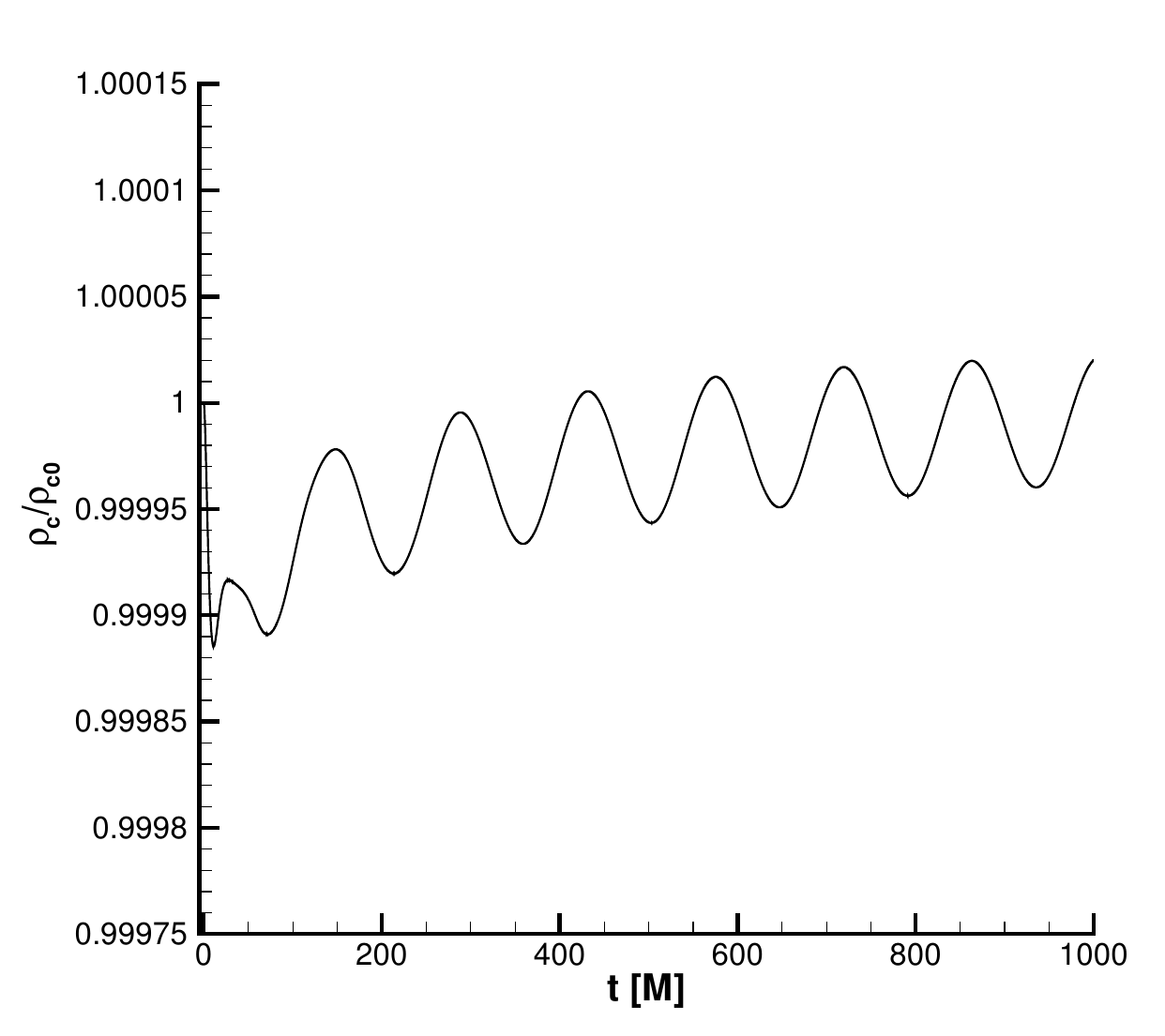}} &
			{\includegraphics[angle=0,width=7.3cm,height=7.3cm]{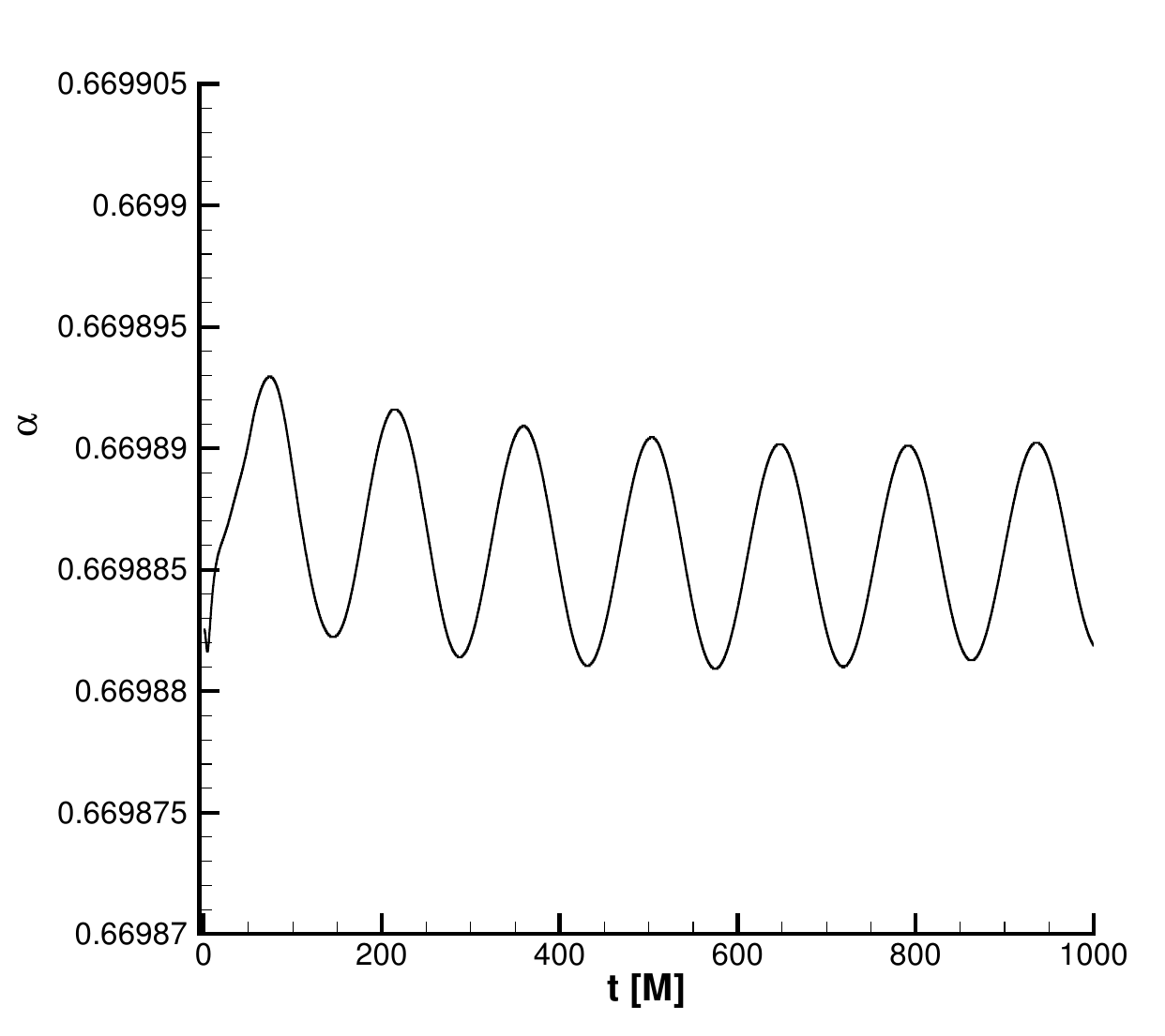}} 
		\end{tabular} 
		\caption{Time evolution of the central mass density (left panel) and of the central lapse (right panel) for the 3D TOV star. 
		} 
		\label{fig:TOV-timeseries}
	\end{center}
\end{figure}
\begin{figure}[!hpb]
	\begin{center}
		\begin{tabular}{cc} 
  		{\includegraphics[angle=0,width=7.3cm,height=7.3cm]{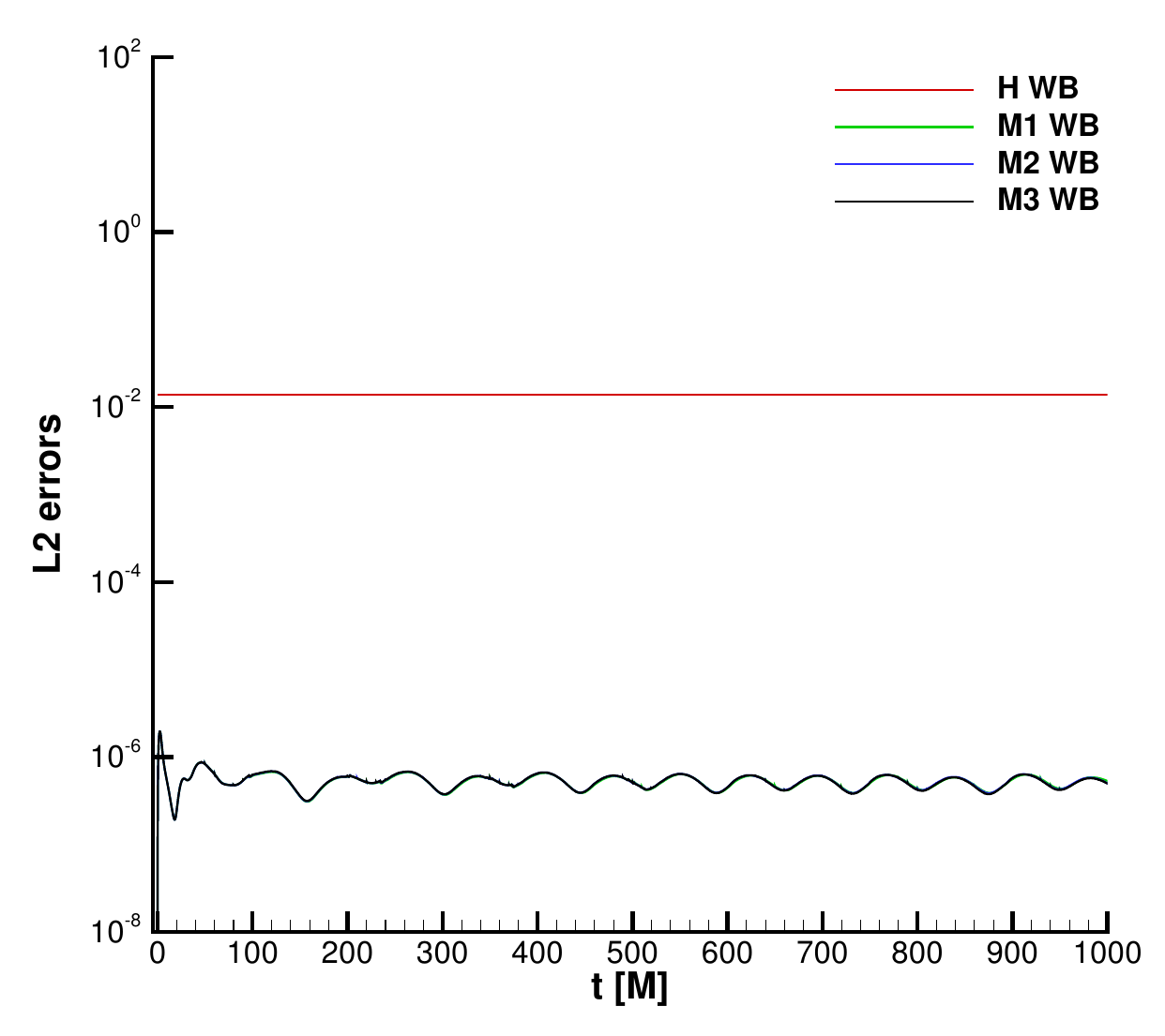}} &
			{\includegraphics[angle=0,width=7.3cm,height=7.3cm]{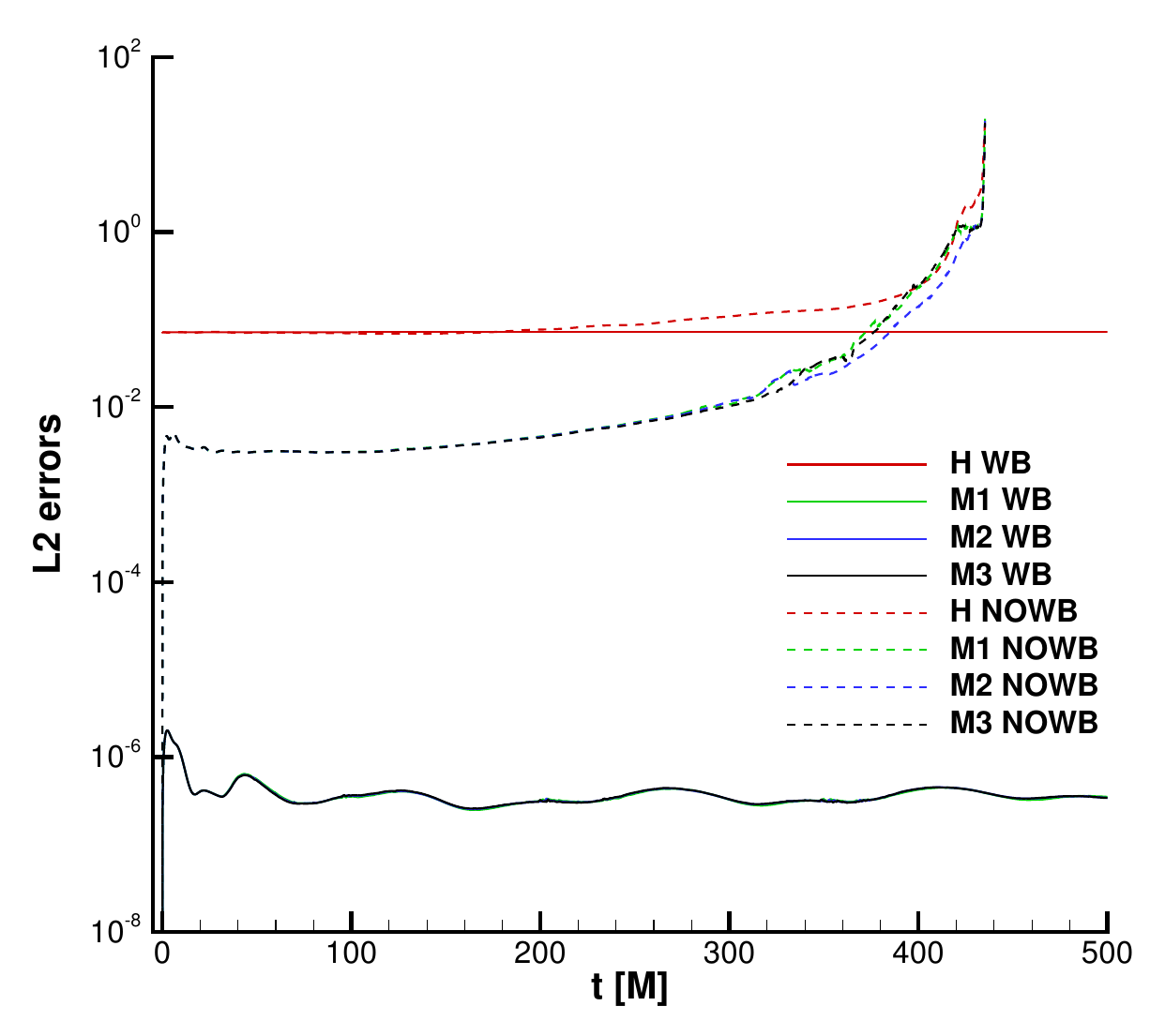}} 
		\end{tabular} 
		\caption{Time evolution of the constraint violations for the stable 3D TOV star. 
		Left panel: well-balanced third order ADER-FV scheme. 
		Right panel: well-balanced third order ADER-DG scheme vs. not well-balanced third order ADER-DG scheme ($N=2$).
		} 
		\label{fig:TOV-Constraint}
	\end{center}
\end{figure}
A  crucial test for numerical relativity, where both the Einstein and the relativistic Euler equations must be accounted for, is represented by 
the time evolution of an equilibrium neutron star. In the non--rotating case, this amounts to solving the so called Tolman--Oppenheimer--Volkoff (TOV) system, which we report here
for completeness~\cite{Tolman,Oppenheimer39b,Rezzolla_book:2013}
\begin{align}
\label{eq:tov1}
\frac{d m}{d r} &= 4\pi r^2e \,,\\
\label{eq:tov2}
\frac{dp}{dr}   &= - \frac{(e+p)(m+4\pi r^3 p)}{r(r-2m)}\,,\\
\label{eq:tov3}
\frac{d \phi}{d r} &= -\frac{1}{e+p}\frac{dp}{dr} \,,
\end{align}
where $m(r)$ is the mass enclosed within the radius $r$, $\phi$ is the unknown metric function in the line element~\eqref{eq:symm_metric}, while
$e^{-2\psi}=1-\frac{2m}{r}$.
The equation of state adopted is that of a polytropic gas, namely $p= K\,\rho^\gamma$.

The TOV system~\eqref{eq:tov1}--\eqref{eq:tov3} constitutes a set of three ODEs, which we have solved using a tenth order accurate discontinuous Galerkin scheme, see~\cite{ADERNSE}.
For high order ADER-DG schemes, in fact, simple initial 
data computed via Runge-Kutta ODE integrators are not accurate enough. 
We have adopted a stable model with parameters which have by now become canonical in numerical relativity~\cite{Font2002}, namely
a central rest mass density 
$\rho_c=1.28\times 10^{-3}$, $K=100$ and $\gamma=2$. 
Having done that, the numerical integration of~\eqref{eq:tov1}--\eqref{eq:tov3}
provides all the radial profiles as well as the remaining physical characteristics of the star, 
i.e. a total mass $M=1.4\,M_{\odot}$ and a radius $R=9.585 \,M_{\odot}= 14.15\,km$. 
When performing the coordinate transformation 
\begin{equation}
\frac{d\bar{r}}{\bar{r}}=\left(1-\frac{2m}{r}\right)^{-1/2}\frac{dr}{r}, 
\end{equation}
see~\cite{bugner}, then the spatial part of the metric~\eqref{eq:symm_metric} becomes conformally flat, namely
\begin{equation}
ds^2 = -e^{2\phi} dt^2 + e^{2\bar{\psi}}( d\bar{r}^2 + \bar{r}^2 d\theta^2  + \bar{r}^2 \sin^2\theta d\phi^2) =-e^{2\phi} dt^2 + e^{2\bar{\psi}}(d\bar{x}^2+d\bar{y}^2+d\bar{z}^2  )\,,
\end{equation}
thus generating a spatial metric that is just $\gamma_{ij}=(r/\bar{r})^2\eta_{ij}$. In the space outside the star, due to Birkoff's theorem,
the spacetime is that of a Schwarzschild solution produced by a mass $M$, i.e.
\begin{equation}
e^{2\phi}=1-\frac{2M}{r},\hspace{1cm}\bar{r}=\frac{1}{2}\left( \sqrt{r^2-2Mr} + r - M\right),
\end{equation}
while all the hydrodynamic variables collapse to zero. For this test problem the full Einstein--Euler 
system is evolved in the domain $\Omega = [-16,+16]^3$ until $t=1000\,M$ using a third order ADER-WENO finite volume scheme with $60^3$ elements. We set the damping coefficients to $\kappa_1 = \kappa_2 = 0.05$. 
We stress that, thanks to our new conversion from the conservative to the primitive variables (see Sect.~\ref{sec:cons2prim}), there is no need to insert a low density atmosphere in the exterior of the neutron star. 
For this test the fluid pressure was initially perturbed by adding 
a small fluctuation $p' = p_0 \exp \left( -\halb \frac{\x^2}{\sigma^2} \right)$ to the pressure obtained from the TOV solution, with amplitude $p_0=10^{-7}$ and halfwidth $\sigma=0.2$.  

To obtain better results, and only for this test, we had to resort to a well-balanced third order ADER-FV scheme \cite{AMR3DCL}, which became necessary for its increased robustness with respect to ADER-DG, especially at the surface of the star. 
Figure~\ref{fig:TOV-profiles} shows the results of our computations, by reporting the 1D-cuts of a few representative quantities at the final time, compared to the
reference equilibrium solution. 
A perfect matching is obtained, apart for 
very small deviations in the profiles of the velocity (along $x$) and in the trace of the extrinsic curvature $K$.
To the best of our knowledge, this is the first time that a numerical relativity code can evolve
a TOV star in a (matter) vacuum atmosphere with $\rho=p=0$.

In addition, in Fig.~\ref{fig:TOV-timeseries} we report the time evolution of the central rest--mass density (left panel, normalized to its initial value)
and of the central lapse (right panel). We just mention briefly that from this  oscillating behavior it is possible to extract the normal modes of oscillation of the neutron star,
comparing them with those obtained through a perturbative analysis and inferring fundamental aspects of neutron star physics \cite{friedman_stergioulas_2013}. 
As we are not interested to enter such details in this work,
we postpone further analysis to future investigations.

Finally, Fig.~\ref{fig:TOV-Constraint} shows the behaviour of the Einstein constraints during the evolution.
The left panel refers to the same simulation reported in Fig.~\ref{fig:TOV-profiles}, and it shows that the $L^2$ norm of the Einstein constraints 
remains low and stationary all along the evolution. 
The right panel refers instead to a second simulation
with the third--order ADER-DG scheme. In this case we have compared the well-balanced (WB) algorithm with the not well-balanced (NOWB) one.
The difference is remarkable, since in the not well-balanced evolution (NOWB) the Einstein constraints start increasing around $t\sim 300$, entering an exponential grow 
which eventually makes the code crash.

\subsection{Two puncture black holes}
\label{sec:two-punctures}

\begin{figure}[!htbp]
\begin{center}
\begin{subfigure}{0.49\linewidth}
	\centering
	\includegraphics[trim=10 10 10 10,clip,width=0.8\textwidth]{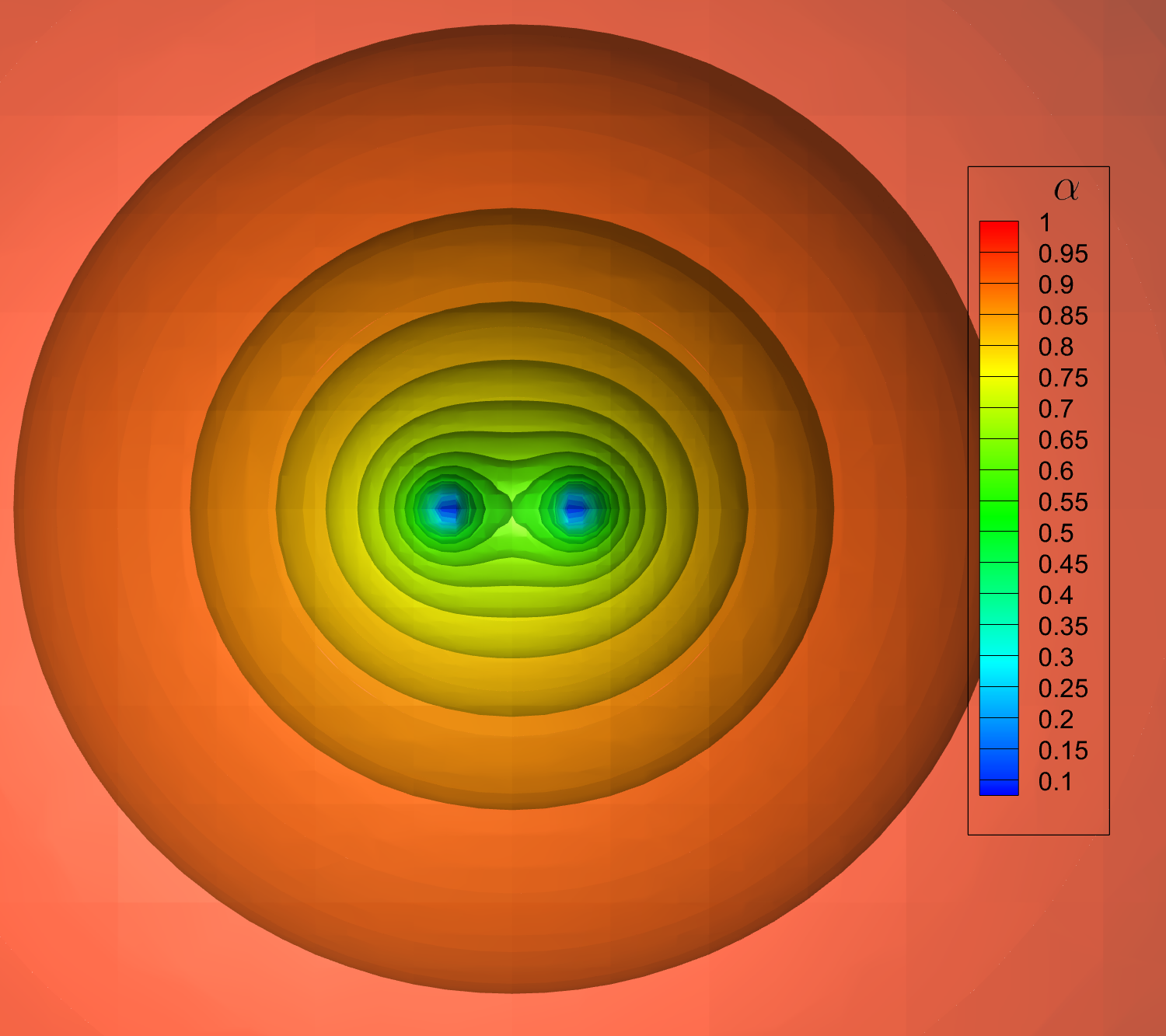}\\[-4pt]
	\caption*{$ t=0 $}
\end{subfigure}
\begin{subfigure}{0.49\linewidth}
	\centering
	\includegraphics[trim=10 10 10 10,clip,width=0.8\textwidth]{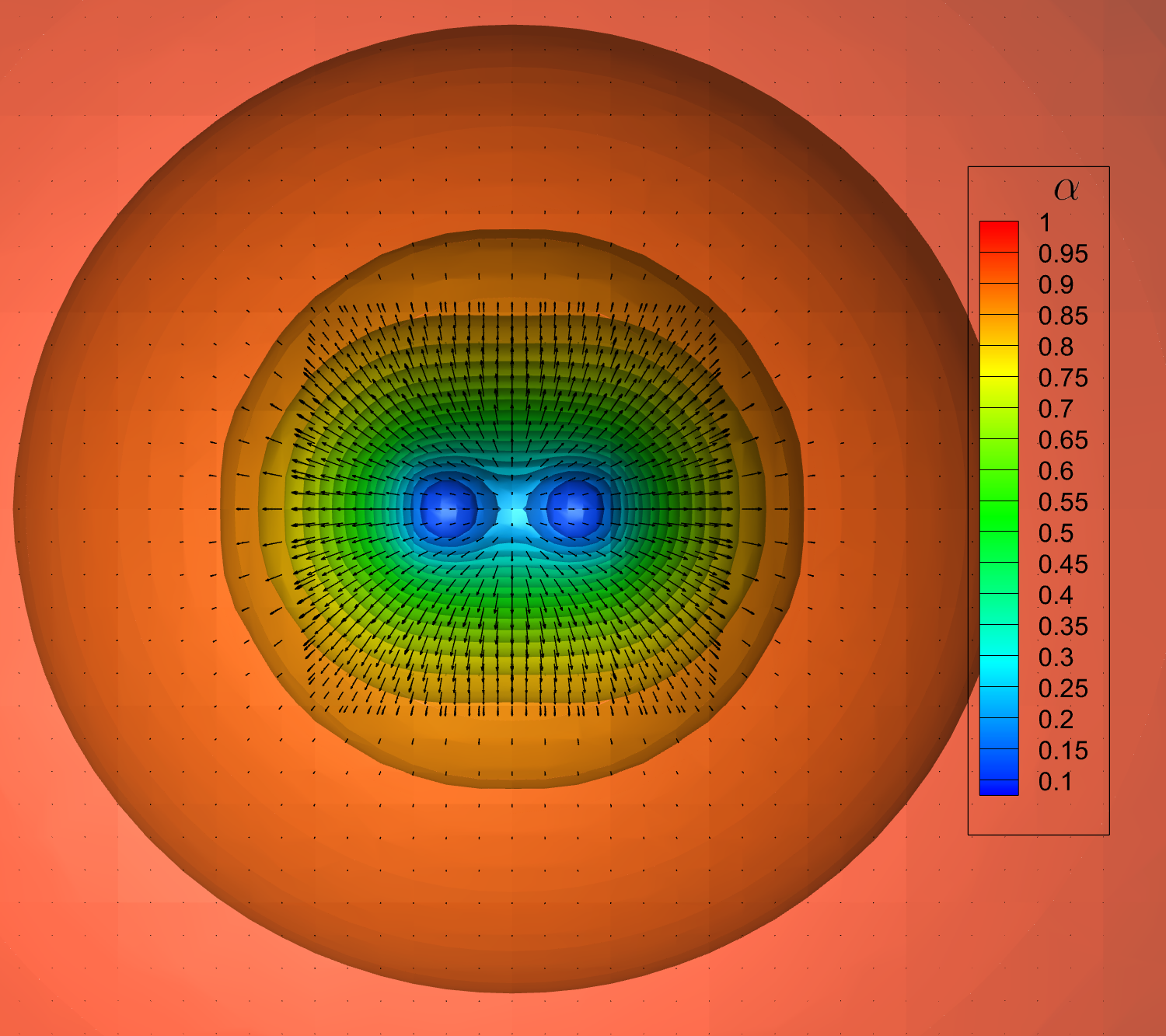}\\[-4pt]
	\caption*{$ t=5 $}
\end{subfigure}		
\par\medskip
\begin{subfigure}{0.49\linewidth}
	\centering
	\includegraphics[trim=10 10 10 10,clip,width=0.8\textwidth]{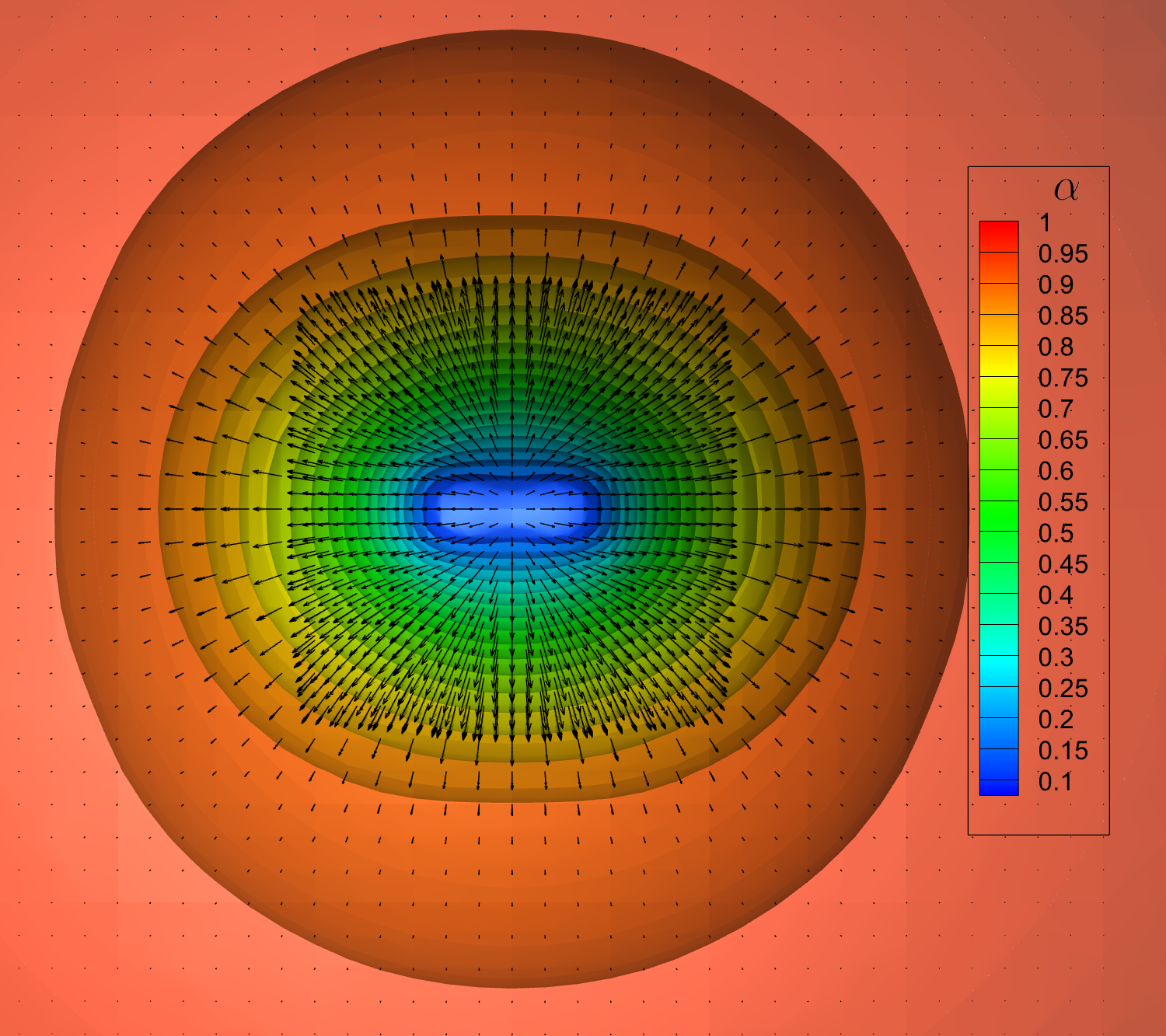}\\[-4pt]
	\caption*{$ t=7 $}
\end{subfigure}
\begin{subfigure}{0.49\linewidth}
	\centering
	\includegraphics[trim=10 10 10 10,clip,width=0.8\textwidth]{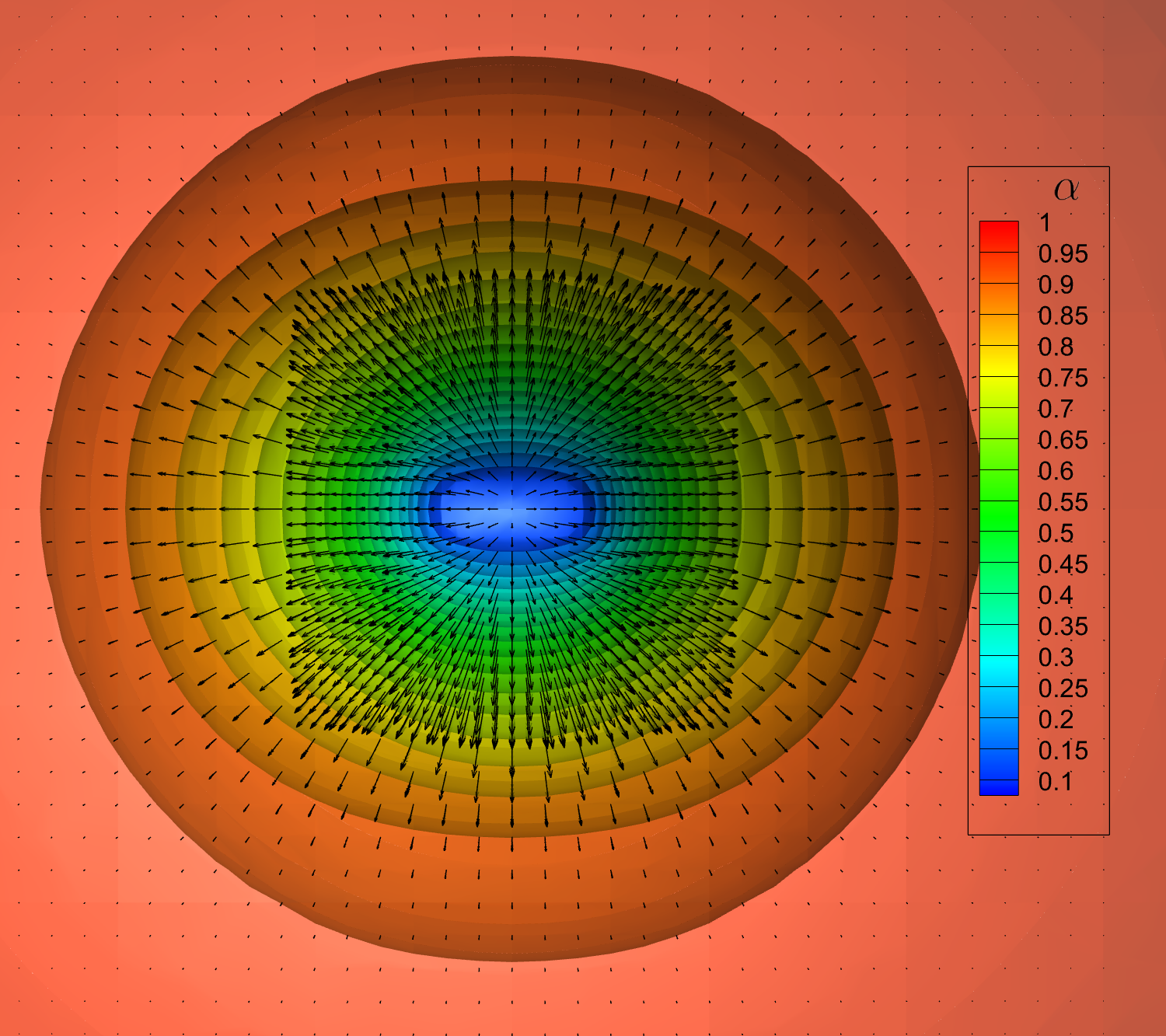}\\[-4pt]
	\caption*{$ t=8 $}
\end{subfigure}		
\par\bigskip
\begin{subfigure}{0.49\linewidth}
	\centering
	\includegraphics[trim=10 10 10 10,clip,width=0.8\textwidth]{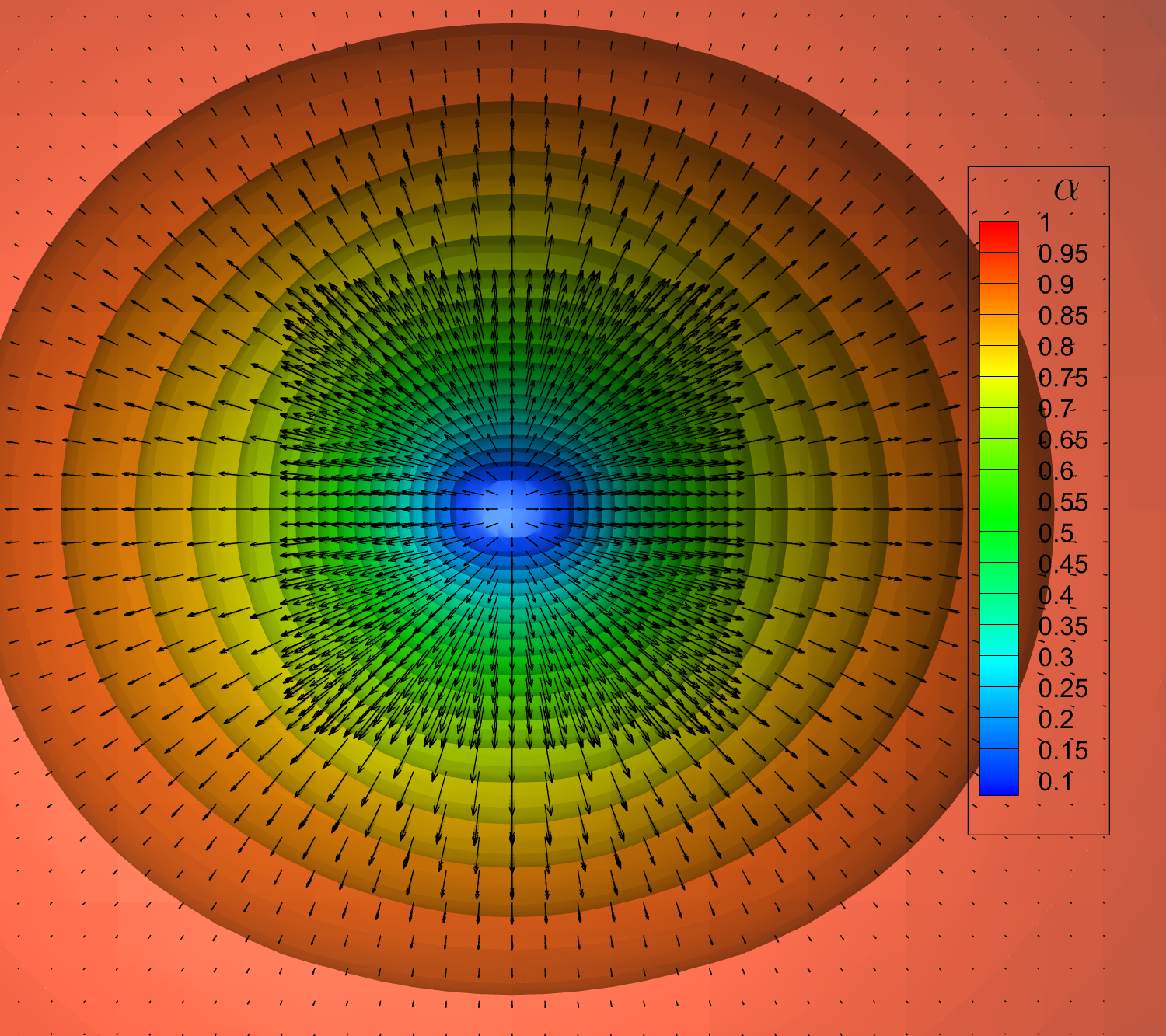}\\[-4pt]
	\caption*{$ t=10 $}
\end{subfigure}
\begin{subfigure}{0.49\linewidth}
	\centering
	\includegraphics[trim=10 10 10 10,clip,width=0.8\textwidth]{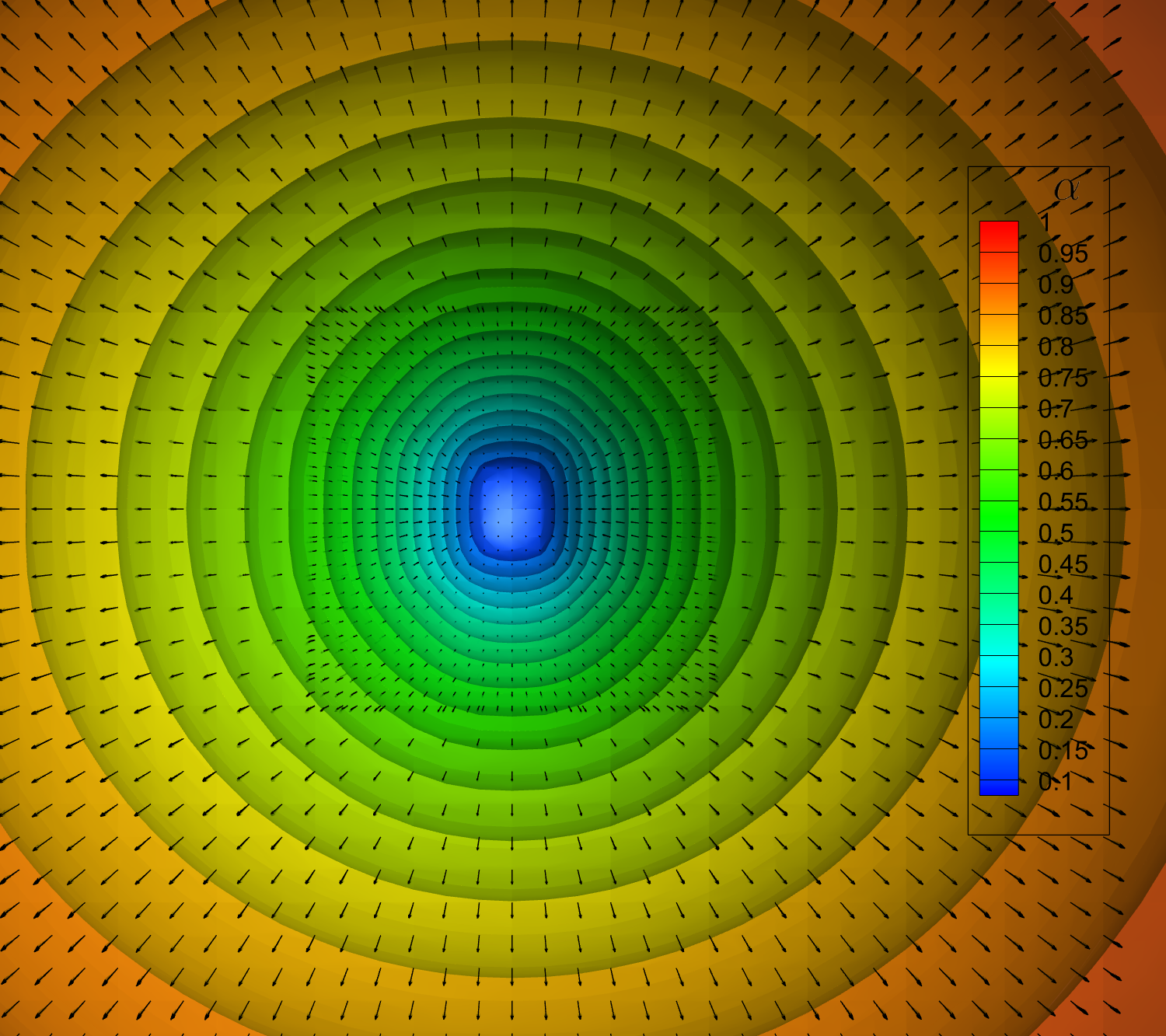}\\[-4pt]
\caption*{$ t=20 $}
\end{subfigure}		
	
		%
		\caption{Contour surfaces of the lapse for the two punctures black holes. The solution
		is reported at six different times: $t=0,5,7,8,10,20\,M$.}  
		\label{fig:2BH3D}
	\end{center}
\end{figure}

\begin{figure}[!hpb]
	\begin{center}
		{\includegraphics[angle=0,width=0.7\textwidth]{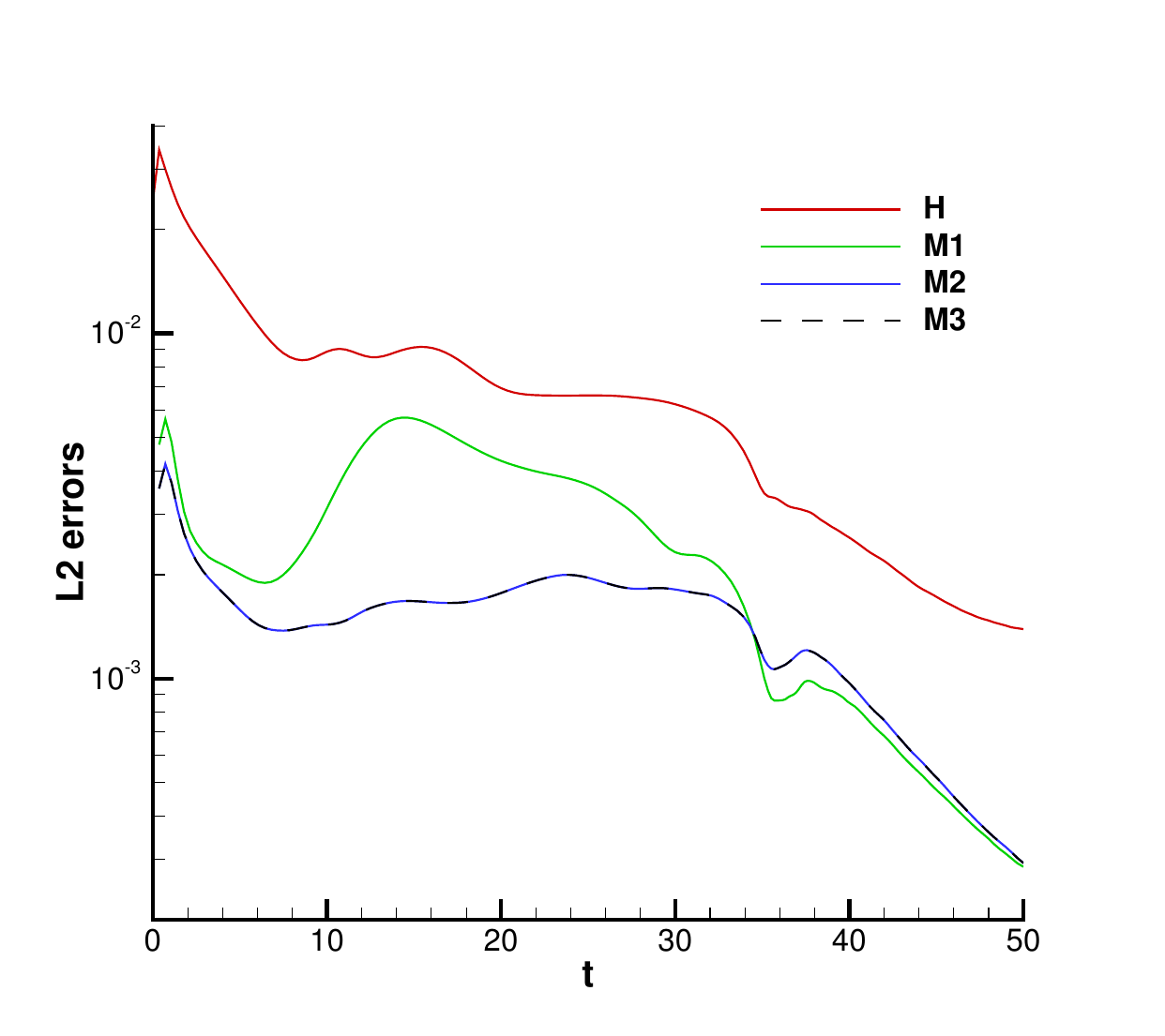}} 
		\caption{Time evolution of the constraint violations for two punctures black holes. 
		} 
		\label{fig:TwoPunctures-Constraint}
	\end{center}
\end{figure}

As a last test we have analysed the head-on collision of two nonrotating black holes, 
which are modelled as two moving punctures. 
The initial conditions can be obtained by the
\texttt{TwoPunctures} initial data code~\cite{Ansorg:2004ds}, and are prescribed as follows:
\begin{itemize}
\item equal black hole masses, $M=1$, with no spin;
\item initial positions given by $\boldsymbol{x}^- =
(-1,0,0)$ and $\boldsymbol{x}^+ = (+1,0,0)$;
\item zero linear momenta;
\item zero initial extrinsic curvature.
\end{itemize}
We have performed this test to the purpose of showing the ability of the DG scheme based on our improved Z4 implementation of the Einstein equations to solve 
moving punctures, irrespective of the possibility of extracting gravitational waves, which will be the subject of a future research. 
The three-dimensional
computational domain is given by $\Omega = [-60;60]^3$ and flat
Minkowski spacetime is imposed as boundary condition everywhere. We use adaptive mesh refinement 
(AMR) with time accurate local time stepping~\cite{AMR3DCL} and one level of refinement with 
refinement factor $\varrho=3$ inside the 
box $[-10,10]^3$. The subcell finite volume limiter is always activated within the box $[-3,3]^3$.   
The numerical relevant parameters  are set as $\kappa_1=0.2$, $\kappa_2=0.2$, $c=0$, $\mu=0.0$.

For this test, the activation of the {\emph {gamma--driver}} is mandatory. In order for the evolution to proceed successfully, we have found that it is necessary
to perform the following actions:
in the inner region the lapse $\alpha$ is flattened as
\begin{equation}
\alpha=\frac{\alpha r^6 + \epsilon \alpha_{min}}{r^6 + \epsilon}\,,
\end{equation}
where $\alpha_{min}=0.01$, $\epsilon=10^{-4}$, in such a way that the spacetime evolution is effectively frozen. Simultaneously, all the metric terms are filtered as
\begin{eqnarray}
f&=&\erf\left (\frac{\gamma_{max}}{\gamma_{ij}} \left[1 + \left(\frac{r}{0.4}\right)^4 \right]   \right),\\
\gamma_{ij}&=&\gamma_{max}(1-f) + \gamma_{ij}f\,,
\end{eqnarray}
so as to avoid metric spikes, but rather reaching a smooth maximum value at $\gamma_{max}\sim 25$. In addition, since there is not an exact solution for this test, the well-balancing
property is switched off completely.

In Figure~\ref{fig:2BH3D} we present the contour iso-surfaces
of the lapse at different times, showing the merger process of the two black holes. In Figure~\ref{fig:TwoPunctures-Constraint} the time evolution 
of the Hamiltonian and momentum constraints are reported, showing a stable evolution of the system until the end of the merger process. To the very best knowledge of the authors, this is the very first stable 3D simulation of a head-on collision of two puncture black holes carried out with a high order DG scheme applied to the first order reformulation of the Z4 system of the Einstein field equation.

\section{Conclusions}
\label{sec.conclusions}

In this paper we have investigated the first--order version of the Z4 formulation of the Einstein--Euler equations, originally proposed by
\cite{Bona:2003fj, Bona:2003qn}, via a new well-balanced discontinuous Galerkin
scheme for \emph{non conservative systems}. We have 
shown substantial advantages with respect to its analogous 
first--order CCZ4 version, already discussed in~\cite{Dumbser2017strongly}.  
Along with an obvious simpler form of the equations, when compared to CCZ4, in the Z4 system  the $Z^\mu$ four vector 
is an evolved quantity, allowing for a direct monitoring of the Einstein constraints violations. 
Strong hyperbolicity has been verified by computing the full set of eigenvectors for a general metric in case of frozen shift.
The new high order well-balanced ADER-DG scheme for conservative and non-conservative systems relies on the framework of path-conservative schemes. The choice of the path is irrelevant in the case of the Einstein field equations, since the non-conservative part of the system concerns only the metric, which cannot develop discontinuities as all associated characteristic fields are linearly degenerate. We have verified the nominal order of convergence of our new scheme up to seventh order in space and time.
Two additional and fundamental features make the new numerical scheme particularly robust and attractive:
\begin{enumerate}
\item The overall scheme is well-balanced, in the sense that it can preserve stationary equilibrium solutions exactly up to machine precision. This has been obtained
in a pragmatic but very effective way by subtracting the discretized equilibrium solution from the evolved one during the simulation. 
For highly dynamical systems, on the other hand, the well-balancing property is not useful and hence not adopted.
\item The conversion from the conservative to the primitive variables, which has been plaguing relativistic hydrodynamic codes for so long, has been made
substantially more robust by the introduction of a special filter function, which avoids division by zero and thus the divergence of the velocity in regimes of very low rest mass densities. To the best of our knowledge, this is the very first time that compact objects like neutron star can be simulated by setting $\rho=0$ outside the object, instead of requiring a numerical atmosphere.  
\end{enumerate}
After these improvements,  
we have been able to reproduce all the standard tests of numerical relativity with unprecedented accuracy in the computation
of stationary solutions. 
In particular, and
to the best of our knowledge, this  is the first time that a stationary black hole (including an extreme Kerr one with $a=0.99$)
has been evolved with a high order DG scheme in three space dimensions within the 3+1 formalism up to $t=1000 M$, and with no limitation to proceed even further.
\textcolor{black}{
	Our new approach could be beneficial for the numerical study of quasi--normal modes (QNM) of oscillations of black holes, which represents a fertile field of research  in high energy astrophysics (see, among the others, \cite{Baibhav2019}.)
}

Second, our new filter in the conversion from the conserved to the primitive variables allowed us to evolve a TOV star in true vacuum, namely with $p=\rho=0$ outside the star.
This new feature is likely to play a major role in future applications of high energy astrophysics where very low density regions are involved.

Finally, at the level of a proof of concept calculation and with no intention yet to compute the gravitational wave emission from a binary system, we have obtained first encouraging preliminary results concerning the head--on collision of two equal masses black holes. This demonstrates the possibility to account for a physical problem that was previously considered off--limits  for the original Z4 formulation. 

Future work will concern the application of the new numerical scheme to the simulation of the inspiral and merger of binary black holes and binary neutron star systems with the calculation of the related gravitational waves.

\section{Acknowledgments}

This work was financially supported by the Italian Ministry of Education, University 
and Research (MIUR) in the framework of the PRIN 2022 project \textit{High order structure-preserving semi-implicit schemes for hyperbolic equations} and via the  Departments of Excellence  Initiative 2018--2027 attributed to DICAM of the University of Trento (grant L. 232/2016).
M.~D. and I.~P. are members of the INdAM GNCS group in Italy.  

E.~G. is member of the CARDAMOM team at the Inria center of the University of Bordeaux and 
gratefully acknowledges the support received from the European Union’s Horizon 2020 Research and Innovation Programme under the Marie Sk\l{}odowska-Curie Individual Fellowship \textit{SuPerMan}, grant agreement No. 101025563, 
and the support and funding received from the European Union with
the ERC Starting Grant ALcHyMiA (No. 101114995). 
Views and opinions expressed are however those of the author(s) only and do not necessarily reflect those of the European Union or the
European Research Council Executive Agency. Neither the European Union nor the granting authority can be held responsible for them. 

We would kindly like to thank Carlos Palenzuela, Luciano Rezzolla and Konrad Topolski for the inspiring discussions.


\appendix

\section{The eigenstructure of the first--order Z4 system}
\label{sec:eigenappendix}
We stress that the Euler and the Einstein sector of the full PDE given by~\eqref{eqn.rho}--\eqref{eqn.D} are coupled only through the source terms, 
since all the metric derivatives arising in the matrix $\partial {\bf F}(\Q)/\partial \Q$, and corresponding to the Euler block,  have been moved 
to the source terms on the right hand side as auxiliary variables. Hence, with no loss of generality, we can analyze the eigenstructure of the Einstein--Euler
system by focusing on the Einstein block, more specifically by setting to zero all the hydrodynamic variables $(D, S_1, S_2, S_3, E)$, whose eigenvectors
are well known.
In addition, assuming the \emph{1+log gauge condition} with zero shift ($\beta^i=0$, $s=0$), excluding the passive quantity $K_0$ from the analysis, and
using  $c=0$, the remaining 55 variables for the state vector $\Q$ relative
to the matter and spacetime evolution are given by
\begin{eqnarray}
\label{eqn.pde.Q}
{\boldsymbol Q}^T &=& \Big(
D, S_1, S_2, S_3, E, \ln{\alpha},
\beta^1, \beta^2, \beta^3,
\gamma_{11}, \gamma_{12}, \gamma_{13},
\gamma_{22}, \gamma_{23}, \gamma_{33},
 K_{11},  K_{12},  K_{13},
 K_{22},  K_{23},  K_{33},
\Theta,
Z_1, Z_2, Z_3,
\nonumber \\
&&
A_1, A_2, A_3,
B^{\,\,1}_1, B^{\,\,1}_2, B^{\,\,1}_3,
B^{\,\,2}_1, B^{\,\,2}_2, B^{\,\,2}_3,
B^{\,\,3}_1, B^{\,\,3}_2, B^{\,\,3}_3,
D_{111}, D_{112}, D_{113},
D_{122}, D_{123}, D_{133},
D_{211}, D_{212}, D_{213},
\nonumber \\
&&
D_{222}, D_{223}, D_{233},
D_{311}, D_{312}, D_{313},
D_{322}, D_{323}, D_{333},
\Big)
\,.
\end{eqnarray}
Under such circumstances, the eigenvalues are given by \\
\begin{equation}
\begin{array}{llll}
\lambda_{1} &=&\sqrt{\gamma^{11}}\alpha\, e  &   \textrm{(multiplicity 1)} \,,  \\
\lambda_{2} &=&-\sqrt{\gamma^{11}}\alpha\,e &  \textrm{(multiplicity 1)} \,, \\
\lambda_{3,\cdots,7}  &=&  \sqrt{\gamma^{11}}\alpha &  \textrm{(multiplicity 5)} \,, \\
\lambda_{8,\cdots,12} &=&- \sqrt{\gamma^{11}}\alpha & \textrm{(multiplicity 5)} \,, \\
\lambda_{13,\cdots,53} &=&  0                       & \textrm{(multiplicity 41)} \,, \\
\lambda_{54} &=& \sqrt{2} \sqrt{\alpha \,\gamma^{11}}  &   \textrm{(multiplicity 1)}   \,, \\
\lambda_{55} &=& -\sqrt{2} \sqrt{\alpha \,\gamma^{11}}  &   \textrm{(multiplicity 1)}\,,
\end{array}
\end{equation}
\\
with corresponding eigenvectors: 
\
\footnotesize
\begin{eqnarray}
%
\boldsymbol{r}_1^T & = & \Big( 0\,, 0\,, 0\,, 0\,, 0\,, 0\,, 0\,, 0\,, 0\,, 0\,, 0\,, 0\,, 0\,, 0\,, 0\,, \sqrt {\gamma^{11}}e\,, 0\,, 0\,, 0\,, 0\,, 0\,, 1/2\,{\frac {{\gamma^{11}}^{3/2}e  {e}^{2}\alpha-2  }{\alpha}}\,, -1/2\,{\frac {\gamma^{11}\,  {e}^{2}\alpha-2  }{\alpha}}\,, 0\,, 0\,, \nonumber \\
&& 2\,{\frac {\gamma^{11}}{\alpha}}\,, 0\,, 0\,, 0\,, 0\,, 0\,, 0\,, 0\,, 0\,, 0\,, 0\,, 0\,, 1\,, 0\,, 0\,, 0\,, 0\,, 0\,, 0\,, 0\,, 0\,, 0\,, 0\,, 0\,, 0\,, 0\,, 0\,, 0\,, 0\,, 0 \Big) \\
%
\boldsymbol{r}_2^T & = &  \Big(  0\,,  0\,,  0\,,  0\,,  0\,,  0\,,  0\,,  0\,,  0\,,  0\,,  0\,,  0\,,  0\,,  0\,,  0\,,  -\sqrt {\gamma^{11}}e\,,  0\,,  0\,,  0\,,  0\,,  0\,,  -1/2\,{\frac {{\gamma^{11}}^{3/2}e ( {e}^{2}\alpha-2 ) }{\alpha}}\,, \nonumber \\
&& -1/2\,{\frac{\gamma^{11}\, ( {e}^{2}\alpha-2 ) }{\alpha}}\,,  0\,,  0\,,  2\,{\frac{\gamma^{11}}{\alpha}}\,,  0\,,  0\,,  0\,,  0\,,  0\,,  0\,,  0\,,  0\,,  0\,,  0\,,  0\,,  1\,,  0\,,  0\,,  0\,,  0\,,  0\,,  0\,,  0\,,  0\,,  0\,,  0\,,  0\,,  0\,,  0\,,  0\,,  0\,,  0\,,  0 \Big) \\
%
\boldsymbol{r}_3^T & = & \Big( 0\,, 0\,, 0\,, 0\,, 0\,, 0\,, 0\,, 0\,, 0\,, 0\,, 0\,, 0\,, 0\,, 0\,, 0\,, -2\,{\frac {\gamma^{12}}{\sqrt {\gamma^{11}}}}\,, \sqrt {\gamma^{11}}\,, 0\,, 0\,, 0\,, 0\,, 0\,, \gamma^{12}\,, -\gamma^{11}\,, 0\,, 0\,, 0\,, 0\,, 0\,, 0\,, 0\,, 0\,, 0\,, 0\,, 0\,, 0\,, \nonumber \\
&&0\,, -2\,{\frac {\gamma^{12}}{\gamma^{11}}}\,,  1\,, 0\,, 0\,, 0\,, 0\,, 0\,, 0\,, 0\,, 0\,, 0\,, 0\,, 0\,, 0\,, 0\,, 0\,, 0\,, 0 \Big) \\
%
\boldsymbol{r}_4^T & = &  \Big( 0\,, 0\,, 0\,, 0\,, 0\,, 0\,, 0\,, 0\,, 0\,, 0\,, 0\,, 0\,, 0\,, 0\,, 0\,, -2\,\frac{\gamma^{13}}{\sqrt {\gamma^{11}}}\,, 0\,, \sqrt{\gamma^{11}}\,, 0\,, 0\,, 0\,, 0\,, \gamma^{13}\,, 0\,, -\gamma^{11}\,, 0\,, 0\,, 0\,, 0\,, 0\,, 0\,, 0\,, 0\,, 0\,, 0\,, 0\,, 0\,, \nonumber \\
&&-2\,\frac{\gamma^{13}}{\gamma^{11}}\,, 0\,, 1\,, 0\,, 0\,, 0\,, 0\,, 0\,, 0\,, 0\,, 0\,, 0\,, 0\,, 0\,, 0\,, 0\,, 0\,, 0 \Big) \\
%
\boldsymbol{r}_5^T & = &  \Big(0\,, 0\,, 0\,, 0\,, 0\,, 0\,, 0\,, 0\,, 0\,, 0\,, 0\,, 0\,, 0\,, 0\,, 0\,, -\frac{\gamma^{22}}{\sqrt {\gamma^{11}}}\,, 0\,, 0\,, \sqrt {\gamma^{11}}\,, 0\,, 0\,, 0\,, \gamma^{22}\,, -\gamma^{12}\,, 
0\,, 0\,, 0\,, 0\,, 0\,, 0\,, 0\,, 0\,, 0\,, 0\,, 0\,, 0\,, 0\,, \nonumber \\
&&-{\frac {\gamma^{22}}{\gamma^{11}}}\,, 0\,, 0\,, 1\,, 0\,, 0\,, 0\,, 0\,, 0\,, 0\,, 0\,, 0\,, 0\,, 0\,, 0\,, 0\,, 0\,, 0\Big)  \\
%
\boldsymbol{r}_6^T & = &  \Big(0\,, 0\,, 0\,, 0\,, 0\,, 0\,, 0\,, 0\,, 0\,, 0\,, 0\,, 0\,, 0\,, 0\,, 0\,, -2\,\frac{\gamma^{23}}{\sqrt {\gamma^{11}}}\,, 0\,, 0\,, 0\,, \sqrt{\gamma^{11}}\,, 0\,, 0\,, 2\,\gamma^{23}\,, -\gamma^{13}\,, -\gamma^{12}\,, 0\,, 0\,, 0\,, 0\,, 0\,, 0\,, 0\,, 0\,, 0\,, 0\,, \nonumber \\
&& 0\,, 0\,, -2\,{\frac {\gamma^{23}}{\gamma^{11}}}\,, 0\,, 0\,, 0\,, 1\,, 0\,, 0\,, 0\,, 0\,, 0\,, 0\,, 0\,, 0\,, 0\,, 0\,, 0\,, 0\,, 0 \Big) \\
%
\boldsymbol{r}_7^T & = & \Big( 0\,, 0\,, 0\,, 0\,, 0\,, 0\,, 0\,, 0\,, 0\,, 0\,, 0\,, 0\,, 0\,, 0\,, 0\,, -{\frac {\gamma^{33}}{\sqrt {\gamma^{11}}}}\,, 0\,, 0\,, 0\,, 0\,, \sqrt {\gamma^{11}}\,, 0\,, \gamma^{33}\,, 0\,, -\gamma^{13}\,, 0\,, 0\,, 0\,, 0\,, 0\,, 0\,, 0\,, 0\,, 0\,, 0\,, 0\,, 0\,, \nonumber \\
&& -{\frac{\gamma^{33}}{\gamma^{11}}}\,, 0\,, 0\,, 0\,, 0\,, 1\,, 0\,, 0\,, 0\,, 0\,, 0\,, 0\,, 0\,, 0\,, 0\,, 0\,, 0\,, 0 \Big) \\
%
\boldsymbol{r}_8^T & = &  \Big(0\,, 0\,, 0\,, 0\,, 0\,, 0\,, 0\,, 0\,, 0\,, 0\,, 0\,, 0\,, 0\,, 0\,, 0\,, {\frac{\gamma^{22}}{\sqrt {\gamma^{11}}}}\,, 0\,, 0\,, -\sqrt{\gamma^{11}}\,, 0\,, 0\,, 0\,, \gamma^{22}\,, -\gamma^{12}\,, 0\,, 0\,, 0\,, 0\,, 0\,, 0\,, 0\,, 0\,, 0\,, 0\,, 0\,, 0\,, 0\,, \nonumber \\
&& -{\frac{\gamma^{22}}{\gamma^{11}}}\,, 0\,, 0\,, 1\,, 0\,, 0\,, 0\,, 0\,, 0\,, 0\,, 0\,, 0\,, 0\,, 0\,, 0\,, 0\,, 0\,, 0 \Big) \\
%
\boldsymbol{r}_9^T & = & \Big(0\,, 0\,, 0\,, 0\,, 0\,, 0\,, 0\,, 0\,, 0\,, 0\,, 0\,, 0\,, 0\,, 0\,, 0\,, {\frac{\gamma^{33}}{\sqrt {\gamma^{11}}}}\,, 0\,, 0\,, 0\,, 0\,, -\sqrt {\gamma^{11}}\,, 0\,, \gamma^{33}\,, 0\,, -\gamma^{13}\,, 0\,, 0\,, 0\,, 0\,, 0\,, 0\,, 0\,, 0\,, 0\,, 0\,, 0\,, 0\,, \nonumber \\
&& -{\frac {\gamma^{33}}{\gamma^{11}}}\,, 0\,, 0\,, 0\,, 0\,, 1\,, 0\,, 0\,, 0\,, 0\,, 0\,, 0\,, 0\,, 0\,, 0\,, 0\,, 0\,, 0\Big) \\
%
\boldsymbol{r}_{10}^T & = &  \Big(0\,, 0\,, 0\,, 0\,, 0\,, 0\,, 0\,, 0\,, 0\,, 0\,, 0\,, 0\,, 0\,, 0\,, 0\,, 2\,{\frac{\gamma^{12}}{\sqrt {\gamma^{11}}}}\,, -\sqrt {\gamma^{11}}\,, 0\,, 0\,, 0\,, 0\,, 0\,, \gamma^{12}\,, -\gamma^{11}\,, 0\,, 0\,, 0\,, 0\,, 0\,, 0\,, 0\,, 0\,, 0\,, 0\,, 0\,, 0\,, \nonumber \\ 
&& 0\,, -2\,{\frac {\gamma^{12}}{\gamma^{11}}}\,, 1\,, 0\,, 0\,, 0\,, 0\,, 0\,, 0\,, 0\,, 0\,, 0\,, 0\,, 0\,, 0\,, 0\,, 0\,, 0\,, 0\Big) \\
%
\boldsymbol{r}_{11}^T & = &  \Big(0\,, 0\,, 0\,, 0\,, 0\,, 0\,, 0\,, 0\,, 0\,, 0\,, 0\,, 0\,, 0\,, 0\,, 0\,, 2\,{\frac{\gamma^{13}}{\sqrt {\gamma^{11}}}}\,, 0\,, -\sqrt {\gamma^{11}}\,, 0\,, 0\,, 0\,, 0\,, \gamma^{13}\,, 0\,, -\gamma^{11}\,, 0\,, 0\,, 0\,, 0\,, 0\,, 0\,, 0\,, 0\,, 0\,, 0\,, 0\,, \nonumber \\
&&0\,, -2\,{\frac {\gamma^{13}}{\gamma^{11}}}\,, 0\,, 1\,, 0\,, 0\,, 0\,, 0\,, 0\,, 0\,, 0\,, 0\,, 0\,, 0\,, 0\,, 0\,, 0\,, 0\,, 0\Big) \\
%
\boldsymbol{r}_{12}^T & = & \Big(0\,, 0\,, 0\,, 0\,, 0\,, 0\,, 0\,, 0\,, 0\,, 0\,, 0\,, 0\,, 0\,, 0\,, 0\,, 2\,{\frac{\gamma^{23}}{\sqrt {\gamma^{11}}}}\,, 0\,, 0\,, 0\,, -\sqrt {\gamma^{11}}\,, 0\,, 0\,, 2\,\gamma^{23}\,, -\gamma^{13}\,, -\gamma^{12}\,, 0\,, 0\,, 0\,, 0\,, 0\,, 0\,, 0\,, 0\,, 0\,, \nonumber \\
&&0\,, 0\,, 0\,, -2\,{\frac {\gamma^{23}}{\gamma^{11}}}\,, 0\,, 0\,, 0\,, 1\,, 0\,, 0\,, 0\,, 0\,, 0\,, 0\,, 0\,, 0\,, 0\,, 0\,, 0\,, 0\,, 0\Big) \\
%
\boldsymbol{r}_{13}^T &  = &   \Big( 0\,, 0\,, 0\,, 0\,, 0\,, 0\,, 0\,, 0\,, 0\,, 0\,, 0\,, 0\,, 0\,, 0\,, 0\,, 0\,, 0\,, 0\,, 0\,, 0\,, 0\,, 0\,, 0\,, 0\,, 0\,, 0\,, 0\,, 0\,, 0\,, 0\,, 0\,, 0\,, 0\,, 0\,, 0\,, 0\,, 0\,, 0\,, 0\,, 0\,, 0\,, 0\,, 0\,, \nonumber \\
&&0\,, 0\,, 0\,,  0\,, 0\,, 0\,, 0\,, 0\,, 0\,, 0\,, 0\,, 1 \Big) \\
%
\boldsymbol{r}_{14}^T & = &  \Big ( 0\,, 0\,, 0\,, 0\,, 0\,, 0\,, 0\,, 0\,, 0\,, 0\,, 0\,, 0\,, 0\,, 0\,, 0\,, 0\,, 0\,, 0\,, 0\,, 0\,, 0\,, 0\,, 0\,, 0\,, 0\,, 0\,, 0\,, 0\,, 0\,, 0\,, 0\,, 0\,, 0\,, 0\,, 0\,, 0\,, 0\,, 0\,, 0\,, 0\,, 0\,, 0\,, 0\,, \nonumber \\
&&0\,, 0\,, 0\,,  0\,, 0\,, 0\,, 1\,, 0\,, 0\,, 0\,, 0\,, 0 \Big)\\
%
\boldsymbol{r}_{15}^T & = & \Big ( 0\,, 0\,, 0\,, 0\,, 0\,, 0\,, 0\,, 0\,, 0\,, 0\,, 0\,, 0\,, 0\,, 0\,, 0\,, 0\,, 0\,, 0\,, 0\,, 0\,, 0\,, 0\,, 0\,, 0\,, 0\,, 0\,, 0\,, 0\,, 0\,, 0\,, 0\,, 0\,, 0\,, 0\,, 0\,, 0\,, 0\,, 0\,, 0\,, 0\,, 0\,, 0\,, 0\,, \nonumber \\
&&0\,, 0\,, 0\,,  0\,, 1\,, 0\,, 0\,, 0\,, 0\,, 1\,, 0\,, 0 \Big) \\
%
\boldsymbol{r}_{16}^T & = & \Big( 0\,, 0\,, 0\,, 0\,, 0\,, 0\,, 0\,, 0\,, 0\,, 0\,, 0\,, 0\,, 0\,, 0\,, 0\,, 0\,, 0\,, 0\,, 0\,, 0\,, 0\,, 0\,, 0\,, 0\,, 0\,, 0\,, 0\,, 0\,, 0\,, 0\,, 0\,, 0\,, 0\,, 0\,, 0\,, 0\,, 0\,, 0\,, 0\,, 0\,, 0\,, 0\,, 0\,, \nonumber \\
&&0\,, 0\,, 0\,,  1\,, 0\,, 0\,, 0\,, 0\,, 0\,, 0\,, 0\,, 0 \Big)\\
%
\boldsymbol{r}_{17}^T & = & \Big( 0\,, 0\,, 0\,, 0\,, 0\,, 0\,, 0\,, 0\,, 0\,, 0\,, 0\,, 0\,, 0\,, 0\,, 0\,, 0\,, 0\,, 0\,, 0\,, 0\,, 0\,, 0\,, 0\,, 0\,, 0\,, 0\,, 0\,, 0\,, 0\,, 0\,, 0\,, 0\,, 0\,, 0\,, 0\,, 0\,, 0\,, 0\,, 0\,, 0\,, 0\,, 0\,, 0\,, \nonumber \\
&&1\,, 0\,, 0\,, 0\,, 0\,, 0\,, 0\,, 0\,, 0\,, 0\,, 0\,, 0 \Big) \\
%
\boldsymbol{r}_{18}^T & = & \Big( 0\,, 0\,, 0\,, 0\,, 0\,, 0\,, 0\,, 0\,, 0\,, 0\,, 0\,, 0\,, 0\,, 0\,, 0\,, 0\,, 0\,, 0\,, 0\,, 0\,, 0\,, 0\,, 0\,, 0\,, 0\,, 0\,, 0\,, 0\,, 0\,, 0\,, 0\,, 0\,, 0\,, 0\,, 0\,, 0\,, 0\,, 0\,, 0\,, 0\,, 0\,, 0\,, 1\,, \nonumber \\
&&0\,, 0\,, 0\,, 0\,, 0\,, 0\,, 0\,, 0\,, 1\,, 0\,, 0\,, 0 \Big)\\
%
\boldsymbol{r}_{19}^T & = & \Big( 0\,, 0\,, 0\,, 0\,, 0\,, 0\,, 0\,, 0\,, 0\,, 0\,, 0\,, 0\,, 0\,, 0\,, 0\,, 0\,, 0\,, 0\,, 0\,, 0\,, 0\,, 0\,, 0\,, 0\,, 0\,, 0\,, 0\,, 0\,, 0\,, 0\,, 0\,, 0\,, 0\,, 0\,, 0\,, 0\,, 0\,, 0\,, 0\,, 0\,, 1\,, 0\,, 0\,, \nonumber \\
&&0\,, 1\,, 0\,, 0\,, 0\,, 0\,, 0\,, 0\,, 0\,, 0\,, 0\,, 0\Big)\\
%
\boldsymbol{r}_{20}^T & = & \Big( 0\,, 0\,, 0\,, 0\,, 0\,, 0\,, 0\,, 0\,, 0\,, 0\,, 0\,, 0\,, 0\,, 0\,, 0\,, 0\,, 0\,, 0\,, 0\,, 0\,, 0\,, 0\,, 0\,, 0\,, 0\,, 0\,, 0\,, 0\,, 0\,, 0\,, 0\,, 0\,, 0\,, 0\,, 0\,, 0\,, 0\,, 0\,, 0\,, 1\,, 0\,, 0\,, 0\,, \nonumber \\
&&0\,, 0\,, 0\,, 0\,, 0\,, 0\,, 0\,, 0\,, 0\,, 0\,, 0\,, 0\Big)\\
%
\boldsymbol{r}_{21}^T & = & \Big( 0\,, 0\,, 0\,, 0\,, 0\,, 0\,, 0\,, 0\,, 0\,, 0\,, 0\,, 0\,, 0\,, 0\,, 0\,, 0\,, 0\,, 0\,, 0\,, 0\,, 0\,, 0\,, 0\,, 0\,, 0\,, 0\,, 0\,, 0\,, 0\,, 0\,, 0\,, 0\,, 0\,, 0\,, 0\,, 0\,, 0\,, 0\,, 1\,, 0\,, 0\,, 0\,, 0\,, \nonumber \\
&&0\,, 0\,, 0\,, 0\,, 0\,, 0\,, 0\,, 0\,, 0\,, 0\,, 0\,, 0 \Big) \\
%
\boldsymbol{r}_{22}^T & = & \Big( 0\,, 0\,, 0\,, 0\,, 0\,, 0\,, 0\,, 0\,, 0\,, 0\,, 0\,, 0\,, 0\,, 0\,, 0\,, 0\,, 0\,, 0\,, 0\,, 0\,, 0\,, 0\,, 0\,, 0\,, 0\,, 0\,, 0\,, 0\,, 0\,, 0\,, 0\,, 0\,, 0\,, 0\,, 0\,, 0\,, 0\,, 1\,, 0\,, 0\,, 0\,, 0\,, 0\,, \nonumber \\ 
&&0\,, 0\,, 0\,, 0\,, 0\,, 0\,, 0\,, 0\,, 0\,, 0\,, 0\,, 0 \Big)\\
%
\boldsymbol{r}_{23}^T & = & \Big( 0\,, 0\,, 0\,, 0\,, 0\,, 0\,, 0\,, 0\,, 0\,, 0\,, 0\,, 0\,, 0\,, 0\,, 0\,, 0\,, 0\,, 0\,, 0\,, 0\,, 0\,, 0\,, 0\,, 0\,, 0\,, 0\,, 0\,, 0\,, 0\,, 0\,, 0\,, 0\,, 0\,, 0\,, 0\,, 0\,, 1\,, 0\,, 0\,, 0\,, 0\,, 0\,, 0\,, \nonumber \\
&&0\,, 0\,, 0\,, 0\,, 0\,, 0\,, 0\,, 0\,, 0\,, 0\,, 0\,, 0\Big)\\
%
\boldsymbol{r}_{24}^T & = & \Big( 0\,, 0\,, 0\,, 0\,, 0\,, 0\,, 0\,, 0\,, 0\,, 0\,, 0\,, 0\,, 0\,, 0\,, 0\,, 0\,, 0\,, 0\,, 0\,, 0\,, 0\,, 0\,, 0\,, 0\,, 0\,, 0\,, 0\,, 0\,, 0\,, 0\,, 0\,, 0\,, 0\,, 0\,, 0\,, 1\,, 0\,, 0\,, 0\,, 0\,, 0\,, 0\,, 0\,, \nonumber \\
&&0\,, 0\,, 0\,, 0\,, 0\,, 0\,, 0\,, 0\,, 0\,, 0\,, 0\,, 0 \Big) \\
%
\boldsymbol{r}_{25}^T & = & \Big( 0\,, 0\,, 0\,, 0\,, 0\,, 0\,, 0\,, 0\,, 0\,, 0\,, 0\,, 0\,, 0\,, 0\,, 0\,, 0\,, 0\,, 0\,, 0\,, 0\,, 0\,, 0\,, 0\,, 0\,, 0\,, 0\,, 0\,, 0\,, 0\,, 0\,, 0\,, 0\,, 0\,, 0\,, 1\,, 0\,, 0\,, 0\,, 0\,, 0\,, 0\,, 0\,, 0\,, \nonumber \\
&&0\,, 0\,, 0\,, 0\,, 0\,, 0\,, 0\,, 0\,, 0\,, 0\,, 0\,, 0 \Big) \\
%
\boldsymbol{r}_{26}^T & = &\Big( 0\,, 0\,, 0\,, 0\,, 0\,, 0\,, 0\,, 0\,, 0\,, 0\,, 0\,, 0\,, 0\,, 0\,, 0\,, 0\,, 0\,, 0\,, 0\,, 0\,, 0\,, 0\,, 0\,, 0\,, 0\,, 0\,, 0\,, 0\,, 0\,, 0\,, 0\,, 0\,, 0\,, 1\,, 0\,, 0\,, 0\,, 0\,, 0\,, 0\,, 0\,, 0\,, 0\,, \nonumber \\
&&0\,, 0\,, 0\,, 0\,, 0\,, 0\,, 0\,, 0\,, 0\,, 0\,, 0\,, 0  \Big) \\
%
\boldsymbol{r}_{27}^T & = &\Big( 0\,, 0\,, 0\,, 0\,, 0\,, 0\,, 0\,, 0\,, 0\,, 0\,, 0\,, 0\,, 0\,, 0\,, 0\,, 0\,, 0\,, 0\,, 0\,, 0\,, 0\,, 0\,, 0\,, 0\,, 0\,, 0\,, 0\,, 0\,, 0\,, 0\,, 0\,, 0\,, 1\,, 0\,, 0\,, 0\,, 0\,, 0\,, 0\,, 0\,, 0\,, 0\,, 0\,, \nonumber \\
&&0\,, 0\,, 0\,, 0\,, 0\,, 0\,, 0\,, 0\,, 0\,, 0\,, 0\,, 0  \Big) \\
%
\boldsymbol{r}_{28}^T & = &\Big( 0\,, 0\,, 0\,, 0\,, 0\,, 0\,, 0\,, 0\,, 0\,, 0\,, 0\,, 0\,, 0\,, 0\,, 0\,, 0\,, 0\,, 0\,, 0\,, 0\,, 0\,, 0\,, 0\,, 0\,, 0\,, 0\,, 0\,, 0\,, 0\,, 0\,, 0\,, 1\,, 0\,, 0\,, 0\,, 0\,, 0\,, 0\,, 0\,, 0\,, 0\,, 0\,, 0\,, \nonumber \\
&&0\,, 0\,, 0\,, 0\,, 0\,, 0\,, 0\,, 0\,, 0\,, 0\,, 0\,, 0  \Big) \\
%
\boldsymbol{r}_{29}^T & = &\Big( 0\,, 0\,, 0\,, 0\,, 0\,, 0\,, 0\,, 0\,, 0\,, 0\,, 0\,, 0\,, 0\,, 0\,, 0\,, 0\,, 0\,, 0\,, 0\,, 0\,, 0\,, 0\,, 0\,, 0\,, 0\,, 0\,, 0\,, 0\,, 0\,, 0\,, 1\,, 0\,, 0\,, 0\,, 0\,, 0\,, 0\,, 0\,, 0\,, 0\,, 0\,, 0\,, 0\,, \nonumber \\
&&0\,, 0\,, 0\,, 0\,, 0\,, 0\,, 0\,, 0\,, 0\,, 0\,, 0\,, 0  \Big) \\
%
\boldsymbol{r}_{30}^T & = &\Big( 0\,, 0\,, 0\,, 0\,, 0\,, 0\,, 0\,, 0\,, 0\,, 0\,, 0\,, 0\,, 0\,, 0\,, 0\,, 0\,, 0\,, 0\,, 0\,, 0\,, 0\,, 0\,, 0\,, 0\,, 0\,, 0\,, 0\,, 0\,, 0\,, 1\,, 0\,, 0\,, 0\,, 0\,, 0\,, 0\,, 0\,, 0\,, 0\,, 0\,, 0\,, 0\,, 0\,, \nonumber \\
&&0\,, 0\,, 0\,, 0\,, 0\,, 0\,, 0\,, 0\,, 0\,, 0\,, 0\,, 0  \Big) \\
%
\boldsymbol{r}_{31}^T & = &\Big( 0\,, 0\,, 0\,, 0\,, 0\,, 0\,, 0\,, 0\,, 0\,, 0\,, 0\,, 0\,, 0\,, 0\,, 0\,, 0\,, 0\,, 0\,, 0\,, 0\,, 0\,, 0\,, 0\,, 0\,, 0\,, 0\,, 0\,, 0\,, 1\,, 0\,, 0\,, 0\,, 0\,, 0\,, 0\,, 0\,, 0\,, 0\,, 0\,, 0\,, 0\,, 0\,, 0\,, \nonumber \\
&&0\,, 0\,, 0\,, 0\,, 0\,, 0\,, 0\,, 0\,, 0\,, 0\,, 0\,, 0  \Big) \\
%
\boldsymbol{r}_{32}^T & = &\Big( 0\,, 0\,, 0\,, 0\,, 0\,, 0\,, 0\,, 0\,, 0\,, 0\,, 0\,, 0\,, 0\,, 0\,, 0\,, 0\,, 0\,, 0\,, 0\,, 0\,, 0\,, 0\,, -1/2\,{\frac {\gamma^{12}}{\gamma^{11}}}\,, 1/2\,, 0\,, -{\frac {\gamma^{12}}{\gamma^{11}}}\,, 1\,, 0\,, 0\,, 0\,, 0\,, 0\,, 0\,, 0\,, 0\,, 0\,, 0\,, \nonumber \\
&&0\,, 0\,, 0\,, 0\,, 0\,, 0\,, 0\,, 0\,, 0\,, 0\,, 0\,, 0\,, 0\,, 0\,, 0\,, 0\,, 0\,, 0  \Big) \\
%
\boldsymbol{r}_{33}^T & = &\Big( 0\,, 0\,, 0\,, 0\,, 0\,, 0\,, 0\,, 0\,, 0\,, 0\,, 0\,, 0\,, 0\,, 0\,, 0\,, 0\,, 0\,, 0\,, 0\,, 0\,, 0\,, 0\,, -1/2\,{\frac {\gamma^{13}}{\gamma^{11}}}\,, 0\,, 1/2\,, -{\frac {\gamma^{13}}{\gamma^{11}}}\,, 0\,, 1\,, 0\,, 0\,, 0\,, 0\,, 0\,, 0\,, 0\,, 0\,, 0\,, \nonumber \\
&&0\,, 0\,, 0\,, 0\,, 0\,, 0\,, 0\,, 0\,, 0\,, 0\,, 0\,, 0\,, 0\,, 0\,, 0\,, 0\,, 0\,, 0  \Big) \\
%
\boldsymbol{r}_{34}^T & = &\Big(0\,, 0\,, 0\,, 0\,, 0\,, 0\,, 0\,, 0\,, 0\,, 0\,, 0\,, 0\,, 0\,, 0\,, 0\,, 0\,, 0\,, 0\,, 0\,, 0\,, 0\,, 0\,, 1/2\,{\frac {-{\gamma^{12}}^{2}\gamma^{33}+{\gamma^{13}}^{2}\gamma^{22}}{\gamma^{11}\,\gamma^{12}}}\,, \nonumber \\
&&-1/2\,{\frac {-\gamma^{11}\,\gamma^{12}\,\gamma^{33}-\gamma^{11}\,\gamma^{13}\,\gamma^{23}+2\,\gamma^{12}\,{\gamma^{13}}^{2}}{\gamma^{11}\,\gamma^{12}}}\,, 1/2\,{\frac {-\gamma^{11}\,\gamma^{12}\,\gamma^{23}-\gamma^{13}\,\gamma^{22}\,\gamma^{11}+2\,{\gamma^{12}}^{2}\gamma^{13}}{\gamma^{11}\,\gamma^{12}}}\,, 0\,, 0\,, 0\,, 0\,, 0\,, 0\,, 0\,, 0\,, 0\,, 0\,, \nonumber \\  
&&0\,, 0\,, 0\,, 0\,, 0\,, 0\,, 0\,, 0\,, 0\,,  -{\frac {{\gamma^{13}}^{2}}{\gamma^{11}\,\gamma^{12}}}\,, 0\,, 0\,, {\frac {\gamma^{13}}{\gamma^{12}}}\,, 1\,, 0\,, 0\,, {\frac {\gamma^{12}}{\gamma^{11}}}\,, 0\,, 0\,, 0  \Big) \\
%
\boldsymbol{r}_{35}^T & = &\Big( 0\,, 0\,, 0\,, 0\,, 0\,, 0\,, 0\,, 0\,, 0\,, 0\,, 0\,, 0\,, 0\,, 0\,, 0\,, 0\,, 0\,, 0\,, 0\,, 0\,, 0\,, 0\,, 1/2\,{\frac {{\gamma^{12}}^{2}\gamma^{33}-{\gamma^{13}}^{2}\gamma^{22}}{\gamma^{11}\,\gamma^{12}}}\,, \nonumber \\
&&-1/2\,{\frac {\gamma^{11}\,\gamma^{12}\,\gamma^{33}+\gamma^{11}\,\gamma^{13}\,\gamma^{23}-2\,\gamma^{12}\,{\gamma^{13}}^{2}}{\gamma^{11}\,\gamma^{12}}}\,, 1/2\,{\frac {\gamma^{11}\,\gamma^{12}\,\gamma^{23}+\gamma^{13}\,\gamma^{22}\,\gamma^{11}-2\,{\gamma^{12}}^{2}\gamma^{13}}{\gamma^{11}\,\gamma^{12}}}\,, 0\,, 0\,, 0\,, 0\,, 0\,, 0\,, 0\,, 0\,, 0\,, 0\,, \nonumber \\
&&0\,, 0\,, 0\,, 0\,, 0\,, 0\,, 0\,, 0\,, 0\,, {\frac {{\gamma^{13}}^{2}}{\gamma^{11}\,\gamma^{12}}}\,, 0\,, 0\,, -{\frac {\gamma^{13}}{\gamma^{12}}}\,, 0\,, 0\,, 0\,, -{\frac {\gamma^{12}}{\gamma^{11}}}\,, 0\,, 1\,, 0  \Big) \\
%
\boldsymbol{r}_{36}^T & = &\Big( 0\,, 0\,, 0\,, 0\,, 0\,, 0\,, 0\,, 0\,, 0\,, 0\,, 0\,, 0\,, 0\,, 0\,, 0\,, 0\,, 0\,, 0\,, 0\,, 0\,, 0\,, 0\,, 1/2\,{\frac {-\gamma^{11}\,\gamma^{13}\,\gamma^{23}+\gamma^{13}\,\gamma^{22}\,\gamma^{11}}{\gamma^{11}\,\gamma^{13}}}\,, \nonumber \\
&&-1/2\,{\frac {-{\gamma^{11}}^{2}\gamma^{23}+\gamma^{11}\,\gamma^{13}\,\gamma^{13}}{\gamma^{11}\,\gamma^{13}}}\,, 1/2\,{\frac {-{\gamma^{11}}^{2}\gamma^{22}+\gamma^{11}\,{\gamma^{13}}^{2}}{\gamma^{11}\,\gamma^{13}}}\,, 0\,, 0\,, 0\,, 0\,, 0\,, 0\,, 0\,, 0\,, 0\,, 0\,, 0\,, 0\,, 0\,, 0\,, 0\,, 0\,, 0\,, 0\,, 0\,, \nonumber \\
&&-{\frac {\gamma^{13}}{\gamma^{13}}}\,, 0\,, 0\,, {\frac {\gamma^{11}}{\gamma^{13}}}\,, 0\,, 0\,, 1\,, 0\,, 0\,, 0\,, 0  \Big) \\
%
\boldsymbol{r}_{37}^T & = &\Big( 0\,, 0\,, 0\,, 0\,, 0\,, 0\,, 0\,, 0\,, 0\,, 0\,, 0\,, 0\,, 0\,, 0\,, 0\,, 0\,, 0\,, 0\,, 0\,, 0\,, 0\,, 0\,, 1/2\,{\frac {-\gamma^{11}\,\gamma^{13}\,\gamma^{23}+\gamma^{13}\,\gamma^{22}\,\gamma^{11}}{\gamma^{11}\,\gamma^{13}}}\,, \nonumber \\
&& -1/2\,{\frac {-{\gamma^{11}}^{2}\gamma^{23}+\gamma^{11}\,\gamma^{13}\,\gamma^{13}}{\gamma^{11}\,\gamma^{13}}}\,, 1/2\,{\frac {-{\gamma^{11}}^{2}\gamma^{22}+\gamma^{11}\,{\gamma^{13}}^{2}}{\gamma^{11}\,\gamma^{13}}}\,, 0\,, 0\,, 0\,, 0\,, 0\,, 0\,, 0\,, 0\,, 0\,, 0\,, 0\,, 0\,, 0\,, 0\,, 0\,, 0\,, 0\,, 0\,, 0\,, \nonumber \\
&&-{\frac {\gamma^{13}}{\gamma^{13}}}\,, 1\,, 0\,, {\frac {\gamma^{11}}{\gamma^{13}}}\,, 0\,, 0\,, 0\,, 0\,, 0\,, 0\,, 0  \Big) \\
%
\boldsymbol{r}_{38}^T & = &\Big( 0\,, 0\,, 0\,, 0\,, 0\,, 0\,, 0\,, 0\,, 0\,, 0\,, 0\,, 0\,, 0\,, 0\,, 0\,, 0\,, 0\,, 0\,, 0\,, 0\,, 0\,, 0\,, 1/2\,{\frac {2\,\gamma^{11}\,\gamma^{13}\,\gamma^{23}-2\,\gamma^{13}\,\gamma^{22}\,\gamma^{11}}{\gamma^{11}\,\gamma^{13}}}\,, \nonumber \\
&&-1/2\,{\frac {2\,{\gamma^{11}}^{2}\gamma^{23}-2\,\gamma^{11}\,\gamma^{13}\,\gamma^{13}}{\gamma^{11}\,\gamma^{13}}}\,, 1/2\,{\frac {2\,{\gamma^{11}}^{2}\gamma^{22}-2\,\gamma^{11}\,{\gamma^{13}}^{2}}{\gamma^{11}\,\gamma^{13}}}\,, 0\,, 0\,, 0\,, 0\,, 0\,, 0\,, 0\,, 0\,, 0\,, 0\,, 0\,, 0\,, 0\,, 0\,, 0\,, 0\,, 1\,, 0\,, 0\,, \nonumber \\
&&2\,{\frac {\gamma^{13}}{\gamma^{13}}}\,, 0\,, 0\,, -2\,{\frac {\gamma^{11}}{\gamma^{13}}}\,, 0\,, 0\,, 0\,, 0\,, 0\,, 0\,, 0  \Big) \\
%
\boldsymbol{r}_{39}^T & = &\Big( 0\,, 0\,, 0\,, 0\,, 0\,, 0\,, 0\,, 0\,, 0\,, 0\,, 0\,, 0\,, 0\,, 0\,, 1\,, 0\,, 0\,, 0\,, 0\,, 0\,, 0\,, 0\,, 0\,, 0\,, 0\,, 0\,, 0\,, 0\,, 0\,, 0\,, 0\,, 0\,, 0\,, 0\,, 0\,, 0\,, 0\,, 0\,, 0\,, 0\,, 0\,, 0\,, 0\,, 0\,, \nonumber \\
&&0\,, 0\,, 0\,, 0\,, 0\,, 0\,, 0\,, 0\,, 0\,, 0\,, 0  \Big) \\
%
\boldsymbol{r}_{40}^T & = &\Big( 0\,, 0\,, 0\,, 0\,, 0\,, 0\,, 0\,, 0\,, 0\,, 0\,, 0\,, 0\,, 0\,, 1\,, 0\,, 0\,, 0\,, 0\,, 0\,, 0\,, 0\,, 0\,, 0\,, 0\,, 0\,, 0\,, 0\,, 0\,, 0\,, 0\,, 0\,, 0\,, 0\,, 0\,, 0\,, 0\,, 0\,, 0\,, 0\,, 0\,, 0\,, 0\,, 0\,, 0\,, \nonumber \\
&&0\,, 0\,, 0\,, 0\,, 0\,, 0\,, 0\,, 0\,, 0\,, 0\,, 0  \Big) \\
%
\boldsymbol{r}_{41}^T & = &\Big( 0\,, 0\,, 0\,, 0\,, 0\,, 0\,, 0\,, 0\,, 0\,, 0\,, 0\,, 0\,, 1\,, 0\,, 0\,, 0\,, 0\,, 0\,, 0\,, 0\,, 0\,, 0\,, 0\,, 0\,, 0\,, 0\,, 0\,, 0\,, 0\,, 0\,, 0\,, 0\,, 0\,, 0\,, 0\,, 0\,, 0\,, 0\,, 0\,, 0\,, 0\,, 0\,, 0\,, 0\,, \nonumber \\
&&0\,, 0\,, 0\,, 0\,, 0\,, 0\,, 0\,, 0\,, 0\,, 0\,, 0  \Big) \\
%
\boldsymbol{r}_{42}^T & = &\Big( 0\,, 0\,, 0\,, 0\,, 0\,, 0\,, 0\,, 0\,, 0\,, 0\,, 0\,, 1\,, 0\,, 0\,, 0\,, 0\,, 0\,, 0\,, 0\,, 0\,, 0\,, 0\,, 0\,, 0\,, 0\,, 0\,, 0\,, 0\,, 0\,, 0\,, 0\,, 0\,, 0\,, 0\,, 0\,, 0\,, 0\,, 0\,, 0\,, 0\,, 0\,, 0\,, 0\,, 0\,, \nonumber \\
&&0\,, 0\,, 0\,, 0\,, 0\,, 0\,, 0\,, 0\,, 0\,, 0\,, 0  \Big) \\
%
\boldsymbol{r}_{43}^T & = &\Big( 0\,, 0\,, 0\,, 0\,, 0\,, 0\,, 0\,, 0\,, 0\,, 0\,, 1\,, 0\,, 0\,, 0\,, 0\,, 0\,, 0\,, 0\,, 0\,, 0\,, 0\,, 0\,, 0\,, 0\,, 0\,, 0\,, 0\,, 0\,, 0\,, 0\,, 0\,, 0\,, 0\,, 0\,, 0\,, 0\,, 0\,, 0\,, 0\,, 0\,, 0\,, 0\,, 0\,, 0\,, \nonumber \\
&&0\,, 0\,, 0\,, 0\,, 0\,, 0\,, 0\,, 0\,, 0\,, 0\,, 0  \Big) \\
%
\boldsymbol{r}_{44}^T & = &\Big( 0\,, 0\,, 0\,, 0\,, 0\,, 0\,, 0\,, 0\,, 0\,, 1\,, 0\,, 0\,, 0\,, 0\,, 0\,, 0\,, 0\,, 0\,, 0\,, 0\,, 0\,, 0\,, 0\,, 0\,, 0\,, 0\,, 0\,, 0\,, 0\,, 0\,, 0\,, 0\,, 0\,, 0\,, 0\,, 0\,, 0\,, 0\,, 0\,, 0\,, 0\,, 0\,, 0\,, 0\,, \nonumber \\
&&0\,, 0\,, 0\,, 0\,, 0\,, 0\,, 0\,, 0\,, 0\,, 0\,, 0  \Big) \\
%
\boldsymbol{r}_{45}^T & = &\Big( 0\,, 0\,, 0\,, 0\,, 0\,, 0\,, 0\,, 0\,, 1\,, 0\,, 0\,, 0\,, 0\,, 0\,, 0\,, 0\,, 0\,, 0\,, 0\,, 0\,, 0\,, 0\,, 0\,, 0\,, 0\,, 0\,, 0\,, 0\,, 0\,, 0\,, 0\,, 0\,, 0\,, 0\,, 0\,, 0\,, 0\,, 0\,, 0\,, 0\,, 0\,, 0\,, 0\,, 0\,, \nonumber \\
&&0\,, 0\,, 0\,, 0\,, 0\,, 0\,, 0\,, 0\,, 0\,, 0\,, 0  \Big) \\
%
\boldsymbol{r}_{46}^T & = &\Big( 0\,, 0\,, 0\,, 0\,, 0\,, 0\,, 0\,, 1\,, 0\,, 0\,, 0\,, 0\,, 0\,, 0\,, 0\,, 0\,, 0\,, 0\,, 0\,, 0\,, 0\,, 0\,, 0\,, 0\,, 0\,, 0\,, 0\,, 0\,, 0\,, 0\,, 0\,, 0\,, 0\,, 0\,, 0\,, 0\,, 0\,, 0\,, 0\,, 0\,, 0\,, 0\,, 0\,, 0\,, \nonumber \\
&&0\,, 0\,, 0\,, 0\,, 0\,, 0\,, 0\,, 0\,, 0\,, 0\,, 0  \Big) \\
%
\boldsymbol{r}_{47}^T & = &\Big( 0\,, 0\,, 0\,, 0\,, 0\,, 0\,, 1\,, 0\,, 0\,, 0\,, 0\,, 0\,, 0\,, 0\,, 0\,, 0\,, 0\,, 0\,, 0\,, 0\,, 0\,, 0\,, 0\,, 0\,, 0\,, 0\,, 0\,, 0\,, 0\,, 0\,, 0\,, 0\,, 0\,, 0\,, 0\,, 0\,, 0\,, 0\,, 0\,, 0\,, 0\,, 0\,, 0\,, 0\,, \nonumber \\
&&0\,, 0\,, 0\,, 0\,, 0\,, 0\,, 0\,, 0\,, 0\,, 0\,, 0  \Big) \\
%
\boldsymbol{r}_{48}^T & = &\Big( 0\,, 0\,, 0\,, 0\,, 0\,, 1\,, 0\,, 0\,, 0\,, 0\,, 0\,, 0\,, 0\,, 0\,, 0\,, 0\,, 0\,, 0\,, 0\,, 0\,, 0\,, 0\,, 0\,, 0\,, 0\,, 0\,, 0\,, 0\,, 0\,, 0\,, 0\,, 0\,, 0\,, 0\,, 0\,, 0\,, 0\,, 0\,, 0\,, 0\,, 0\,, 0\,, 0\,, 0\,, \nonumber \\
&&0\,, 0\,, 0\,, 0\,, 0\,, 0\,, 0\,, 0\,, 0\,, 0\,, 0  \Big) \\
%
\boldsymbol{r}_{49}^T & = &\Big( 0\,, 0\,, 0\,, 0\,, 1\,, 0\,, 0\,, 0\,, 0\,, 0\,, 0\,, 0\,, 0\,, 0\,, 0\,, 0\,, 0\,, 0\,, 0\,, 0\,, 0\,, 0\,, 0\,, 0\,, 0\,, 0\,, 0\,, 0\,, 0\,, 0\,, 0\,, 0\,, 0\,, 0\,, 0\,, 0\,, 0\,, 0\,, 0\,, 0\,, 0\,, 0\,, 0\,, 0\,, \nonumber \\
&&0\,, 0\,, 0\,, 0\,, 0\,, 0\,, 0\,, 0\,, 0\,, 0\,, 0  \Big) \\
%
\boldsymbol{r}_{50}^T & = &\Big( 0\,, 0\,, 0\,, 1\,, 0\,, 0\,, 0\,, 0\,, 0\,, 0\,, 0\,, 0\,, 0\,, 0\,, 0\,, 0\,, 0\,, 0\,, 0\,, 0\,, 0\,, 0\,, 0\,, 0\,, 0\,, 0\,, 0\,, 0\,, 0\,, 0\,, 0\,, 0\,, 0\,, 0\,, 0\,, 0\,, 0\,, 0\,, 0\,, 0\,, 0\,, 0\,, 0\,, 0\,, \nonumber \\
&&0\,, 0\,, 0\,, 0\,, 0\,, 0\,, 0\,, 0\,, 0\,, 0\,, 0  \Big) \\
%
\boldsymbol{r}_{51}^T & = &\Big( 0\,, 0\,, 1\,, 0\,, 0\,, 0\,, 0\,, 0\,, 0\,, 0\,, 0\,, 0\,, 0\,, 0\,, 0\,, 0\,, 0\,, 0\,, 0\,, 0\,, 0\,, 0\,, 0\,, 0\,, 0\,, 0\,, 0\,, 0\,, 0\,, 0\,, 0\,, 0\,, 0\,, 0\,, 0\,, 0\,, 0\,, 0\,, 0\,, 0\,, 0\,, 0\,, 0\,, 0\,, \nonumber \\
&&0\,, 0\,, 0\,, 0\,, 0\,, 0\,, 0\,, 0\,, 0\,, 0\,, 0  \Big) \\
%
\boldsymbol{r}_{52}^T & = &\Big( 0\,, 1\,, 0\,, 0\,, 0\,, 0\,, 0\,, 0\,, 0\,, 0\,, 0\,, 0\,, 0\,, 0\,, 0\,, 0\,, 0\,, 0\,, 0\,, 0\,, 0\,, 0\,, 0\,, 0\,, 0\,, 0\,, 0\,, 0\,, 0\,, 0\,, 0\,, 0\,, 0\,, 0\,, 0\,, 0\,, 0\,, 0\,, 0\,, 0\,, 0\,, 0\,, 0\,, 0\,, \nonumber \\
&&0\,, 0\,, 0\,, 0\,, 0\,, 0\,, 0\,, 0\,, 0\,, 0\,, 0  \Big) \\
%
\boldsymbol{r}_{53}^T & = &\Big( 1\,, 0\,, 0\,, 0\,, 0\,, 0\,, 0\,, 0\,, 0\,, 0\,, 0\,, 0\,, 0\,, 0\,, 0\,, 0\,, 0\,, 0\,, 0\,, 0\,, 0\,, 0\,, 0\,, 0\,, 0\,, 0\,, 0\,, 0\,, 0\,, 0\,, 0\,, 0\,, 0\,, 0\,, 0\,, 0\,, 0\,, 0\,, 0\,, 0\,, 0\,, 0\,, 0\,, 0\,, \nonumber \\
&&0\,, 0\,, 0\,, 0\,, 0\,, 0\,, 0\,, 0\,, 0\,, 0\,, 0  \Big) \\
%
\boldsymbol{r}_{54}^T & = &\Big( 0\,, 0\,, 0\,, 0\,, 0\,, 0\,, 0\,, 0\,, 0\,, 0\,, 0\,, 0\,, 0\,, 0\,, 0\,, 1/2\,{\frac {\sqrt {2}\sqrt {\alpha\,\gamma^{11}}}{\gamma^{11}}}\,, 0\,, 0\,, 0\,, 0\,, 0\,, 0\,, 0\,, 0\,, 0\,, 1\,, 0\,, 0\,, 0\,, 0\,, 0\,, 0\,, 0\,, \nonumber \\
&&0\,, 0\,, 0\,, 0\,, 1/2\,{\frac {\alpha}{\gamma^{11}}}\,, 0\,, 0\,, 0\,, 0\,, 0\,, 0\,, 0\,, 0\,, 0\,, 0\,, 0\,, 0\,, 0\,, 0\,, 0\,, 0\,, 0  \Big) \\
%
\boldsymbol{r}_{55}^T & = &\Big( 0\,, 0\,, 0\,, 0\,, 0\,, 0\,, 0\,, 0\,, 0\,, 0\,, 0\,, 0\,, 0\,, 0\,, 0\,, -1/2\,{\frac {\sqrt {2}\sqrt {\alpha\,\gamma^{11}}}{\gamma^{11}}}\,, 0\,, 0\,, 0\,, 0\,, 0\,, 0\,, 0\,, 0\,, 0\,, 1\,, 0\,, 0\,, 0\,, 0\,, 0\,, 0\,, 0\,, \nonumber \\
&&0\,, 0\,, 0\,, 0\,, 1/2\,{\frac {\alpha}{\gamma^{11}}}\,, 0\,, 0\,, 0\,, 0\,, 0\,, 0\,, 0\,, 0\,, 0\,, 0\,, 0\,, 0\,, 0\,, 0\,, 0\,, 0\,, 0  \Big) 
\end{eqnarray}
\

\bibliographystyle{plain}
\bibliography{./referencesZ4_B.bib}

\end{document}